\documentclass{ws-rv}
\usepackage{lscape}
\usepackage{ws-rv-van}     
\usepackage{ws-rv-thm}     
\usepackage{subfigure}     
\makeindex

\renewcommand{\vec}[1]{\mbox{\boldmath$#1$}}
\def\beq{\begin{equation}}
\def\eeq{\end{equation}}
\def\be{\begin{equation}}
\def\ee{\end{equation}}

\begin{document}

\chapter[Composite Fermions @ 30]{Thirty Years of Composite Fermions and Beyond
\label{ra_ch1}}

\author[J. K. Jain]{J. K. Jain}

\address{Physics Department, 104 Davey Laboratory, Pennsylvania State University, University Park, Pennsylvania 16802, USA
}

\begin{abstract} 
This chapter appears in {\it Fractional Quantum Hall Effects: New Developments}, edited by B. I. Halperin and J. K. Jain (World Scientific, 2020). The chapter begins with a primer on composite fermions, and then reviews three directions that have recently been pursued. It reports on theoretical calculations making detailed quantitative predictions for two sets of phenomena, namely spin polarization transitions and the phase diagram of the crystal. This is followed by the Kohn-Sham density functional theory of the fractional quantum Hall effect. The chapter concludes with recent applications of the parton theory of the fractional quantum Hall effect to certain delicate states. 
\end{abstract}


\body

\tableofcontents

\section{The mystery of the fractional quantum Hall effect}

\begin{figure}[t]
\begin{center}
\includegraphics[width=0.8\textwidth]{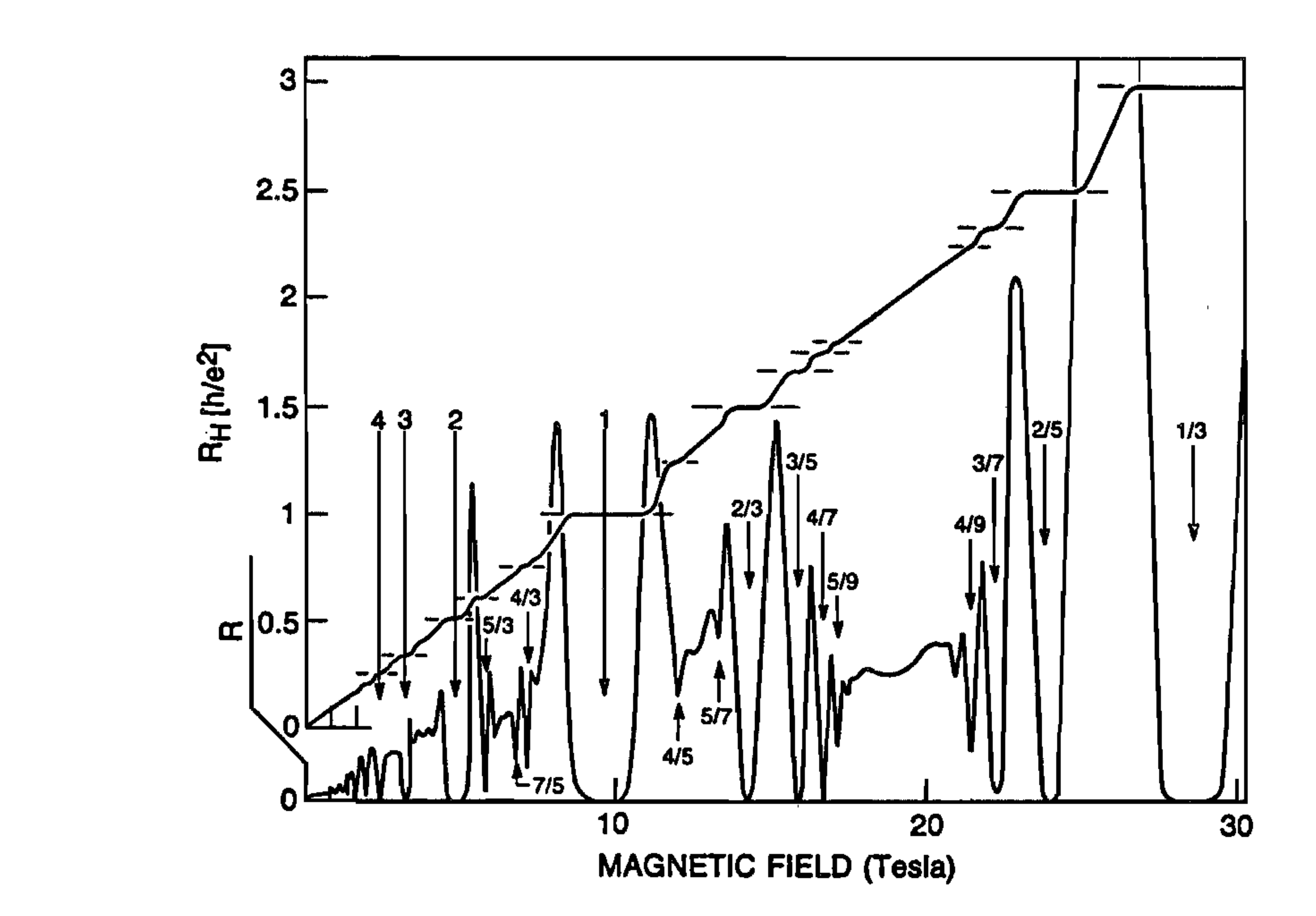}
\caption{The Hall and the longitudinal resistances, $R_{\rm H}$ and $R_{\rm L}$, respectively. The fractions associated with the plateaus (or the resistance minima) are indicated. Source: H. L. Stormer and D. C. Tsui, ``Composite fermions in the fractional quantum hall effect," in {\em Perspectives in Quantum Hall Effects}, pp. 385-421 (Wiley-VCH Verlag GmbH, 2007)~\cite{Stormer07}.}
\end{center}
\label{fig:stormer98}
\end{figure}

The fractional quantum Hall effect (FQHE) is among the most stunning manifestations of quantum mechanics at macroscopic scales (Fig.~\ref{fig:stormer98}).  It occurs when electrons are driven into an extreme quantum corner by confining them to two dimensions, cooling them down to very low temperatures, and exposing them to a strong magnetic field. The term FQHE does not refer to a single observation but encompasses a myriad of non-trivial states and phenomena. A fractional quantum Hall (FQH) state is characterized by a precisely quantized plateau in the Hall resistance at $R_{\rm H}=h/fe^2$, where $f$ is a fraction, approximately centered at the Landau level (LL) filling factor $\nu= f$. (The nominal number of filled LLs, called the filling factor, is given by $\nu=\rho\phi_0/B$, where  $\rho$ is the density, $\phi_0=hc/e$ is called the flux quantum, and $B$ is the magnetic field. See Appendix~\ref{sec:symmetric}.) The plateau in $R_{\rm H}$ is accompanied by a minimum in longitudinal resistance $R_{\rm L}$, which vanishes as $R_{\rm L}\sim e^{-\Delta/2k_BT}$  as the temperature tends to zero, indicating the presence of a gap $\Delta$ in the excitation spectrum. To date, close to 100 fractions have been observed in the best quality samples. The number of FQH states is greater than the number of observed fractions because, in general, many distinct FQH states can occur at a given fraction, differing in their spin polarization, valley polarization or some other quantum number. Experimentalists have measured the energy gaps, collective modes, spin polarizations, spin wave excitations, transport coefficients, thermal Hall effect, etc. for many of these FQH states as a function of density, quantum well width, temperature, and the Zeeman energy. Measurements have been performed in two-dimensional and also bilayer systems made of a variety of materials, such as GaAs, AlAs and ZnO quantum wells, heterostructures, and graphene. The FQHE is a data rich field. 

To bring out the non-triviality of these observations it is helpful to introduce the ``minimal" model Hamiltonian for the FQHE:
\begin{equation}
H=\sum_{j<k=1}^{N}{1 \over |\vec{r}_j-\vec{r}_k|}\;\; ({\rm LLL}\;\;{\rm subspace)} \;,
\label{idealH}
\end{equation}
which describes a two-dimensional system of electrons confined to the lowest LL (LLL).
We have used the magnetic length $l=\sqrt{\hbar c/eB}$ as the unit of length and $e^2/\epsilon l$ as the unit of energy ($\epsilon$ is the dielectric constant of the background material), and suppressed the term representing interaction  with a uniform positively charged background. In writing Eq.~\ref{idealH} we have assumed $\nu<1$ and the limit of very high magnetic field, $\kappa\equiv (e^2/\epsilon l)/\hbar\omega_c \rightarrow 0$, where $\hbar \omega_c=\hbar eB/m_b c$ is the cyclotron energy ($m_b$ is the electron band mass). In this limit the interaction is unable to cause LL mixing and, hence, electrons are strictly confined to the LLL.  This Hamiltonian, which is to be solved within the Hilbert space of the LLL states,\footnote{For states in a different LL, this Hamiltonian needs to be solved within the Hilbert space of that LL. The matrix elements of the Coulomb interaction depend on the LL index and thus produce different behaviors in different LLs.} has been stripped off of all features that are inessential to the FQH physics. In particular, the quantum-well width, LL mixing and disorder have all been set to zero in Eq.~\ref{idealH}; these cause quantitative corrections but are not necessary for the phenomenon of the FQHE.  For the same reason, the periodic potential due to the lattice has also been neglected, which is justified because the magnetic length, which controls the size of the wave function, is large compared to the lattice constant. The minimal Hamiltonian clarifies, in essence, that the physics of FQHE is governed by the Coulomb interaction alone. It is also noteworthy that the minimal model contains no free parameters, i.e., all sample specific parameters (e.g. the dielectric constant) can be absorbed into the measurement units. The FQHE is actually the most strongly correlated state in the world: the strength of correlations is measured by the ratio of the interaction energy to the kinetic energy, and the latter is absent here. 

At the most fundamental level, the puzzle of the FQHE may be stated as follows. In the absence of interaction, all configurations (that is, all Slater determinant basis functions) of electrons in the LLL are degenerate ground states. There are very many of them. Even for a small system, say $N=100$ electrons at $\nu=1/3$, the number of degenerate ground states is ${300\choose 100}\sim 10^{83}$, which is on the order of the number of quarks in the entire Universe. With so many choices, the electrons in the LLL are enormously frustrated. At the same time, the observed phenomenology is telling us that the system is on the verge of a spectacular non-perturbative reorganization as soon as the repulsive Coulomb interaction is turned on. In particular, the observation of FQHE implies that at certain special filling factors nature conspires to eliminate the astronomical degeneracy to yield unique, non-degenerate ground states, which are certain entangled linear superpositions of all of the basis functions. 
This raises many questions. What is the organizing principle? What is the mechanism of the FQHE? What makes certain filling factors special? What is unique about the ground states at these fractions? What are their wave functions, and what physics do they represent? What are their excitations? What role does the spin degree of freedom play? What is the quantitative theory? How do gaps depend on the filling factor? What are the neutral collective modes and their dispersions? ... Finally, what other surprising phenomena lurk around the corner?

It turns out that we theorists can add to the wealth of FQHE data by performing our own experiments on the computer. A system on the computer is fully defined by two integers\footnote{This statement refers to the so-called spherical geometry, in which electrons move on the surface of a sphere subject to a radial magnetic field. In the periodic (torus) geometry, the aspect ratio (defined by the modular parameter) and the quasi-periodic boundary conditions are additional variables.}: the number of electrons ($N$) and the number of magnetic flux quanta ($2Q$) to which they are exposed.  The dimension of the Hilbert space is finite for a given $(N, 2Q)$ system (assuming the LLL constraint), and when it is not too large, a brute force diagonalization can be performed to obtain the exact eigenstates and eigenenergies. This information exists for hundreds of $(N, 2Q)$ systems, typically with $N <18-20$ for today's computer, producing tens of thousands of exact eigenstates and eigenenergies. While the laboratory experiments present us with a few correlation and response functions, the computer experiments deliver the complete genomes of miniature FQH systems in the form of long lists of numbers that represent projections of all eigenstates along all directions in the very large Hilbert space. The availability of exact solutions for small systems is a powerful feature of the FQHE, because it allows a detailed and unbiased testing of any candidate theory.

The reader will surely not be surprised to learn that an exact analytical solution of Eq.~\ref{idealH}, which gives all eigenfunctions and eigenenergies for all filling factors, does not exist. It is a certain bet that such a solution will never be found\footnote{It is possible to construct short range model interactions that produce certain simple FQH wave functions as exact zero-energy ground states~\cite{Haldane83}. See the Chapter by Steve Simon for examples. It should be noted, however, that these model interactions are constructed for already known wave functions; they are not solvable for excited states; and different model interactions are needed for different wave functions.}. That may not worry a practitioner of condensed matter physics. After all, a satisfactory understanding of certain other systems of interacting electrons has been achieved without an exact solution. There is an important difference from these other systems, however. To illustrate, let us take the example of a weakly-coupled superconductor. Its understanding relies fundamentally on the availability of a  ``normal sate," namely the Fermi sea, which is obtained when we switch off the interaction between electrons. This provides a unique and well-defined starting point. The minimal model for superconductivity, due to Bardeen, Cooper and Schrieffer (BCS), considers electrons with a weak attractive interaction (with strength small compared to the Fermi energy), and explains superconductivity as a pairing instability of the Fermi sea as a result of this interaction. This instability involves a rearrangement of electrons only in a narrow sliver near the Fermi energy. In contrast, there is no normal state for the FQHE. Switching off the interaction produces not a unique state but a large number of degenerate ground states. The FQHE cannot be understood as an instability of a known state. The absence of a natural starting point coupled with the fact that the Coulomb interaction is not small compared to any other energy scale makes the FQH problem intractable to the usual perturbative or quasi-perturbative treatments. 

How do we proceed, then? As always, the goal of theory is to identify the simple underlying principles that provide a unified explanation of the complex behavior displayed by the interacting system. These principles should provide an intuitive understanding of the qualitative features of the phenomenology, and at the same time guide us toward a quantitative theory that is necessary for a detailed confirmation. Section~\ref{sec:cftheory} describes the unfolding of many important experimental facts and theoretical ideas in the 1980s that led to the postulate that nature relieves the frustration, i.e. eliminates the degeneracy of the partially occupied LLL, by creating a new kind of topological particles called composite fermions, which themselves can be taken as weakly interacting for many purposes. (In other words, the non-perturbative role of  the repulsive interaction is to produce composite fermions; the rest is perturbative.) Section~\ref{sec:cftheory}  provides a pedagogical introduction to the foundations of the composite fermion (CF) theory as well as its prominent verifications.  Section~\ref{sec:laboratory} reports on detailed quantitative comparisons of the experimentally observed phase diagram of the spin polarization and the interplay between the crystal phase and the FQHE with theoretical calculations including the effects of finite quantum well width and LL mixing.  Section \ref{sec:KS*} shows how the Kohn-Sham density functional theory can be formulated for the strongly correlated FQH state by exploiting the CF physics. The chapter concludes in Section \ref{sec:parton} with the parton theory of the FQHE, which produces states beyond the CF theory, including many non-Abelian states (i.e. states that support quasiparticles obeying non-Abelian braid statistics). This section also gives a brief account of recent work indicating that some of these are plausible candidates for certain delicate states observed in higher GaAs or graphene LLs and in the LLL in wide quantum wells.

\section{Composite fermions: A primer}
\label{sec:cftheory}

This section contains an introduction to the essentials of the CF theory. A newcomer to the field may find it useful for the remainder of this chapter, and, perhaps, also for some other chapters in the book.

\subsection{Background}
\label{sec:background}

The birth of the field was announced by the discovery of the integer quantum Hall effect (IQHE) by von Klitzing in 1980~\cite{Klitzing80}, which, in hindsight, marked the beginning of the topological revolution in modern condensed matter physics. Von Klitzing observed that the Hall resistance is precisely quantized at $R_{\rm H}=h/ie^2$, where $i$ is an integer, with the plateau occurring in the vicinity of filling factor $\nu\approx i$.  The quantization is exact as far as we now know, and the equality of the resistance on the $i=1$ plateau in different samples has been established to an extremely high precision (a few parts in ten billion). The longitudinal resistance $R_{\rm L}$ shows a minimum at $\nu=i$, behaving as $R_{\rm L}\sim \exp(-\Delta/2k_{\rm B}T)$ as a function of temperature $T$. A gap $\Delta$ can be extracted from the temperature dependence of the longitudinal resistance. The most remarkable aspect of the IQHE is the universality of the quantization, which is utterly oblivious to the details such as which two-dimensional (2D) system is being used, what is the sample size or geometry, what band structure electrons occupy, what is their effective mass, or the nature or strength of disorder. The IQHE was not predicted, but was almost immediately explained by Laughlin~\cite{Laughlin81} in 1981 as a consequence of the formation of LLs combined with disorder induced Anderson localization of states. Soon thereafter in 1982, Thouless {\em et al.}~\cite{Thouless82} related the Hall conductance to a topological quantity known as the Chern number, and a few years later Haldane~\cite{Haldane88} showed that bands with non-zero Chern numbers do not require a uniform external magnetic field. These works later served as inspiration for the field of topological insulators.

With the IQHE explained, the story seemed complete, and Tsui, Stormer and Gossard~\cite{Tsui82} set out to look for the Wigner crystal~\cite{Wigner34}. These authors' aim was to expose electrons to such high magnetic fields that they are all forced into the LLL. With their kinetic energy thus quenched, it is left entirely to the Coulomb repulsion to determine their state. What else could the electrons do but form a crystal~\cite{Lozovik75}? In 1982 Tsui, Stormer and Gossard discovered instead a Hall plateau quantized at $R_{\rm H}=h/(1/3)e^2$. This was not anticipated by any theory. 

Laughlin again made a quick breakthrough in 1983~\cite{Laughlin83}. He began by noting that a general  LLL wave function must have the form $\Psi=F(\{ z_j \})\exp(-\sum_{i}|z_{i}|^{2}/4)$, where $z_j=x_j-iy_j$ represents the coordinates of the $j$th electron as a complex number and $F(\{ z_j \})$ is a holomorphic function of $z_j$'s that is antisymmetric under exchange of two particles. (See Appendix~\ref{sec:symmetric}.) He then considered a Jastrow form $F(\{ z_j \})=\prod_{j<k}f(z_j-z_k)$, which builds in pairwise correlations and has been found to be useful in the studies of helium superfluidity. Imposing the conditions of antisymmetry under particle exchange and a well defined total angular momentum fixes $f(z_j-z_k)=(z_j-z_k)^m$, where $m$ is an odd integer. That leads to the wave function  
\be
\Psi_{1/m}=\prod_{1\leq j<k\leq N}(z_{j}-z_{k})^m\;
\exp\left[-\frac{1}{4}\sum_{i}|z_{i}|^{2}\right]\;\;.
\label{Laughlin}
\ee
This wave function describes a state at $\nu=1/m$, and has been found to be an excellent representation of the exact ground state at $\nu=1/3$ obtained in computer studies (results shown below). Laughlin postulated that it represents an incompressible state, i.e. it takes a non-zero energy to create an excitation of this state. With a flux insertion argument, he showed that the elementary excitation of this state has a fractional charge of magnitude $e/m$ relative to the ground state (this argument actually relies only on the incompressibility of the state, not on the microscopic physics of incompressibility). He further wrote an ansatz wave function for the positively charged quasihole located at $z_0$ as
\be
\Psi^{\rm quasihole}_{1/m}=\prod_{k=1}^N(z_k-z_0) \Psi_{1/m}.
\label{Laughlinquasihole}
\ee
Laughlin also suggested a wave function for the negatively charged quasiparticle, but a better wave function for it is now available.

At this stage in early 1983 the story again seemed both elegant and complete. It only remained to test the Laughlin wave function, to measure the fractional charge of the excitations, and to look for a plateau quantized at $1/5$. Subsequent exploration showed, however, that the 1/3 plateau was only the tip of the iceberg. Over the next few years, as experimentalists improved the conditions by removing dirt and thermal fluctuations, a deluge of new fractions revealed a large structure that was not a part of Laughlin's theory.

In a parallel development, the concept of particles obeying fractional braid statistics in two dimensions was being pursued, which subsequently played an important role in the theory of the FQHE.  This possibility was introduced by Leinaas and Myrheim~\cite{Leinaas77}, and by Wilczek~\cite{Wilczek82} who christened these particles anyons. These particles are defined by the property that a closed loop of one particle around another has a non-trivial path-independent phase associated with it. (This is referred to as statistics because an exchange of two particles can be viewed as half a loop of one particle around another followed by a rigid translation.) Anyons can be defined only in two dimensions, because here, if one removes particle coincidences (say, by assuming an infinitely strong hard core repulsion), then each particle sees punctures at the positions of all other particles, and a closed path that encloses another particle cannot be continuously deformed into a path that does not. Wilczek~\cite{Wilczek82} modeled anyons as charged bosons or fermions with gauge flux tubes bound to them carrying a flux $\alpha \phi_0$; the statistical phase then arises as the Aharonov-Bohm (AB) phase due to the bound flux. The list of particles in a particle-physics text book does not contain any anyons, but nothing precludes the possibility that certain emergent particles in a strongly correlated condensed-matter system may behave as anyons. Nature seemed to oblige almost immediately. Halperin~\cite{Halperin84} proposed that Laughlin's quasiholes are realizations of anyons, which was confirmed by Arovas, Schrieffer and Wilczek in an explicit Berry phase calculation~\cite{Arovas84}. 

In what is known as the hierarchy theory, Haldane~\cite{Haldane83} and Halperin~\cite{Halperin84} 
sought to understand the general FQH states based on the paradigm of the Laughlin sates. The Laughlin fraction $\nu=1/m$ serves as the point of departure. As the filling factor is varied away from $\nu=1/m$, quasiparticles or quasiholes are created. A natural approach, in the spirit of the Landau theory of Fermi liquids, is to view the system in the vicinity of $\nu=1/m$ in terms of a state of these quasiparticles or quasiholes. The hierarchy approach considers the possibility that these may form their own Laughlin-like states to produce new daughter incompressible states, which would happen provided that the interaction between the quasiparticles or quasiholes is repulsive with the short-distance part dominating. Beginning with the daughter states, their own quasiparticles or quasiholes (which have different charges and braid statistics than those of the $\nu=1/m$ state) may produce, again provided that their interaction has the appropriate form, grand-daughter FQH states. A continuation of this family tree ad infinitum suggests the possibility, in principle, of FQHE at all odd denominator fractions. 

Important ideas were proposed to address the question of what makes the Laughlin wave function special. A key property of this wave function is that it has no wasted zeros, that is, when viewed as the function of a single coordinate, say $z_1$, all of the zeros of the polynomial part of the wave function $\prod_{j<k}(z_j-z_k)^m$ are located on the other particles. This follows from the fundamental theorem of algebra: a simple power counting shows that the wave function, viewed as a function of one coordinate, is a polynomial of degree $m(N-1)$, i.e. has $m(N-1)$ zeros, which are all accounted for by the $m$ zeros on each of the remaining $N-1$ particles.\footnote{The property of ``no wasted zeros" cannot be satisfied for fractions other than $\nu=1/m$. For example, an electron in the wave function of the $\nu=2/5$ state sees, neglecting order one corrections, 5N/2 zeros, only $N$ of which are located at the other electrons.} Because of the holomorphic property of the wave function, each zero is actually a vortex, that is, it has a phase $2\pi$ associated for any closed loop around it. Building upon this observation and Wilczek's flux attachment idea, Girvin and MacDonald~\cite{Girvin87} introduced a singular gauge transformation that attaches an odd number ($m$) of gauge flux quanta to each electron to convert the Laughlin wave function into a bosonic wave function that is everywhere real and non-negative and also has algebraic off-diagonal long-range order. Zhang, Hansson and Kivelson~\cite{Zhang89} formulated a Chern-Simons (CS) field theory for the $\nu=1/m$ state in which the singular gauge transformation is implemented through a CS term. In a mean field approximation, the effect of the external magnetic field is canceled by the $m$ flux quanta bound to the bosons, thus producing a system of bosons in a zero effective magnetic field; the FQHE of electrons at $\nu=1/m$ thus appears as a Bose-Einstein condensation of these bosons~\cite{Zhang89}.

\subsection{Postulates of the CF theory}

The motivation for the CF theory came from the following observation: If you mentally erase all numbers in Fig.~\ref{fig:stormer98}, you will notice that it is impossible to tell the FQHE from the IQHE. All plateaus look qualitatively identical. This observation suggests a deep connection between the FQHE and the IQHE. Can the well understood IQHE serve as the paradigm for understanding the FQHE? This question inspired the proposal that a new kind of fermions are formed, and their IQHE manifests as the FQHE of electrons~\cite{Jain89,Jain07}.

\begin{figure}[t]
\begin{center}
\includegraphics[width=5.0in]{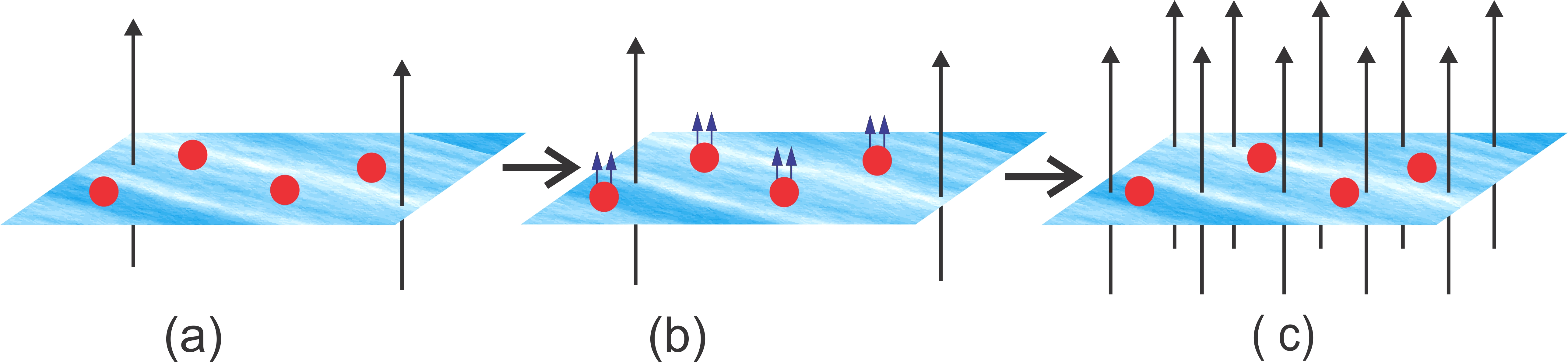}
\end{center}
\caption{Deriving FQHE from the IQHE through composite-fermionization.  We  (a) begin with 
an integer quantum Hall state at $\nu^*=n$, (b) attach two magnetic flux quanta to each electron to convert it into a composite fermion, and (c) spread out the attached flux to obtain electrons in a higher magnetic field. If the gap does not close during the flux smearing process, it produces a FQH state at $\nu=n/(2n+1)$.  More generally, allowing the initial magnetic field to be positive or negative, i.e. $\nu^*=\pm n$, and attaching $2p$ flux quanta produces FQHE at $\nu=n/(2pn\pm 1)$.
\label{fig:jeon-map1}}
\end{figure}

The intuitive idea, explained in Fig.~\ref{fig:jeon-map1}, is as follows~\cite{Jain89}. Let us begin with the integer quantum Hall (IQH) state of non-interacting electrons at $\nu^*=\pm n$ in a magnetic field $B^*=\rho\phi_0 / \nu^*$.  The sign of $B^*$ indicates whether it is pointing in the positive or negative $z$ direction.  Now we attach to each electron an infinitely thin, massless magnetic solenoid carrying $2p$ flux quanta pointing in $+z$ direction.  The bound state of an electron and $2p$ flux quanta is called a composite fermion\footnote{The bound state of an electron and a flux is a model of an anyon~\cite{Wilczek82}. When the flux is an even integer number of flux quanta, the bound state comes a full circle into a fermion.}. The flux added in this manner is unobservable. To see this, consider the Feynman path integral calculation of the partition function, which receives contributions from all closed paths in the configuration space for which the initial and the final positions of electrons are identical, although the paths may involve fermion exchanges, which produces an additional sign $(-1)^P$ for $P$ pairwise exchanges. The excess or deficit of an integral number of flux quanta through any closed path changes the phases only by an integer multiple of $2\pi$ and thus leaves the phase factors unaltered, and the fermionic nature of particles guarantees that the phase factors of paths involving particle exchanges also remain invariant. The new problem defined in terms of composite fermions is thus identical (or dual) to the original problem of non-interacting electrons at $B^*$. The middle panel of Fig.~\ref{fig:jeon-map1} thus represents the $\nu^*=\pm n$ integer quantum Hall (IQH) state of composite fermions in magnetic field $B^*$. (The quantities corresponding to composite fermions are conventionally marked by an asterisk or the superscript CF.)

This exact reformulation prepares the problem for a mean-field approximation that was not available in the original 
language.  Let us adiabatically (i.e., slowly compared to $\hbar/\Delta$, where $\Delta$ is the gap) smear the flux attached to each electron until it becomes a part of the uniform magnetic field. At the end, we obtain particles moving in an enhanced magnetic field
\be
B=B^*+ 2p\rho\phi_0,
\ee
which is identified with the real applied magnetic field. This implies
\be
\nu=\frac{n}{2pn\pm 1}\;,
\label{jains}
\ee
where $\pm$ corresponds to the CF filling $\nu^*=\pm n$.
If the gap does not close during the flux smearing process, i.e., if there is no phase transition, then we have obtained a candidate incompressible state at a fractional filling factor. To be sure, we know from general considerations that the system must undergo a complex evolution through the flux smearing process. The cyclotron energy gap of the IQHE must somehow evolve into an entirely interaction induced gap, and the wave function of $n$ filled LLs into a LLL wave function. The electron mass, which is not a parameter of the LLL problem, is not simply renormalized but must be altogether eliminated during the above process. A satisfactory quantitative description of the evolution of the interacting ground state as the attached flux is spread from point flux to a uniform magnetic field is not known.

To make further progress, we abandon the idea of theoretically implementing the flux smearing process, but rather use the above physics as an inspiration to make an ansaz directly for the final state. A mean field theory suggests~\cite{Jain89} 
\begin{equation}
\Psi^{\rm MF}_{\nu={n\over 2pn\pm  1}}=\prod_{j<k} \left( {z_j-z_k\over |z_j-z_k|}  \right)^{2p}\Phi_{\pm n}\;,
\end{equation}
where the multiplicative factor is a pure phase factor associated with $2p$ flux quanta bound to electrons. Here $\Phi_{- n}=[\Phi_{n}]^*$ is the wave function of $n$ filled LLs in a negative magnetic field, and the magnetic length in the gaussian factor of $\Phi_{\pm n}$ is chosen so as to ensure that the wave function $\Psi$ describes a state at the desired filling factor. A little thought shows that this wave function has serious deficiencies: it does not build good correlations, as can be seen from the fact that 
$|\Psi^{\rm MF}_{\nu={n\over 2pn\pm 1}}|=|\Phi_{\pm n}|$; it has a large admixture with higher LLs; and for $\nu=1/(2p+1)$ it produces the wave function $\Psi_{1/(2p+1)}\sim \prod_{j<k} (z_j-z_k)^{2p+1}/|z_j-z_k|^{2p}$, where we have used $\Phi_1\sim \prod_{j<k} (z_j-z_k)$ (suppressing the ubiquitous Gaussian factors for notational ease), rather than the Laughlin wave function. Many of these problems are eliminated by dropping the denominator~\cite{Jain89}, which does not alter the topological structure. That gives:
\begin{equation}
\Psi^{\rm unproj}_{\nu={n\over 2pn\pm 1}}=\prod_{j<k} \left( {z_j-z_k}  \right)^{2p}\Phi_{\pm n}\;.
\label{jainwf1}
\end{equation}
This wave function explicitly builds good correlations for repulsive interactions, because the configurations wherein two particles approach close to one another have probability vanishing as $r^{4p+2}$, where $r$ is the distance between them, and are thus strongly suppressed. For $\nu=1/(2p+1)$ we recover the Laughlin wave function $\Psi_{1/(2p+1)}\sim \prod_{j<k} (z_j-z_k)^{2p+1}$, but with the new physical interpretation as the $\nu^*=1$ IQH state of composite fermions. Going from $\Psi^{\rm MF}$ to $\Psi^{\rm unproj}$ also significantly reduces admixture with higher LLs, producing wave functions that are predominantly in the LLL as measured by their kinetic energy~\cite{Trivedi91,Kamilla97b}. Because strictly LLL wave functions are convenient for many purposes, we project $\Psi^{\rm unproj}$ explicitly into the LLL to obtain 
\be
\Psi_{\nu={n\over 2pn\pm 1}}={\cal P}_{\rm LLL}\prod_{j<k} \left( {z_j-z_k}  \right)^{2p}\Phi_{\pm n}\;,
\label{jainwf2}
\ee
with the hope that the nice correlations in the unprojected wave function will survive LLL projection.

Further generalizing to arbitrary filling factors, we obtain the final expression\footnote{The wave functions in Eqs.~\ref{jainwf1}, \ref{jainwf2} and \ref{jainwf} are sometimes referred  to as the Jain states and the fractions in Eq.~\ref{jains} as the Jain sequences.}:
\begin{equation}
\Psi^{\alpha}_{\nu={\nu^*\over 2p\nu^*\pm 1}}={\cal P}_{\rm LLL}
\prod_{j<k} (z_j-z_k)^{2p}
\Phi^{\alpha}_{\pm \nu^*}
\label{jainwf}
\end{equation} 
where $\alpha$ labels different eigenstates (not to be confused with the statistics parameter), and $\nu$ is related to the CF filling $\nu^*$ by
\begin{equation}
\nu={\nu^*\over 2p\nu^*\pm 1}\;.
\label{nunu*}
\end{equation}

Eq.~\ref{jainwf} may be taken as the defining postulate of the CF theory. While the line of reasoning leading to it was physically motivated, the wave functions $\Psi^{\alpha}_{\nu}$ are mathematically rigorously defined and allow us to make detailed predictions that can be tested against experiments. It is also possible, in principle, to unpack these wave functions to obtain explicit expansions of all eigenstates along all basis functions and compare with exact computer results.

 Eq.~\ref{jainwf} encapsulates the remarkable assertion of the CF theory, namely that all low-lying eigenstates at arbitrary filling factors in the LLL can be compactly represented by the single equation, which contains no adjustable parameters, and which, as discussed next, reveals in a transparent fashion the emergence of new topological particles that experience a reduced magnetic field. 

\underline{Reading the physics from the wave functions in Eq.~\ref{jainwf}:} 
To see what physics Eq.~\ref{jainwf} represents, let us inspect it afresh, pretending ignorance of the physical motivation that led to it. Disregarding the LLL projection for the moment, there are two important ingredients in the wave function: the Jastrow factor $\prod_{j<k} (z_j-z_k)^{2p}$ and the IQH wave function $\Phi_{\pm \nu^*}$. (i) The Jastrow factor attaches $2p$  vortices to electrons. [A particle, say $z_1$, sees $2p$ vortices at the positions of all other particles, due to the factor $\prod_{j=2}^N(z_1-z_j)^{2p}$.] The bound state of an electron and $2p$ quantized vortices is interpreted as an emergent particle, namely the composite fermion. (ii)  Because the vortices are being attached to electrons in the state $\Phi_{\pm\nu^*}$, the right hand side is naturally interpreted as a state of composite fermions at $\pm\nu^*$. (iii) The relation $\nu=\nu^*/(2p\nu^*\pm 1)$ can be derived from the wave function by determining the angular momentum of the outermost occupied orbit. (iv) The effective magnetic field for composite fermions arises because the Berry phases induced by the bound vortices partly cancel the AB phases due to the external magnetic field. The Berry phase associated with a closed loop of a composite fermion enclosing an area $A$ is given by the sum $-2\pi BA/\phi_0+2\pi 2p N_{\rm enc}$, where the first term is the AB phase of an electron going around the loop, and the second term is the Berry phase of $2p$ vortices going around $N_{\rm enc}$ electrons inside the loop. Interpreting the sum as an effective AB phase $-2\pi B^*A/\phi_0$ produces, with $N_{\rm enc}=\rho A$, the effective magnetic field $B^*=B-2p\rho\phi_0$. (v) The composite fermions are said to be non-interacting because the only role of the interaction is to bind vortices to electrons through the Jastrow factor $\prod_{j<k} (z_j-z_k)^{2p}$ to create composite fermions, and $\Phi_{\pm \nu^*}$ on the right hand side of Eq.~\ref{jainwf} is the wave function of non-interacting fermions. (vi) We can also see that a composite fermion is a topological particle, because a vortex is a topological object, defined through the property that a closed loop of any electron around it produces a Berry phase of $2\pi$, independent of the shape or the size of the loop.  (vi) We finally come to ${\cal P}_{\rm LLL}$.  The LLL projection renormalizes composite fermions in a very complex manner, producing extremely complicated wave functions. We postulate that the projected wave functions are adiabatically connected to the unprojected ones, and therefore describe the same physics. In other words, we assume that LLL projection does not cause any phase transition. While the physics of vortex binding is no longer evident after LLL projection, it is possible to test many qualitative features of the formation of composite fermions with the LLL theory, e.g. the similarity of the spectrum to that of non-interacting fermions at $B^*$.

\begin{figure}[t]
\includegraphics[width=5.0in]{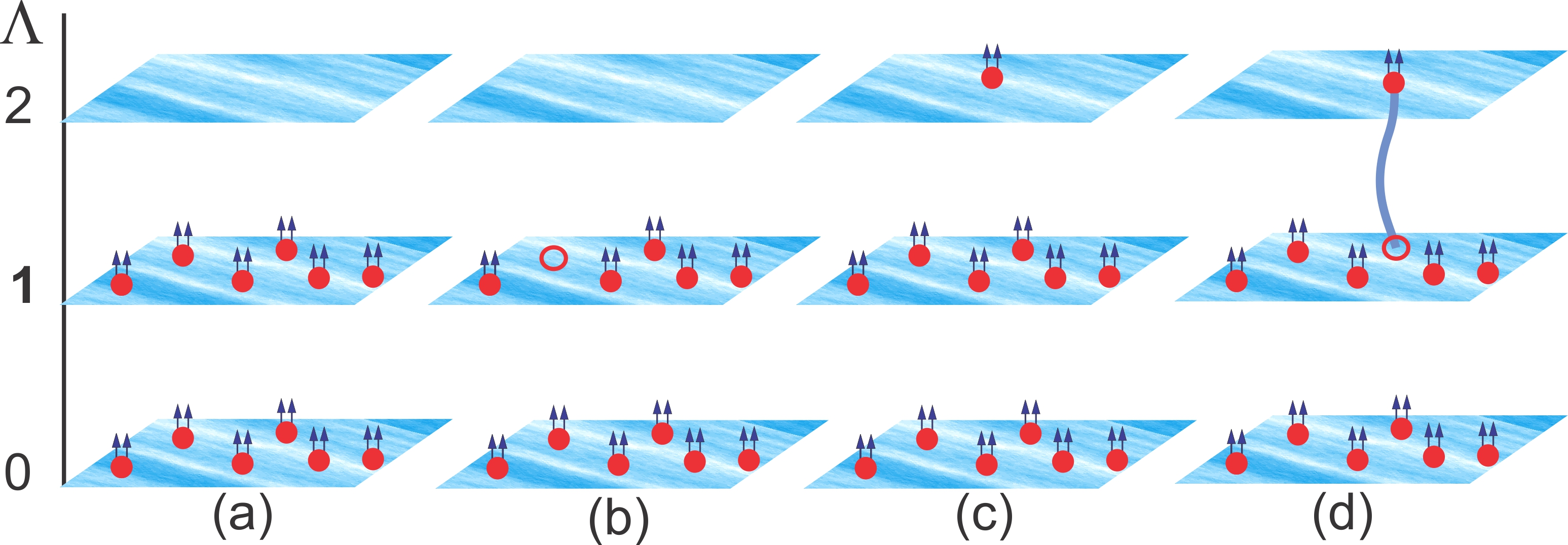}
\caption{Schematic $\Lambda$ level diagrams for: (a) an incompressible ground state; (b) a quasihole, i.e., a missing composite fermion; (c) a quasiparticle, i.e., an additional composite fermion; and (d) a neutral exciton. The $\nu=2/5$ FQH state is taken for illustration, which maps into $\nu^*=2$ of composite fermions.
\label{fig:Jeonel-CF}}
\end{figure}

To summarize: Interacting electrons in the LLL capture $2p$ quantized vortices each to turn into composite fermions. Composite fermions experience an effective magnetic field $B^*=B-2p\rho\phi_0$, because, as they move about, the vortices bound to them produce Berry phases that partly cancel the effect of the external magnetic field.  Composite fermions form their own Landau-like levels, called $\Lambda$ levels ($\Lambda$Ls), in the reduced magnetic field, and fill $\nu^*$ of them. (Recall that all of this physics occurs in the LLL of electrons. The LLL of electrons effectively splits into $\Lambda$Ls of composite fermions.) The occupation of $\Lambda$L orbitals is defined by analogy to the occupation of LL orbitals at $\nu^*$.  See Fig.~\ref{fig:Jeonel-CF} as an example. This physics is described by the electronic wave function in Eq.~\ref{jainwf}, where the right hand side is interpreted as the wave function non-interacting composite fermions at filling factor $\nu^*$.

It ought to be noted that no real flux quanta are bound to electrons. The flux quantum in Fig.~\ref{fig:jeon-map1} is to be understood as a model for a quantum vortex. While the model of composite fermions as point fluxes bound to electrons is not to be taken literally, it is topologically correct and widely used due to its pictorial appeal and the fact that it yields the correct $B^*$. In the same vein, an external magnetometer will always measure the field $B$, not $B^*$. The effective field $B^*$ is internal to composite fermions, and  composite fermions themselves must be used to measure it.

\underline{Construction of CF spectra at arbitrary $\nu$:}  Suppose we are asked to construct the low-energy spectrum at an arbitrary filling factor $\nu$. We first choose the positive even integer $2p$ in Eq.~\ref{nunu*} so as to obtain the largest possible value of $|\nu^*|$. We then construct the basis $\{\Phi^{\beta}_{\pm \nu^*}\}$ of all states, labeled by $\beta$, with the lowest kinetic energy. We multiply each basis function by $\prod_{j<k} (z_j-z_k)^{2p}$, 
project it into the LLL, and postulate that $\{{\cal P}_{\rm LLL}\Phi^{\beta}_{\pm \nu^*}\prod_{j<k} (z_j-z_k)^{2p}\}$ gives us the (in general non-orthogonal) basis for the lowest band of eigenstates of interacting electrons at $\nu$. For many important cases, this produces unique wave functions with no free parameters. 
For example, the ground state at $\nu=n/(2pn+1)$ is related to the ground state at $\nu^*=n$, whose wave function is the Slater determinant: 
\begin{equation}
\Phi_n={\rm Det}[\eta_\alpha(\vec{r}_k)]\;,
\end{equation}
where $\eta_\alpha(\vec{r})$ are given in Appendix~\ref{sec:symmetric}. Using the projection method of Refs.~\refcite{Jain97,Jain97b},  the wave function for the ground state at $\nu=n/(2pn+1)$ can be expressed, quite remarkably, as a single Slater determinant:
\begin{equation}
\Psi_{n\over 2pn+1}={\cal P}_{\rm LLL}{\rm Det}[\eta_\alpha(\vec{r}_k)] \prod_{j<k}(z_j-z_k)^{2p}\equiv {\rm Det}[\eta_\alpha^{\rm CF}(\vec{r}_k)].
\end{equation}
The elements of this determinant,
\begin{equation}
\eta_\alpha^{\rm CF}(\vec{r}_k)\equiv {\cal P}_{\rm LLL}\eta_\alpha(\vec{r}_k)\prod_{i(i\neq k)}(z_k-z_i)^p,
\end{equation}
can be evaluated analytically~\cite{Jain07} and are interpreted as ``single-CF orbitals." The single Slater determinant form for the incompressible states is not only conceptually pleasing but is what enables calculations for systems with 100-200 (or more) particles, for which it would be impossible to store projections on individual Slater determinant basis functions. The wave functions for a single quasiparticle, a single quasihole, and the neutral excitations of the $\nu=n/(2pn\pm 1)$ states, which are images of analogous excitations of the $|\nu^*|=n$ IQH states (see Fig.~\ref{fig:Jeonel-CF}), are also uniquely given by the CF theory, with no adjustable parameters. In these cases, it only remains to obtain the expectation value of the Coulomb interaction, which requires evaluation of a $2N$ dimensional integral, easily performed by the Monte Carlo method. For general fillings, when the topmost partially occupied $\Lambda$L has many composite fermions, the CF basis consists of many states, and it is necessary to diagonalize the Coulomb interaction in the CF basis. That can be accomplished numerically by a process called CF diagonalization~\cite{Mandal02}. (The dimension of the CF basis is exponentially small compared to that of the full LLL Hilbert space.) Basis functions for excited bands can be similarly constructed by composite-fermionizing states in the excited kinetic energy bands at $|\nu^*|$.

The above wave functions are written for electrons in the disk geometry. Other useful geometries are the spherical geometry~\cite{Haldane83} and the periodic (or the torus) geometry~\cite{Haldane85}. Wave functions for composite fermions in the spherical geometry were constructed almost three decades ago (see Ref.~\refcite{Jain07} and references therein), and recently that has been accomplished also for the torus geometry~\cite{Hermanns13,Pu17}. We will not show in this article, for simplicity, the wave functions for the spherical and torus geometries; an interested reader can find them in the literature.

The CF theory naturally gives wave functions.  Many other quantities of interest can be obtained from the wave functions, such as energy gaps, dispersions, pair correlation function, static structure factor, entanglement spectrum, charge and braid statistics of the excitations, etc. Efficient numerical methods for LLL projection~\cite{Jain97,Jain97b} and CF diagonalization~\cite{Mandal02} have been developed, which allow treatment of large systems. Because all wave functions are confined, by construction, to the LLL, the energy differences depend only on the Coulomb interaction and have no dependence on the electron mass.

\underline{Chern-Simons field theory and conformal field theory:} A complementary approach for treating composite fermions is through the CS field theory of composite fermions formulated by Lopez and Fradkin~\cite{Lopez91}, and by Halperin, Lee and Read (HLR)~\cite{Halperin93} (see Halperin's chapter). It has proved very successful in making detailed contact with experiments, especially for the low-energy long-wave length properties of the compressible state at and in the vicinity of the half filled Landau level. Conformal field theory based approaches are reviewed by Hansson {\em et al.}~\cite{Hansson17} and also in the chapter by Simon.

\subsection{Qualitative verifications}
\label{sec:verifications}

The title of a 1993 article by Kang, Stormer {\em et al.}~\cite{Kang93} posed the question: ``How Real Are Composite Fermions?"  

It was natural to question composite fermions. After all, they are are very complex, nonlocal objects. Even a single composite fermion is a collective bound state of all electrons, because all electrons participate in the creation of a vortex. One may wonder: Are such bound states really formed? If they are, in what sense do they behave as particles? Do they have the standard traits that we have come to associate with particles, such as charge, spin, statistics, etc.? To what extent is it valid to treat them as weakly interacting? How can they be observed? How can we verify that they see an effective magnetic field and form LL-like $\Lambda$Ls? These are all important questions, which can ultimately be answered only by putting predictions of the CF theory to the test against experiments and exact computer calculations.

Fortunately, the CF theory leads to many predictions, because weakly interacting fermions exhibit an enormously rich phenomenology. We only need to flip through a standard condensed matter physics textbook to remind ourselves of all of the well studied phenomena and states of electrons, and predict analogous phenomena and states for composite fermions. Let us begin with an account of how the qualitative consequences of composite fermions match up with the experimental phenomenology. Quantitative tests of the CF theory are considered in the next subsection.

\begin{center}
\begin{figure}[t]
\includegraphics[width=5.0in]{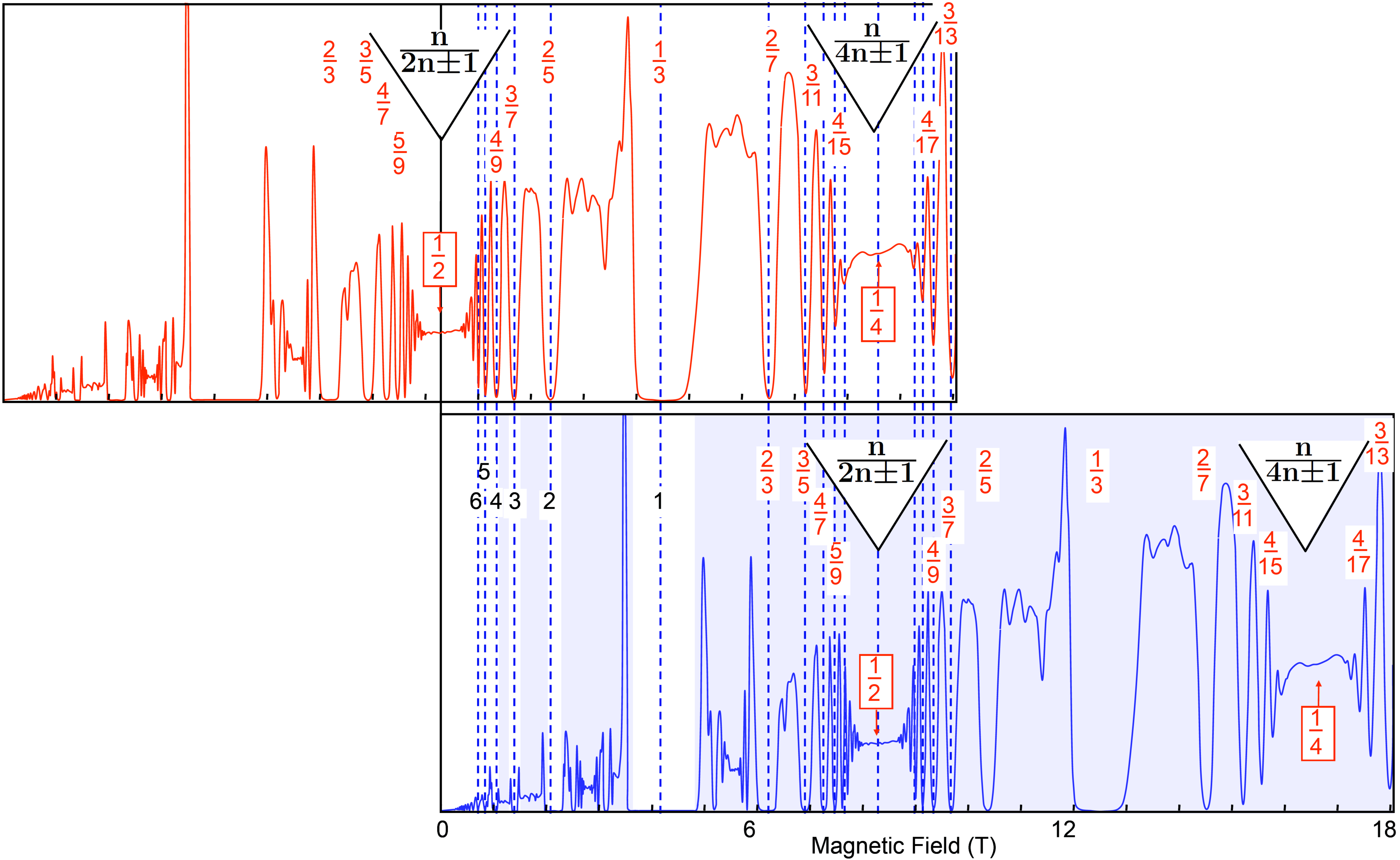}
\caption{In the upper panel, the FQHE trace is plotted as a function of $B^*=B-2\rho\phi_0$, which is the effective magnetic field seen by composite fermions carrying two vortices. A correspondence of the FQHE  around $\nu=1/2$ can be seen with the IQHE of electrons in the lower lower panel. The filling factor $\nu=1/2$ maps into zero magnetic field; the fractional filling factors $n/(2n+1)$ into inter fillings $n$; and the fractional filling factors $n/(4n\pm 1)$ around $\nu=1/4$ map into simpler fractions $n/(2n\pm 1)$. The fractions $n/(4n\pm 1)$ can also be mapped into integers by plotting the top panel as a function of $B^*=B-4\rho\phi_0$, the magnetic field seen by composite fermions carrying four vortices. Each panel is taken from Pan {\em et al.}\cite{Pan03}. 
Source: H. L. Stormer, private communication~\cite{Stormer}. 
\label{fqhe}}
\end{figure}
\end{center}

The most immediate evidence for the formation of composite fermions 
can be seen in Fig.~\ref{fqhe}, due to Stormer~\cite{Stormer}. Here the upper panel is plotted as a function of the effective magnetic field $B^*$ seen by composite fermions carrying two vortices, which simply amounts to shifting the upper panel leftward by an amount $\Delta B=2\rho\phi_0$. A close correspondence between the data in the upper panel and the lower panel is evident. This is a powerful demonstration of emergence of particles in the LLL that behave as fermions in an effective magnetic field $B^*=B-2\rho\phi_0$, which is the defining property of composite fermions, and of the formation of Landau-like $\Lambda$-levels inside the LLL of electrons. 

An important corollary of the above correspondence is the explanation of the FQHE as the IQHE of composite fermions. The fractions $\nu=n/(2n+1)=1/3$, 2/5, 3/7, etc. map into the integers $\nu^*=n=1$, 2, 3, etc. A schematic view of the 2/5 state is shown in Fig.~\ref{fig:Jeonel-CF}(a).  If one attached a mirror image of the lower panel for negative magnetic fields, one would see that the fractions $\nu=n/(2n-1)=2/3$, 3/5, 4/7, $\cdots$ align with integers $\nu^*=-2, -3, -4$ $\cdots$ (in negative magnetic field). The fractions $n/(4n\pm 1)$ in the upper panel map into simpler fractions $n/(2n\pm 1)$ of composite fermions carrying two vortices, but they can also be understood as $\nu^*=\pm n$ IQHE of composite fermions carrying four vortices, as can be confirmed by plotting the upper panel as a function of $B^*$ seen by composite fermions carrying four vortices, which would amount to shifting it leftward by $\Delta B=4\rho\phi_0$. The fractions $n/(2pn\pm 1)$ and their hole partners $1-n/(2pn\pm 1)$ indeed are the prominently observed fractions in the LLL.\footnote{The states at $\nu=1-n/(2pn\pm 1)$ can be understood by formulating the original problem in terms of holes in the LLL, and then making composite fermions by attaching vortices to holes and placing them in $\nu^*=n$ IQH states.} There is  evidence~\cite{Stormer,Pan02,Pan03,Jain14} for ten members of the sequences $\nu=n/(2n\pm 1)$ and six members of the sequences $\nu=n/(4n\pm 1)$. The IQHE of composite fermions produces only odd denominator fractions; this can be traced back to the fermionic nature of composite fermions, which requires $2p$ to be an even integer. The CF theory thus provides a natural explanation for the fact that most of the observed fractions have odd denominators. The weak residual interaction between composite fermions can (and does) produce further fractions, including those with even denominators, but these are expected to be more delicate, just as the FQHE of electrons is weaker than their IQHE.

Notably, the CF theory obtains all fractions of the form $\nu=n/(2pn\pm 1)$ and $\nu=1-n/(2pn\pm 1)$ on the same conceptual footing. The earlier dichotomy of ``Laughlin states" and ``other states" may therefore be dispensed with; drawing such a distinction would be akin to differentiating between the $\nu=1$ and the other IQH states.

The excitations of all FQH states are simply excited composite fermions. The lowest energy positively or negatively charged excitation of the $\nu=n/(2pn\pm 1)$ state is a missing composite fermion in the $n^{\rm th}$ $\Lambda$L or an additional composite fermion in the $(n+1)^{\rm th}$ $\Lambda$L, as shown in Fig.~\ref{fig:Jeonel-CF}(b-c). These are sometimes referred to as a quasihole or a quasiparticle. The neutral excitation is a particle-hole pair, i.e. an exciton, of composite fermions (Fig.~\ref{fig:Jeonel-CF}d). The activation gap deduced from the temperature dependence of the longitudinal resistance is identified with the energy required to create a far separated pair of quasiparticle and quasihole. As seen in the next subsection, the microscopic CF theory provides an accurate estimate for the energy gaps, but some insight into their qualitative behavior may be obtained by introducing a phenomenological mass for composite fermions and interpreting the gap as the cyclotron energy of composite fermions~\cite{Halperin93}. The CF cyclotron energy at $\nu=n/(2pn\pm 1)$ is written as $\hbar\omega_c^*=\hbar {eB^*\over m^* c}=\hbar {eB\over (2pn\pm 1)m^* c}\equiv {C\over 2pn\pm 1}\; {e^2\over \epsilon l}$. The last equality follows because all energy gaps in a LLL theory must be determined by the Coulomb energy alone, and implies that the CF mass behaves as $m^*\sim \sqrt{B}$. Direct calculation of gaps along $\nu=n/(2n+1)$ for $n\leq 7$ using the microscopic CF theory~\cite{Halperin93,Scarola02} has found that the gaps, quoted in units of $e^2/\epsilon l$, are approximately proportional to $1/(2n+1)$, with best fit for a system with zero thickness given by $C=0.33$~\cite{Halperin93}. This corresponds to a CF mass of $m^*=0.079 \sqrt{B[T]} \; m_e$ for parameters appropriate for GaAs, where $B[T]$ is quoted in Tesla and $m_e$ is the electron mass in vacuum. The experimentally measured activation gaps deduced from the Arrhenius behavior of the longitudinal resistance are found to behave as ${C'\over 2pn\pm 1}\; {e^2\over \epsilon l}-\Gamma$, where $\Gamma$ is interpreted as a disorder induced broadening of $\Lambda$Ls~\cite{Du93,Manoharan94}. The CF mass can be deduced from the slope; not unexpectedly, its value depends somewhat on finite thickness, LL mixing and disorder. Neutral excitons of composite fermions have been investigated extensively in light scattering experiments~\cite{Pinczuk93,Kang00,Kukushkin00,Kang01,Dujovne03,Dujovne05,Kukushkin09,Rhone11,Wurstbauer11}.

So far we have assumed that the magnetic field is so high that all electrons, or composite fermions, are fully spin polarized, i.e., effectively spinless. The spin physics of the FQHE is explained in terms of spinful composite fermions~\cite{Wu93,Park98}. Now the integer filling of composite fermions is given by $\nu^*=n=n_\uparrow+n_\downarrow$, where $n_\uparrow$ and $n_\downarrow$ are the number of filled $\Lambda$Ls of spin up and spin down composite fermions. This immediately leads to detailed predictions for the allowed spin polarizations for the various FQH states as well as their energy ordering. Transitions between differently spin polarized states can be caused by varying the Zeeman energy, and are understood in terms of crossings of $\Lambda$Ls with different spins. These considerations also apply to the valley degree of freedom. Spin / valley polarizations of the FQH states have been determined as a function of the spin / valley Zeeman energy, and the $\Lambda$L fan diagram for composite fermions has been constructed~\cite{Du95,Du97,Kukushkin99,Bishop07,Padmanabhan09,Feldman13}. Section~\ref{sec:spin} is devoted to the phase diagram of spin polarization of the FQH states.

A striking experimental fact is the absence of FQHE at $\nu=1/2$. As seen in Fig.~\ref{fqhe}, $\nu=1/2$ in the upper panel aligns with zero magnetic field of the lower panel. In an influential paper, HLR predicted~\cite{Halperin93} that the 1/2 state is a Fermi sea of composite fermions in $B^*=0$. Extensive verifications of the CF Fermi sea (CFFS) and its Fermi wave vector now exist~\cite{Willett93,Kang93,Goldman94,Smet96,Smet98,Willett99,Smet99,Kamburov12,Gokmen10,Kamburov13}.  The semiclassical cyclotron orbits in the vicinity of $\nu=1/2$ have been measured by surface acoustic waves~\cite{Willett93}, magnetic focusing~\cite{Goldman94,Smet96}, and commensurability oscillations in periodic potentials~\cite{Kang93,Smet98,Smet99,Willett99,Kamburov12}. These are considered direct observations of composite fermions. The measured cyclotron radius is consistent with $R_c^*  ={\hbar k^*_F }/{eB^*}$ with $k^*_F=\sqrt{4\pi \rho}$, as appropriate for a fully polarized CF Fermi sea. The CF cyclotron radius is much larger than, and thus clearly distinguishable from, the radius of the orbit an electron would execute in the external magnetic field. The temperature dependence of the spin polarization of the 1/2 state measured by NMR experiments is consistent with that of a Fermi sea of non-interacting fermions~\cite{Kukushkin99,Melinte00}. Shubnikov-de Haas oscillations of composite fermions have been observed and analyzed to yield the CF mass and quantum scattering times~\cite{Du94,Leadley94}. The cyclotron resonance of composite fermions has been observed by microwave radiation, with a wave vector defined by surface acoustic waves~\cite{Kukushkin02,Kukushkin07}. The CFFS is discussed in further detail in the chapters by Halperin and Shayegan. 

In summary, when filtered through the prism of composite fermions, the exponentially large number of choices that were available to electrons disappear, giving way to a host of unambiguous predictions, which have been confirmed by extensive experimental studies. These predictions may appear obvious, even inevitable, once you accept composite fermions, but they are non-trivial from the vantage point of electrons, and would not have been evident without the knowledge of composite fermions.

\subsection{Quantitative verifications against computer experiments}

Let us next come to the quantitative tests of the CF theory. At the time of originally proposing the wave functions in Eqs.~\ref{jainwf1}-\ref{jainwf} relating the FQHE to IQHE through composite fermions, the author believed that they were toy models that would describe the correct phase but did not expect them to be accurate representations of the actual Coulomb states. After all, these wave functions are in general enormously complicated after projection into the LLL. Extensive computer calculations in subsequent years proved otherwise. 

This subsection presents comparisons of results from two independent calculations. The first is a brute force diagonalization of the Coulomb Hamiltonian within the LLL Hilbert space, which produces exact eigenenergies and eigenfunctions. The second constructs wave functions of the CF theory and obtains their exact energy expectation values\footnote{This calculation often uses the Monte Carlo method which involves statistical uncertainty, but several significant figures can be obtained exactly with currently available computational resources.}. Neither of the calculations contains any adjustable parameters. 

\begin{landscape}
\begin{figure}[t]
\includegraphics[width=7.5in]{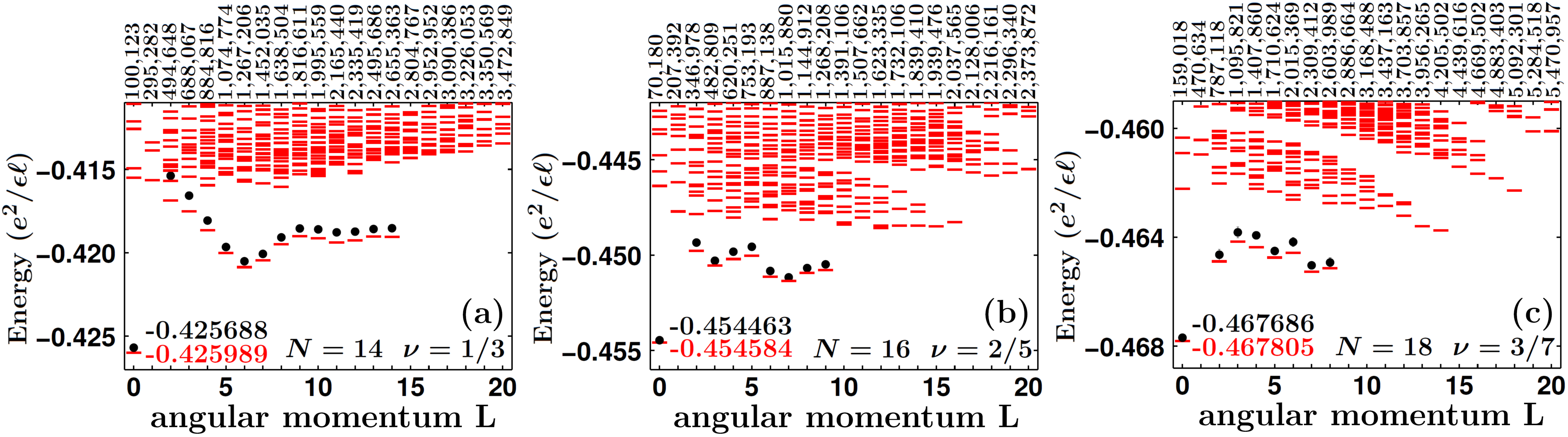}
\caption{These figures show a comparison between the energies (per particle) predicted by the CF theory (dots) and the exact Coulomb energies (dashes), both obtained without any adjustable parameters. Panels (a)-(c) show spectra for $(N, 2Q)=(14, 39)$, $(16,36)$, and $(18, 37)$, which are finite size representations of the 1/3, 2/5 and 3/7 states. The wave function for ground state at $\nu=1/3$ is the same as the Laughlin wave function.  Source: A. C. Balram, A. W\'ojs, and J. K. Jain, Phys. Rev. B.
{\bf 88}, 205312 (2013)\cite{Balram13}, and J. K. Jain, 
Annu. Rev. Condens. Matter Phys. {\bf 6}, 39-62, (2015)~\cite{Jain15}.}
\label{comp1}
\end{figure}
\end{landscape}

\begin{figure}[t]
\includegraphics[width=0.47\textwidth]{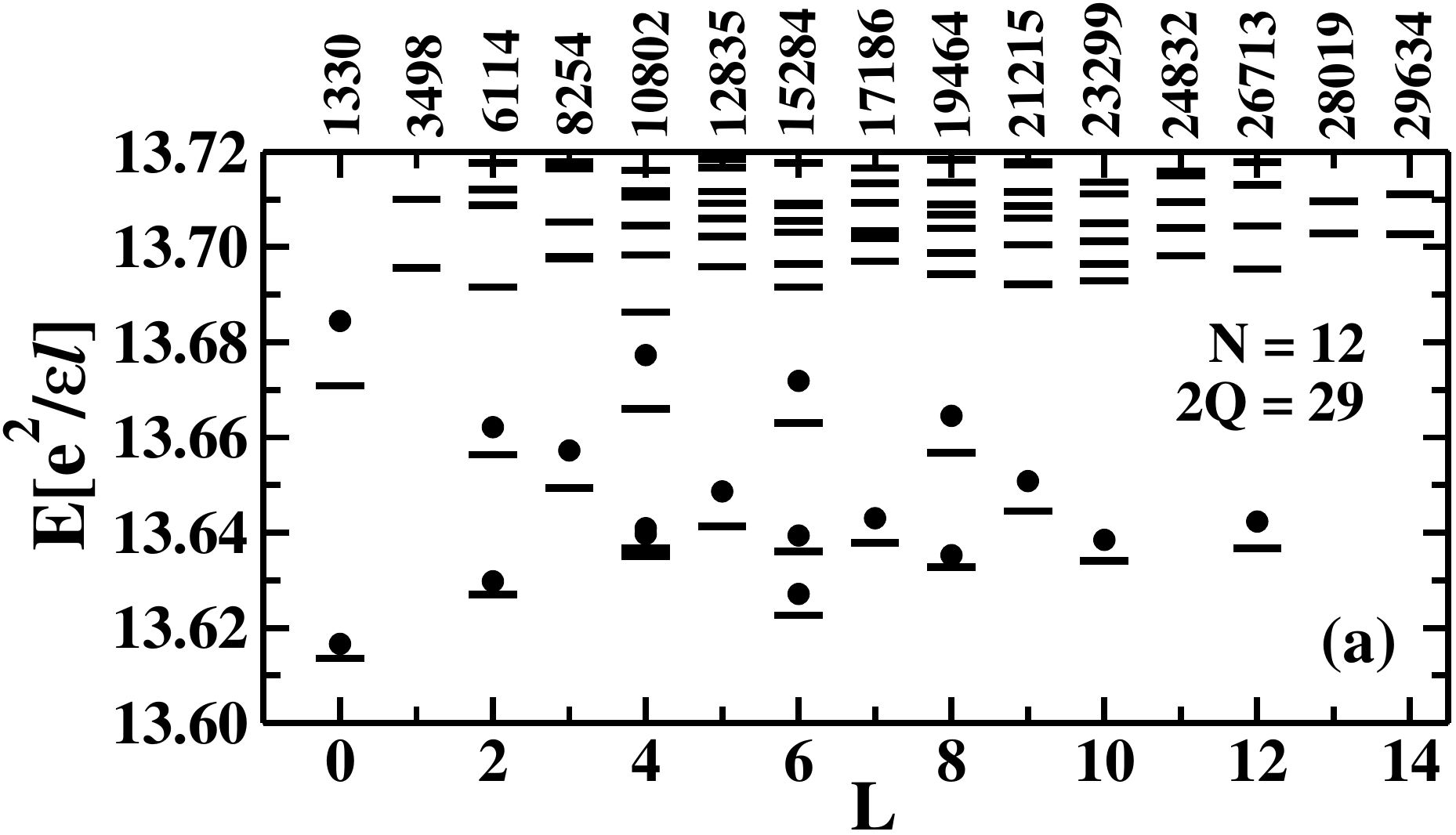}
\includegraphics[width=0.47\textwidth]{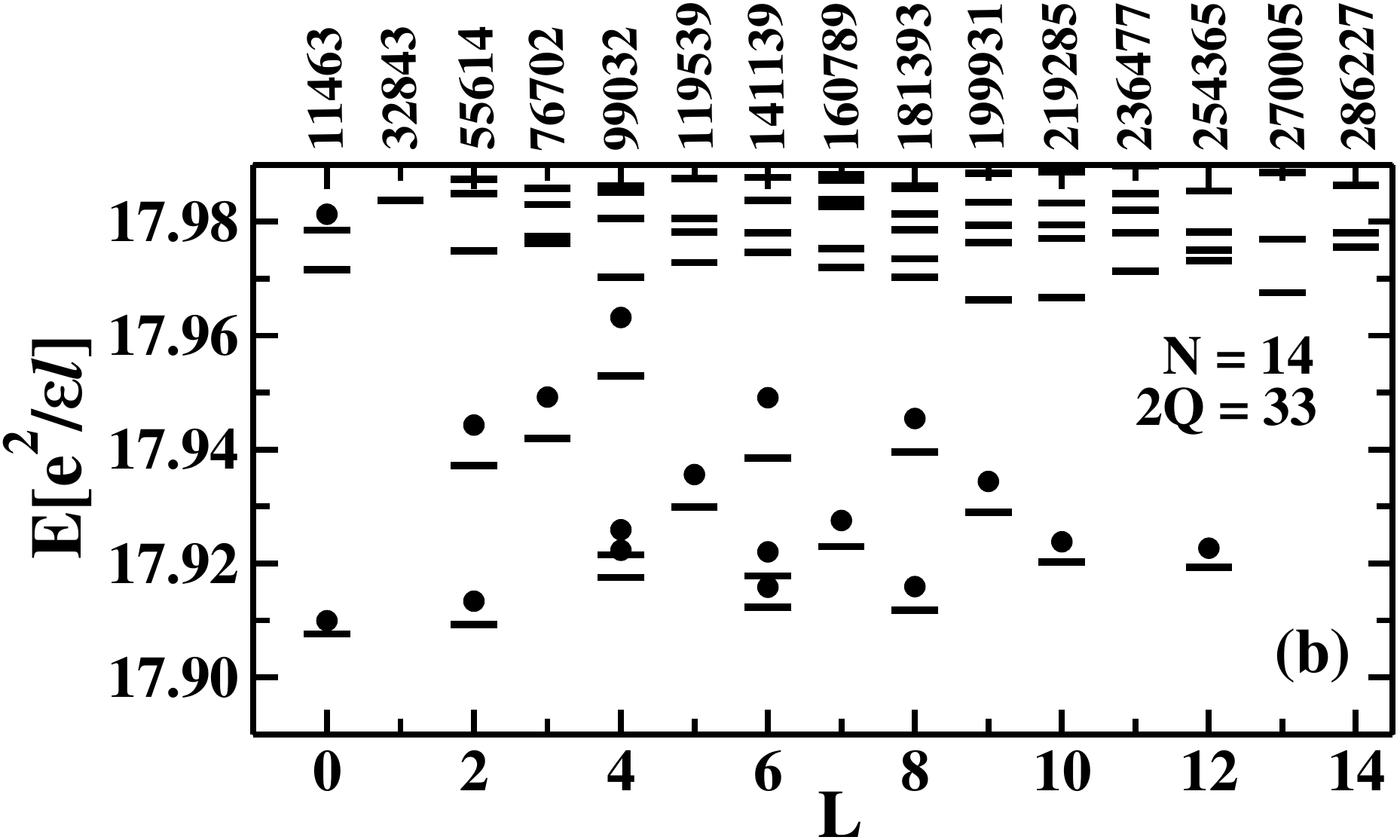}
\caption{Comparison of exact Coulomb spectra (dashes) with the prediction of CF theory (dots) for $(N,2Q)=(12, 29)$ and $(14, 33)$. The dimensions of the Hilbert space in the individual $L$ sectors are shown at the top. Source: S. Mukherjee, S. S. Mandal, A. W\'ojs, and J. K. Jain,  Phys. Rev. Lett. {\bf 109}, 256801 (2012)~\cite{Mukherjee12}.\label{comp3}}
\end{figure}

\begin{center}
\begin{figure}[t]
\begin{center}
\includegraphics[width=0.46\textwidth]{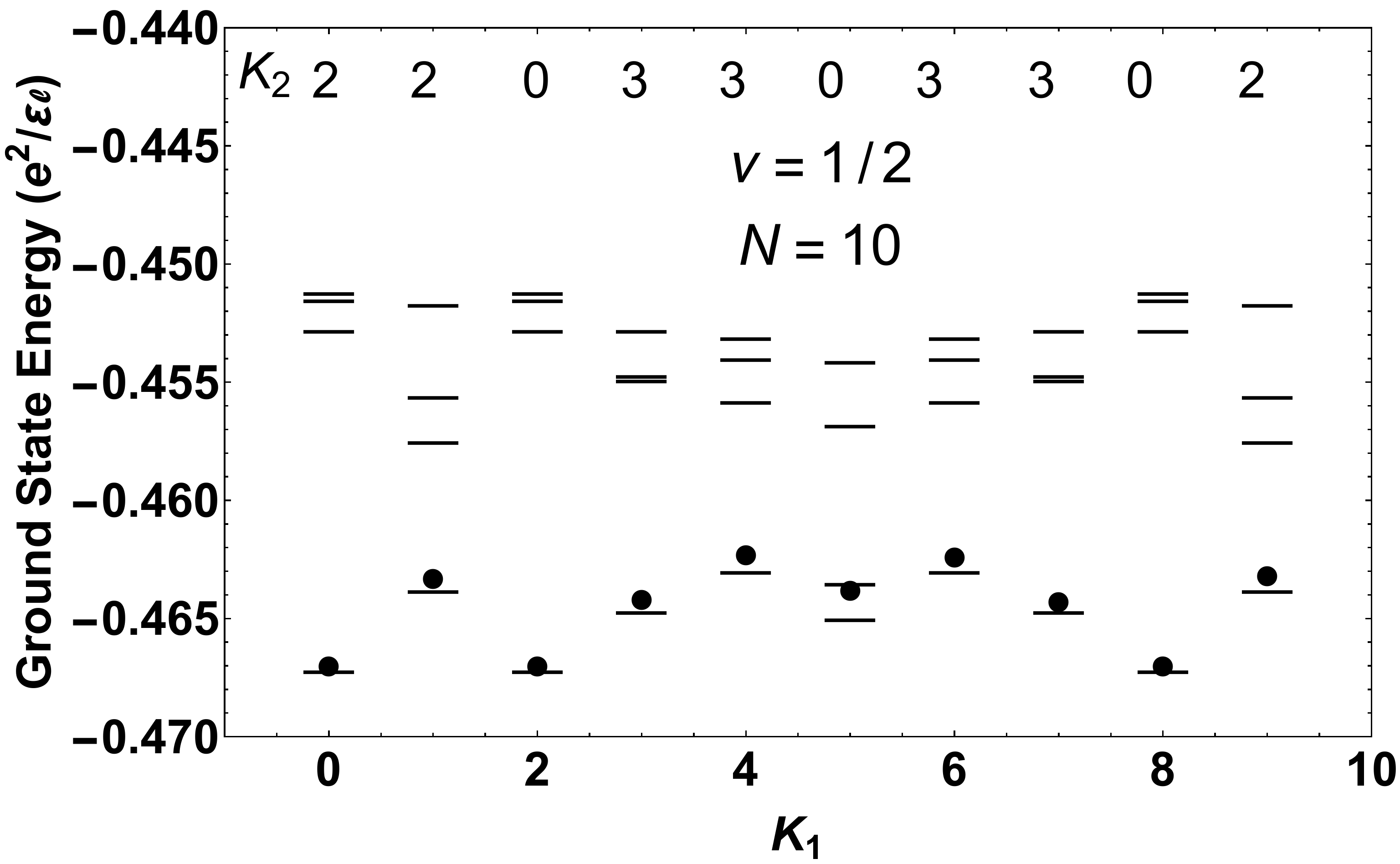} 
\includegraphics[width=0.47\textwidth]{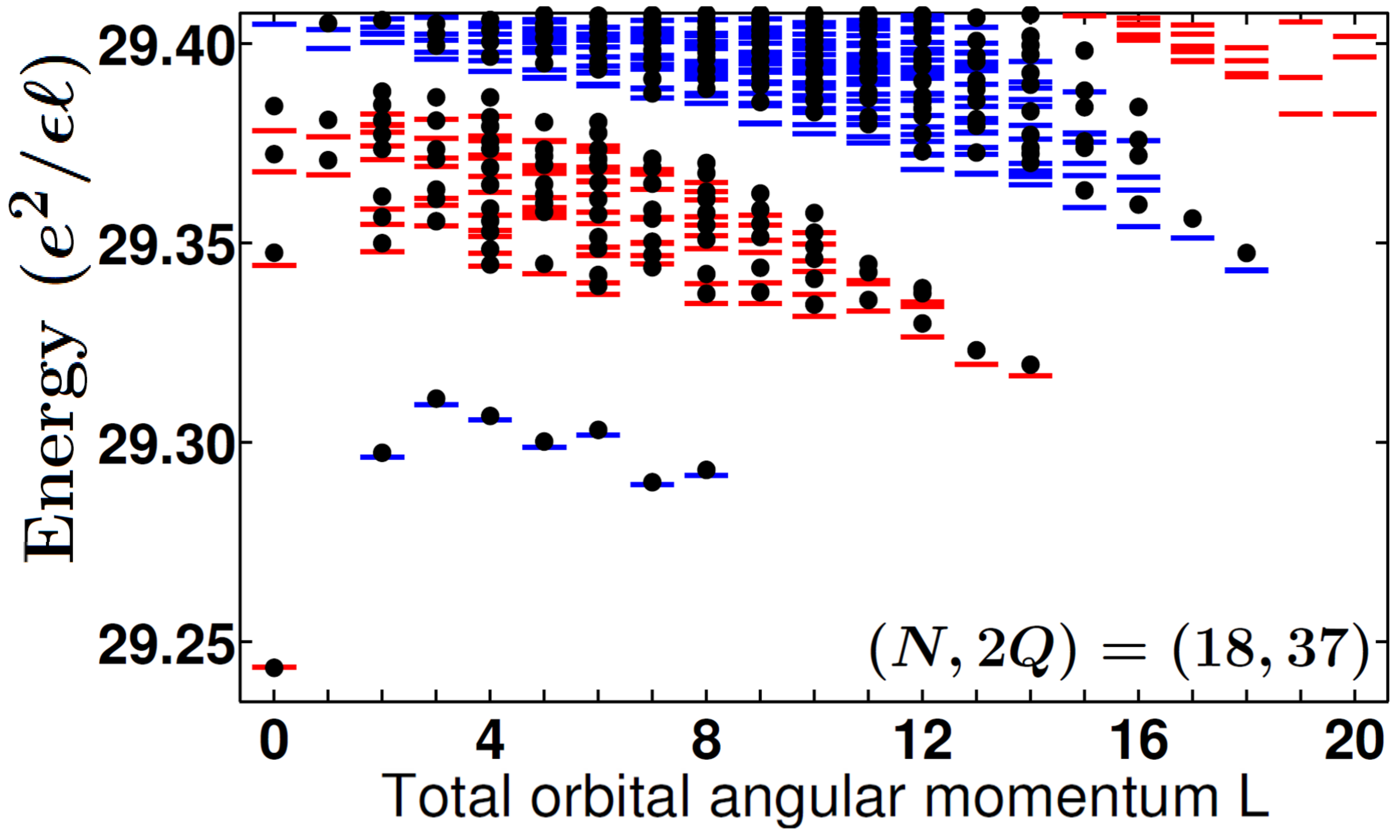}
\end{center}
\caption{Left panel: Comparison between the exact spectra (dashes) and the CF spectra (dots) for $N=10$ particles at $\nu=1/2$ in the periodic torus geometry. The momentum $K_1$ is given on the x-axis. For each $K_1$, the $K_2$ (shown at the top) is the momentum of the lowest energy state.  For the torus geometry, the spectra for $K_1$ and $K_1+N$ are identical.  Right panel: Comparison of the CF and exact spectra for the lowest four bands at $\nu=3/7$ in the spherical geometry. The alternating bands in the exact spectra are shown in different colors for contrast. Source: S. Pu, M. Fremling, and J. K. Jain,  Phys. Rev. B. {\bf 98}, 075304 (2018)~\cite{Pu18}; A. C. Balram, A. W\'ojs, and J. K. Jain, Phys. Rev. B.
{\bf 88}, 205312 (2013)\cite{Balram13}. 
}
\label{comp2}
\end{figure}
\end{center}

A convenient geometry is the spherical geometry~\cite{Haldane83} where $N$ electrons move on the surface of a sphere subjected to a total flux of $2Q\phi_0$, where $2Q$ is quantized to be an integer. Figs.~\ref{comp1}, \ref{comp3}, \ref{comp2} show typical comparisons between the CF theory (dots) and exact results (dashes). To gain a better appreciation, we recall certain basic facts about the spherical geometry. An electron in the $j$th LL ($j=0, 1, \cdots$, with $j=0$ labeling the LLL) has an orbital angular momentum $|Q|+j$. The degeneracy of the $j$th LL is $2(|Q|+j)+1$, corresponding to the different z components of the angular momentum. For a many electron system, the total orbital angular momentum $L$ is a good quantum number, used to label the eigenstates. For a non-interacting system, it is straightforward to determine all of the possible $L$ values for a given $(N, 2Q)$ system. To analyze the exact spectra of interacting electrons in terms of composite fermions, we need to make use of the result that the CF theory relates the interacting electrons system $(N, 2Q)$ to the non-interacting CF system $(N, 2Q^*)$ with 
\be
2Q^*=2Q-2p(N-1)\;.
\ee
This relation follows from the spherical analog of Eq.~\ref{jainwf}, $\Psi_{2Q}={\cal P}_{\rm LLL}\Phi_{2Q^*}\Phi_1^{2p}$, by noting that the flux of the product is the sum of fluxes ($\Phi_1$ occurs at $2Q_1=N-1$), and that the flux remains invariant under LLL projection. Intuitively, the relation between $2Q$ and $2Q^*$ can be understood from the observation that for any given composite fermion, all of the other $N-1$ composite fermions reduce the flux by $2p\phi_0$ each. A corollary of this relation is that the incompressible states do not occur at $2Q=\nu^{-1}N$ but rather at $2Q=\nu^{-1}N-{\cal S}$, where ${\cal S}$ is called the ``shift." For the IQH state at $\nu=n$, the shift is simply ${\cal S}=n$, which follows from the fact that the degeneracy of the $j$th LL is $2|Q|+2j+1$. According to the CF theory, the incompressible FQH state at $\nu={n\over 2pn\pm 1}$ occurs at shift ${\cal S}=2p\pm n$, 
because the shift of the product $\Phi_{\pm n}\Phi_1^{2p}$ is the sum of the shifts, which is preserved under LLL projection.  The shift is $N$-independent, and in the thermodynamic limit we recover $\lim_{N\rightarrow\infty}N/2Q=\nu$ irrespective of the value of the shift. It is noted that different candidate states for a given filling factor may produce different shifts.

Let us now see what features of the exact spectra are explained by the CF theory by taking some concrete examples. Fig.~\ref{comp1} shows exact Coulomb spectra (dashes) for some of the largest systems for which exact diagonalization has been performed. Each dash represents a multiplet of $2L+1$ degenerate eigenstates. The energies (per particle) include the electron-background and background-background interaction.  Only the very low energy part of the spectrum is shown. The total number of independent multiplets at each $L$ is shown at the top. Each eigenstate in this figure is thus a linear superposition of $\sim$one hundred thousand to several million independent basis functions. All sates would be degenerate in the absence of the Coulomb interaction. The emergence of certain well defined bands at low energies is a manifestation of non-perturbative physics arising from  interaction.

The interacting electrons systems $(N, 2Q)=(14, 39)$, $(16,36)$, and $(18, 37)$ map into CF systems $(N, 2Q^*)=(14, 13)$, $(16,6)$, and $(18, 3)$.  The ground states correspond to 1, 2 and 3 filled $\Lambda$Ls, and thus have $L=0$, precisely as seen in the exact spectra. The lowest energy (neutral) excitations for $(N, 2Q^*)=(14, 13)$, $(16,6)$, and $(18, 3)$ consist of a pair of CF-hole and CF-particle with angular momenta 6.5 and 7.5, 4 and 5, and 3.5 and 4.5, respectively. These produce states at $L=1, 2, \cdots L_{\rm max}$ with $L_{\rm max}= 14$, 9 and 8. The $L$ quantum numbers of the lowest excited branch in the exact spectra agree with this prediction, except that there is no state at $L=1$. It turns out that when one attempts to construct the wave function for the CF exciton at $L=1$, the act of LLL projection annihilates it~\cite{Dev92}, bringing the CF prediction into full agreement with the quantum numbers seen in the exact spectra.

Going beyond the qualitative explanation of the origin and the structure of the bands, the CF theory gives parameter free wave functions for the ground states and

\noindent the lowest energy neutral excitations at all fractions $\nu=n/(2pn\pm 1)$, obtaining by composite-fermionizing the corresponding wave functions at $\nu^*=n$. The dots show the expectation values of the Coulomb interaction for these wave functions.  The energies of the ground states agree to within $\sim$ 0.07\%., 0.03\% and 0.04\% for the 1/3, 2/5 and 3/7 systems shown in the figures. Further, the CF theory reproduces the qualitative features of the exact dispersion of the neutral exciton (the wave vector of the neutral exciton is given by $k=L/l\sqrt{Q}$) and predicts its energy (relative to the ground state) with a few \% accuracy.
The CF theory provides a similarly accurate account of the fractionally charged quasiparticle and quasihole for all fractions $\nu=n/(2n\pm 1)$. These are either an isolated CF particle in an otherwise empty $\Lambda$L or an isolated CF hole in an otherwise filled $\Lambda$L, as depicted in Figs.~\ref{fig:Jeonel-CF} (b) and (c).  The CF hole in an otherwise full lowest $\Lambda$L reproduces Laughlin's wave function for the quasihole of the $1/(2p+1)$ state, albeit from a different physical principle.

Fig.~\ref{comp3} shows comparisons away from the special fillings~\cite{Mukherjee12}. The quantum numbers of the states in the low energy band identifiable in the exact spectra are identical to those for non-interacting fermions at $2Q^*$. As an example, consider the electron system $(N,2Q)=(12,29)$ (left panel of Fig.~\ref{comp3}) which maps into the CF system $(N, 2Q^*)= (12,7)$. Here, the lowest energy configurations have filled lowest $\Lambda$L (accommodating $2|Q^*|+1=8$ composite fermions), and four composite fermions in the second $\Lambda$L, each with angular momentum $|Q^*|+1=9/2$. The predicted total angular momenta $L$ (for fermions) are given by ${9\over 2}\otimes {9\over 2} \otimes {9\over 2}\otimes {9\over 2} = 0^2 \oplus 2^2 \oplus 3 \oplus 4^3 \oplus 5 \oplus 6^3\oplus 7 \oplus 8^2 \oplus 9 \oplus 10 \oplus 12 $, which match exactly with the $L$ multiplets seen in the lowest band in the left panel of Fig.~\ref{comp3}. A similar calculation successfully predicts the $L$ quantum numbers of the lowest band of $(N,2Q)=(14,33)$ (right panel of Fig.~\ref{comp3}).  Diagonalization of the Coulomb interaction in the reduced CF basis produces the dots in Fig.~\ref{comp3}. 

The CF Fermi sea at $\nu=1/2$ is obtained in the $n\rightarrow \infty$ limit of the $n/(2n\pm 1)$ fractions in the spherical geometry, or by composite-fermionizing the wave function of the electron fermi sea in the torus geometry~\cite{Rezayi94,Rezayi00,Shao15,Pu17,Wang19,Geraedts18,Fremling18,Pu18}. The left panel of Fig.~\ref{comp2} shows the exact spectrum at $\nu=1/2$ in the torus geometry, along with the CF energies for the lowest energy states in several momentum sectors~\cite{Pu18}.

Higher bands are often not clearly identifiable in the exact spectra, presumably because of the broadening induced by the residual interaction between composite fermions. Interestingly, more and more bands become visible as we we go to higher CF fillings. For example, four reasonably well defined bands can be seen at $\nu=3/7$ in Fig.~\ref{comp1}. The CF theory gives a good account of the higher bands as CF kinetic energy bands, which involve excitations of one or more composite fermions across one or several $\Lambda$Ls. Fig.~\ref{comp2} shows a comparison between the CF theory and the exact spectrum for four lowest bands. A subtle point is that while the one-to-one correspondence between the FQHE spectra of $(N,2Q)$ and the IQHE spectra of $(N,2Q^*)$ is perfect for the lowest band for all LLL spectra studied so far, it is imperfect for higher bands, where the IQHE spectra have a slightly greater number of states. However, when one constructs wave functions by taking the IQH states, multiplying by the Jastrow factor and then performing LLL projection, the last step annihilates many of the states, and, remarkably, the surviving linearly independent states provide a faithful account of the bands seen in the exact FQHE spectra~\cite{Wu95,Balram13,Meyer16}. (What mathematical structure underlies such elimination of states is not yet understood.) The dots in Fig.~\ref{comp2} are obtained by a diagonalization of the Coulomb interaction in the CF basis derived from all IQH states with energies up to 3 $\hbar\omega_c$. Balram {\em et al.}~\cite{Balram13} have performed an extensive study of the higher bands of many systems, showing that the correct counting for higher bands can be obtained by projecting out states certain excitons of the $(N,2Q^*)$ systems.

The CF theory allows, in principle, a systematic improvement of energies by allowing mixing with higher $\Lambda$Ls. An example can be seen in right panel of Fig.~\ref{comp2}, where the ground state and the single exciton energies have improved substantially compared to those in Fig.~\ref{comp1}. In practice, the accuracy of the zeroth order CF theory is sufficient for most purposes because corrections due to other effects (e.g. LL mixing or finite width) are larger.

In summary, for all LLL systems studied by exact diagonalization, the CF theory faithfully predicts the structure of the lowest band (i.e. the number of states and their quantum numbers). It never misses any state, nor does it ever predict any false states.  Furthermore, it predicts the eigenfunctions and eigenenergies almost exactly\footnote{This shows that even though the microscopic wave functions in Eqs.~\ref{jainwf1},\ref{jainwf2},\ref{jainwf} are motivated by the physics of weakly interacting composite fermions, they incorporate the knowledge of inter-CF interactions.}.  In other words, all low energy wave functions obtained in exact diagonalization studies of electrons in the LLL can be succinctly and accurately synthesized into a single, parameter-free equation, Eq.~\ref{jainwf}. These studies prove, at the most microscopic level possible, the formation of composite fermions and the relation between the FQHE and the IQHE that they entail.

\subsection{Remarks}

We close the section with some remarks.

\underline{Universality of wave functions:} 
As noted above, the wave functions in Eq.~\ref{jainwf} contain no free parameters for the ground states as well as the charged and neutral excitations at $\nu=n/(2pn\pm 1)$.\footnote{For $\nu\neq n/(2pn\pm 1)$, the basis functions for the lowest band contain no free parameters, although their mixing and splittings depend on the specific form of the interaction.} How is it then possible that these wave functions so accurately represent the eigenstates of the Coulomb interaction? What if one were to choose some other interaction? Insight into this issue comes from numerical diagonalization studies that demonstrate that the actual eigenfunctions at these fractions are surprisingly insensitive to the detailed form of the interaction so long as it is sufficiently strongly repulsive at short distances. Luckily, the Coulomb interaction in the LLL belongs in that limit. In that sense, the FQHE wave functions in the LLL are universal.\footnote{This may be contrasted with the Hartree-Fock Fermi-liquid and the BCS wave functions that explicitly depend on the interaction.} The good luck continues in that the CF theory captures precisely that limit. FQHE can also occur when the short range part of the repulsive interaction is not strong, as, for example, is the case for Coulomb interaction in the second LL; the wave functions for many second LL FQH states are more sensitive to the form of the interaction, and the agreement with candidate wave functions is not as decisive as that in the LLL.

\underline{Observation of $\Lambda$Ls:}  Electrons and their LLs were known prior to the discovery of the IQHE. In contrast, the FQHE was discovered first, and its similarity to the IQHE gave a clue into the existence of composite fermions and their $\Lambda$Ls. While the LLs can be derived for a single electron, composite fermions and their $\Lambda$Ls provide a single-particle-like interpretation of the inherently many body wave functions of interacting electrons in the LLL. The formation of $\Lambda$Ls within the LLL of electrons can be seen in a variety of ways. In computer calculations, the low-energy spectrum of interacting electrons in the LLL at $\nu$ splits into bands that have a one-to-one correspondence with the kinetic-energy bands of non-interacting electrons at $\nu^*$, and the eigenfunctions of interacting electrons at $\nu$ are related to those of non-interacting electrons at $\nu^*$ through composite-fermionization. In experiments, the $\Lambda$Ls appear remarkably similarly as the LLs, for example, through peaks in the longitudinal resistance R$_{\rm L}$ (see Fig.~\ref{fig:stormer98}).

\underline{Use of higher LLs:} One may ask why the path to the FQH wave functions in the LLL should pass through IQH wave functions involving higher LLs. We begin by noting that there is no fundamental reason to insist on strictly LLL wave functions in the first place. While restricting the Hilbert space to the LLL is convenient for computer calculations, it is not a necessary condition for FQHE. LL mixing is always present in experiments, indicating that the phase diagram of the FQHE extends to regions with non-zero LL mixing. The job of theory is to identify a point inside the FQH phase where the physics is the simplest, and approach the physical point perturbatively starting from there. The CF theory demonstrates that allowing a small admixture with higher LLs makes it possible to construct wave functions that reveal the physics of the FQHE in a transparent manner. The LLL projections of these wave functions accurately represent the exact Coulomb solutions, but are extremely complicated and could not have been guessed directly within a LLL theory. Finally, it ought to be stated that the use of higher LLs is not merely a technical matter but is intimately tied to the CF physics and the analogy between the FQHE and the IQHE.

\underline{Particle-hole symmetry:} When we restrict to the Hilbert space of the LLL, the Hamiltonian with a two-body interaction satisfies an exact symmetry called the particle-hole (PH) symmetry. This refers to the fact that the PH transformation $c_j\rightarrow h_j^\dagger, c_j^\dagger\rightarrow h_j$, which relates the state at $\nu$ to a state at $1-\nu$, leaves the interaction Hamiltonian invariant modulo an overall additive term. In other words, the eigenspectra at $\nu$ and $1-\nu$ are identical (apart from a constant overall shift) when plotted in units of $e^2/\epsilon l$, and the eigenstates are exactly related by PH transformation. In particular, at $\nu=1/2$, unless PH symmetry is spontaneously broken, the Fermi sea wave function must be equal to its PH conjugate. PH symmetry cannot be defined in the presence of LL mixing. It should  be noted that PH symmetry is not a necessary condition for the observation of the FQHE and the CFFS, given that real experiments always involve some LL mixing, which causes no (measurable) correction to the value of the quantized Hall resistance.

The interplay between the emergence of composite fermions and the PH symmetry of electrons has attracted attention in recent years. It has led, on the one hand, Son to propose an effective theory that views composite fermions as Dirac particles~\cite{Son15}, and, on the other, to improved calculations within the CS field theory of HLR. These developments are discussed in the chapter by Halperin. How about the microscopic theory of composite fermions as defined by the LLL-projected wave functions in Eq.~\ref{jainwf}? PH symmetry is neither imposed on these wave functions nor a priori evident, but explicit calculations have  demonstrated that they satisfy PH symmetry to an extremely high degree\footnote{From the fact that $\nu=n/(2n+1)$ maps into $\nu^*=n$ whereas its hole partner $\nu=1-n/(2n+1)=(n+1)/(2n+1)$ into $\nu^*=-(n+1)$, it may appear that the CF theory does not respect PH symmetry. That is not correct. The state obtained from composite-fermionization of $\nu^*=-(n+1)$ is equivalent to the hole partner of the state obtained from composite-fermionization of $\nu^*=n$ in all topological aspects. Their edge physics are identical as are their mean-field gaps (see Supplemental Material of Ref.~\refcite{Balram15b}). Furthermore, the explicit wave functions constructed in the two approaches have almost perfect overlap~\cite{Wu93,Davenport12}. Incidentally, as discussed in Section~\ref{sec:spin}, the mapping of $\nu=n/(2n+1)$ and $\nu=(n+1)/(2n+1)$ into $\nu^*=n$ and $\nu^*=-(n+1)$, respectively, is crucial for explaining the qualitatively different spin physics at these filling factors. For example, $\nu=1/3$, which maps into $\nu^*=1$, is predicted to be always fully spin polarized, whereas $\nu=2/3$, which maps into $\nu^*=-2$, is predicted to admit both fully spin polarized and spin singlet states, depending on whether the Zeeman energy is larger or smaller than the CF cyclotron energy. Both spin singlet and fully spin polarized states have been observed at $\nu=2/3$; the nature of these states and the phase transition between them are quantitatively well explained by the CF theory.} for both the FQH states~\cite{Wu93,Davenport12,Balram16b} and the CFFS~\cite{Rezayi00,Fremling18,Pu18}. This is a corollary of the fact that these wave functions are very close to the Coulomb eigenstates, which satisfy the PH symmetry exactly. The wave functions in Eq.~\ref{jainwf} are constructed by composite-fermionizing the IQH states and Fermi sea of non-relativistic electrons.

\underline{New emergent structures due to inter-CF interaction:} As always, explanation of finer and finer features of experimental observations requires increasingly more sophisticated theoretical models and approximations. The model of non-interacting electrons explains the most robust phenomenon, namely the IQHE, but the interaction between electrons causes new structure, namely the FQHE. Analogously, the model of non-interacting composite fermions explains FQHE at $\nu=n/(2pn\pm 1)$ and $\nu=1-n/(2pn\pm 1)$, which exhaust a large majority of the observed fractions, but not all. Certain fractions require a consideration of the residual interaction between composite fermions, which is complex but can be determined within the CF theory~\cite{Lee02,Jain07}. The FQH states at $\nu=4/11$ and $\nu=5/13$ are examples of FQHE of composite fermions~\cite{Pan02,Samkharadze15b,Pan15}. Another example of new physics arising from the inter-CF interaction is the 5/2 state, which is believed to occur because of a p-wave pairing instability of the CF Fermi sea~\cite{Moore91,Read00} (see the chapters by Halperin and Heiblum and Feldman). One may ask how pairing can arise in a model with purely repulsive interaction. It arises because the objects forming pairs are not electrons but composite fermions. The interaction between composite fermions, which is a complex function of the interaction between electrons, is weak, and nothing really forbids it from being attractive. Explicit calculations indicate that at $\nu=5/2$, the binding of two vortices by electrons over-screens the repulsive Coulomb interaction between electrons to produce a weakly attractive interaction between composite fermions~\cite{Scarola00}. (In contrast, the inter-CF interaction remains repulsive~\cite{Scarola00} at $\nu=1/2$, where the interaction between electrons is more strongly repulsive than that at $\nu=5/2$.) Certain other paired states of composite fermions are considered in Section~\ref{sec:parton}.

\underline{FQHE in graphene:} In recent years, graphene has produced extensive FQHE. For the Dirac electrons of graphene, LLs occur for positive and negative energies, have a spacing proportional to $\sqrt{|n|}$ where $n$ is the LL index, and the $n=0$ LL is located at zero energy. When one restricts the Hilbert space to a specific LL, the LLs of Dirac electrons differ from those of non-relativistic electrons in two aspects. First, there is additional degeneracy in graphene because of two valleys. The valley degree of freedom can be accommodated into the CF theory in the same manner as the spin. Second, the Coulomb matrix elements are in general different from those in the LLs of non-relativistic electrons. It turns out that for the $n=0$ LL, the Coulomb matrix elements for Dirac and non-relativistic electrons are identical (for a strictly 2D system). The observed FQHE in the graphene $n=0$ LL corresponds precisely to what is expected from the CF theory. The Coulomb matrix elements in the $n=1$ graphene LL are different from those of the $n=1$ LL of non-relativistic electrons and closer to those of the $n=0$ LL. Indeed, the FQHE in the $n=1$ graphene LL is also explained nicely in terms of non-interacting composite fermions. The status of FQHE in graphene is reviewed in the chapter by Dean, Kim, Li and Young. 

\underline{The role of topology in FQHE:}  It is useful to ask the question~\cite{Zee10,Tong16}: What can we say about the properties of a FQH state without knowing its microscopic origin? Here one assumes a gapped state at a certain filling factor and asks what quantum field theory would produce a non-zero Hall conductance. Electrons, being high energy objects, are not a part of this theory, which, as any effective field theory, deals with the low-energy physics. This line of reasoning naturally leads to CS theories with emergent gauge fields~\cite{Zee10,Tong16}. These theories make precise predictions for certain quantities that are of topological origin, i.e. are invariant under continuous changes of the Hamiltonian so long as no phase boundary is breached (which is why their calculation does not require a microscopic understanding). In particular, the CS theories reveal the existence of quasiparticles with fractional charge and fractional braid statistics~\cite{Zee10,Tong16}.

The current chapter focuses on the microscopic mechanism of the FQHE. You may recall seeing an animated GIF in a continuous loop, perhaps in a physics department colloquium, showing a coffee mug adiabatically metamorphosing into a doughnut and back, to drive home the fact that the two share the same genus-one topology. The coffee mug and the doughnut are of course different objects, as even a topologist may ascertain by performing the experiment, with care, of biting hard or pouring hot coffee into them. A master chef ready to prepare a doughnut will need to know, aside from its toroidal shape, the various ingredients as well as the recipe for how to put them together. We are similarly concerned in this chapter with the microscopic ingredients of the FQHE (composite fermions) and how they are assembled into various states (IQHE, Fermi sea, crystal, etc.) to produce the phenomenology. We are concerned with microscopic wave functions and calculation of measurable quantities. It turns out, nonetheless, that topology lies at the front and center of the CF theory, for the simple reason that composite fermions themselves are topological particles. The attached vortices endow composite fermions with a U(1) topological character, which, in turn, manifests directly through the effective magnetic field experienced by composite fermions. The effective magnetic field has been measured and is responsible for the explanation or prediction of the vast body of unexpected phenomenology of the FQHE. All of the qualitative phenomenology of composite fermions thus has topological origin. In fact, the FQHE is doubly topological. Recall that IQHE is topological because electrons fill topological bands (LLs) characterized by non-zero Chern numbers. In FQHE, topological particles (composite fermions) fill topological bands ($\Lambda$Ls). The two topological quantum numbers characterizing a FQH state are $2p$, the CF vorticity, and $n$, the number of filled $\Lambda$Ls. It is worth stressing that while all topological properties of the FQHE can be derived starting from the CF theory, the existence of composite fermions and their effective magnetic field, which relate to the microscopic origin of the FQHE, cannot be derived from the purely topological perspective mentioned in the preceding paragraph.

\underline{Fractional charge and fractional braid statistics:} An attentive reader may have noticed that the above explanations of the FQHE and other related phenomena make no mention of fractional charge and fractional braid statistics. That composite fermions are fermions is beyond question. Their fermionic nature is central to the explanations of the FQHE as the IQHE of composite fermions and of the 1/2 state as the Fermi sea of composite fermions.  Furthermore, computer calculations confirm, beyond doubt, that the quasiparticles and quasiholes are nothing but excited composite fermions or the holes they leave behind, as depicted in Fig.~\ref{fig:Jeonel-CF}. At the same time, the existence of fractional charge and fractional braid statistics for the quasiparticles or quasiholes can be inferred from no more than the assumption of a gap at a fractional filling factor; in fact, the allowed values for them can be derived without an understanding of the microscopic origin of the FQHE\footnote{\label{fnstatistics}The value of the filling factor puts constraints on the allowed values for the charge and braid statistics of the quasiparticles~\cite{Su86}. Assuming an incompressible state at $\nu=n/(2pn\pm 1)$, adiabatic insertion of a unit flux  produces, \`a la Laughlin\cite{Laughlin83}, an excitation of charge $en/(2pn\pm 1)$. This in general is a collection of several elementary quasiparticles. Assuming that we have a single type of elementary quasiparticles, the requirement that an integer number of them also produce an electron gives $e^*=e/[k(2pn\pm 1)]$, where $k$ is an arbitrary integer. The simplest choice corresponds to $k=1$. Braid statistics of the elementary quasiparticles can be deduced analogously from general considerations~\cite{Su86}.}.
In spite of the appearances, there is no contradiction. The fractional charge and fractional braid statistics can be derived within the CF theory as follows. Consider the state at a filling factor $\nu^*=n$ with two additional composite fermions in the $(n+1)^{\rm st}$ $\Lambda$L. One may seek an effective formulation of the problem in terms of only two particles by integrating out all composite fermions in the the lower filled $\Lambda$Ls. This must be done with care, however, because the two composite fermions in the $(n+1)^{\rm st}$ $\Lambda$L are topologically correlated with the composite fermions in lower filled $\Lambda$Ls as well (i.e. see $2p$ vortices on them). The effect of the lower filled $\Lambda$Ls is to ``screen" both the charge and the braid statistics of the composite fermions in the $(n+1)^{\rm st}$ $\Lambda$L.  There are several ways within the CF theory to derive~\cite{Jeon04,Jain07} the fractional charge $e^*=e/(2pn\pm 1)$ and braid statistics parameter $\alpha=2p/(2pn\pm 1)$ for the quasiparticles of the $\nu=n/(2pn\pm 1)$ FQH state.  These  are the simplest  values allowed by general considerations, and are also in agreement with those produced previously by the hierarchy theory~\cite{Halperin84}.

The CF theory goes beyond these quantum numbers and gives a precise microscopic account of the quasiparticles and quasiholes of all $\nu=n/(2pn\pm 1)$ states, which allows us to calculate their density profiles, energies, interactions, dispersions, {\em etc}. Most remarkably, the CF theory reveals that the quasiparticles of all $\nu=n/(2pn\pm 1)$ FQH states are, in a deep sense, the same objects, namely composite fermions, which are also the particles that form the ground states. Composite fermions remain sharply defined even when the concept of fractional charge and fractional braid statistics ceases to be meaningful, e.g. at $\nu=1/2$ (where the state is compressible), or when a $\Lambda$L is sufficiently populated that the composite fermions in that $\Lambda$L are strongly overlapping.

\section{Quantitative comparison with laboratory experiments}
\label{sec:laboratory}

Given the accuracy of the CF theory as seen in computer experiments, we can dispense with exact diagonalization and study systems of composite fermions. With the help of convenient numerical methods for LLL projection~\cite{Jain97,Jain97b} and CF diagonalization~\cite{Mandal02}, we can go to large systems (with as many as 200 composite fermions or more) to explore phenomena that are not accessible in exact diagonalization studies, and also to obtain thermodynamic limits for various quantities of experimental interest. Numerous observables, such as excitation gaps, dispersions of the neutral CF exciton, dispersions of spin waves, phase diagrams of various states as a function of parameters, have been calculated (see Refs.~\refcite{Jain07,Jain15} for a review).  A priori, one should expect a few percent agreement between theory and experiment (which can be systematically further improved if so desired). That indeed would have been the case had we been dealing with a phenomenon in atomic or high energy physics, but the FQH systems, in spite of being among the most pristine and the best characterized of all condensed matter systems, present additional complications. Unlike experiments in atomic or high energy physics, FQH experiments in different laboratories and different samples produce different numbers, because the experimental results are modified by features that were set to zero in computer studies mentioned in the previous section, namely finite quantum well width, LL mixing and disorder. These must be included in the theoretical calculation for a precise quantitative comparison. It is somewhat ironic that we have an extremely accurate quantitative understanding of the nontrivial part of the physics, namely the FQHE, but our understanding of the corrections due to finite width, LL mixing and disorder is less precise. That is the reason why quantitative comparisons with experiments, while decent, do not reflect the full potential of the CF theory.

This section is devoted to recent calculations~\cite{Zhang16,Zhao18} that incorporate the effects of finite width and LL mixing (Sections~\ref{sec:LDA} and \ref{sec:DMC}) to the best extent currently possible. Because we do not include disorder, we focus on thermodynamic quantities that are not expected to be very sensitive to disorder, as opposed to quantities such as excitations gaps that are more strongly affected by disorder. 

Section~\ref{sec:spin} considers transitions between differently spin polarized FQH states. These are understood, physically, as $\Lambda$L crossing transitions as the Zeeman energy is varied relative to the CF cyclotron energy. The critical Zeeman energies at which these transitions are observed are a direct measure of the differences between the Coulomb energies of the competing states. Comparisons with experiments show that after incorporating finite width and LL mixing corrections, the CF theory obtains these energy differences, which are on the order of 1\% of the individual energies, with a few percent accuracy. These calculations also shed light on the dissimilarities observed between the behaviors at $\nu=n/(2n\pm 1)$ and $\nu=2-n/(2n\pm 1)$.

Section~\ref{sec:crystal} deals with the competition between the liquid and the crystal phases as a function of filling factor and LL mixing. It provides evidence that the crystal phase is not an ordinary, featureless Wigner crystal of electrons but contains a series of crystals of composite fermions with different vorticity. The essential theoretical picture is that as the filling factor is lowered, at some point composite fermions begin to bind fewer than the maximal number of vortices available to them and use the remaining freedom to form a crystal of composite fermions. Given how favorable the CF correlations are, it should not be surprising that nature would exploit them even in the crystal phase to find the lowest energy state.  In particular, theoretical calculations show that the crystal of composite fermions with two attached vortices is energetically favored over the FQH state of composite fermions with four attached vortices for a narrow range of filling factors between $\nu=1/5$ and $\nu=2/9$, thus explaining the observed insulating phase between the 1/5 and 2/9 FQH liquid states. The CF crystal beats the FQHE here by a mere $\sim$0.0005 $e^2/\epsilon l$ per particle, which is an indication of the theoretical accuracy required to capture the physics of the re-entrant crystal phase. Calculations further show that the enhanced LL mixing in low-density p-doped GaAs quantum wells  also stabilizes a crystal in between $\nu=1/3$ and 2/5, as seen experimentally. 

One may ask: Given that the underlying CF physics is already well established, why expend a substantial amount of effort toward calculating numbers very precisely? The reason, from a general perspective, is that progress in physics often relies on a precise quantitative understanding of experiments, which prepares the ground for new discoveries. Significant quantitative deviations between theory and experiment are inevitably found as more accurate tests are performed and as new regimes are explored, pointing to new physics. In the context of the FQHE, an additional motivation for seeking a precise microscopic understanding of experiments is simply that we can. The FQHE is a rare example of a highly nontrivial strongly-correlated state for which it has been possible to achieve a detailed microscopic description in the quantum chemistry sense. Given that an understanding of the role of interactions is a primary goal of modern condensed matter physics, it appears to be of value to push the comparison between theory and experiment in FQHE to its limits. 

\subsection{Finite width corrections: Local density approximation}
\label{sec:LDA}

The nonzero transverse width of GaAs-${\mathrm{Al}}_{\mathrm{x}}$${\mathrm{Ga}}_{1\mathrm{-}\mathrm{x}}$As heterojunctions and quantum wells can be incorporated into theory by using an effective 2D interaction given by:
\begin{equation}
V^{\text{eff}}(r) = \frac{e^{2}}{\epsilon} \int dz_{1} \int dz_{2} \frac{|\xi(z_{1})|^{2} |\xi(z_{2})|^{2}}{[r^{2} + (z_{1} - z_{2})^{2}]^{1/2}},
\end{equation}
where $\xi(z)$ is the transverse wave function, $z_{1}$ and $z_{2}$ denote the real coordinates perpendicular to the 2D plane ($z$ here is not to be confused with the complex in-plane coordinate introduced previously), and $r =\sqrt{ (x_{1} - x_{2})^{2}+ (y_{1} - y_{2})^{2}}$. The interaction $V^{\text{eff}}(r)$ is less repulsive at short distances than the ideal 2D interaction $e^{2}/\epsilon r$.  We need a model for $\xi(z)$. At zero magnetic field, a realistic $\xi(z)$ for any given density and quantum well width can be obtained by solving the Schr\"odinger and Poisson equations self-consistently in the density functional theory with the exchange-correlation functional treated in a local density approximation (LDA)~\cite{Ortalano97}. (For an earlier model, see Ref.~\refcite{Zhang86}.) The resulting $V^{\text{eff}}(r)$ depends on both quantum well width and the electron density. It is customary to assume that $\xi(z)$ remains unaffected by the application of a magnetic field perpendicular to the 2D plane.

\subsection{LL mixing: fixed phase diffusion Monte Carlo method}
\label{sec:DMC}

The parameter $\kappa=(e^2/\epsilon l)/\hbar\omega_c$, the ratio of the Coulomb interaction to the cyclotron energy,  provides a measure of LL mixing.  It is related to the standard parameter $r_s$ of electrons (namely the interparticle separation in units of the Bohr radius) through $\kappa=(\nu/2)^{1/2}r_s$. LL mixing is suppressed in the limit $\kappa\rightarrow 0$. For small values of $\kappa$, the effect of LL mixing can be treated in a perturbative approach~\cite{MacDonald84,Melik-Alaverdian95,Murthy02,Bishara09,Wojs10, Sodemann13, Simon13,Peterson13,Peterson14} that modifies the 2D interaction. However, the reliability of the perturbative treatment for typical experiments is unclear, given that $\kappa\sim 0.8-2$ for n-doped GaAs and $\kappa\sim 2-20$ in p-doped GaAs systems. A lack of quantitative understanding of LL mixing has been an impediment to the goal of an accurate comparison between theory and experiment.

We treat the effect of LL mixing through the nonperturbative method of fixed-phase diffusion Monte Carlo (DMC) calculations~\cite{Ortiz93,Melik-Alaverdian97, Melik-Alaverdian01}.  This is a generalization of the powerful DMC method~\cite{Reynolds82,Foulkes01} for obtaining the ``exact" ground state energies for certain interacting systems. We give here a brief account of the method; more details can be found in the literature. 

Let us assume that the ground state wave function is real and non-negative, as is the case for bosons. The Schr\"odinger equation for imaginary time ($t\rightarrow i t$)
\be
-{\partial\over \partial t} \Psi({\mathcal R},t)=(H-E_T) \Psi({\mathcal R},t)
\ee
can then be viewed as a diffusion equation with the wave function $\Psi({\mathcal R},t)$ interpreted as the density of the diffusing particles. Here ${\mathcal R}$ collectively represents the coordinates of all the particles and $E_T$ is a conveniently chosen energy offset. Let us now begin with an initial trial function $\Psi({\mathcal R},t=0)$ which can be expressed in terms of the exact eigenstates $\Phi_\alpha$ as $\Psi({\mathcal R},t=0)=\sum_\alpha C_\alpha \Phi_\alpha$. Its evolution in imaginary time is given by 
\be
\Psi({\mathcal R},t)=\sum_\alpha C_\alpha  e^{-(E_\alpha-E_T)t}\Phi_\alpha\rightarrow C_0 e^{-(E_0-E_T)t}\Phi_0\;\;{\rm for}\; t\rightarrow \infty\;.
\ee
Thus, in the large imaginary time limit the evolution operator projects out the ground state provided it has non-zero overlap with the initial trial wave function. DMC is a stochastic projector method for implementing this scheme through an importance sampling method using a trial or guiding wave function. In the absence of a potential, we have the distribution of random walkers (or diffusing Brownian particles) in the $2N$ dimensional configuration space. In the presence of a potential, the most effective method is through a branching (or a birth / death) algorithm in which either a walker dies with some probability in regions of high potential energy, or new walkers are created in regions of low potential energy, according to certain rules. The probability distribution of the walkers converges to the ground state in the limit $t\rightarrow\infty$. The energy offset $E_T$ controls the population of the walkers; one adjusts $E_T$ to keep the walker population at around 100 - 1000. 
The energy offset must be adjusted to the ground state energy to obtain a stationary distribution. Alternatively the energy can be obtained from an average of the so-called local energy.  One typically keeps the acceptance ratio at around 99\%.

The DMC method cannot be applied directly to FQH systems, which, due to the broken time-reversal symmetry, have complex valued eigenfunctions. For such systems, an approximate strategy known as the fixed-phase DMC was introduced by Ortiz, Ceperley and Martin (OCM)~\cite{Ortiz93} which searches for the ground state in a restricted subspace. (The fixed phase DMC is closely related to the fixed node DMC used for real wave functions~\cite{Melton17}.) Following OCM, we substitute $\Psi(\mathcal R)=\Phi(\mathcal R)e^{i\varphi(\mathcal R)}$ where $\Phi(\mathcal R)=|\Psi(\mathcal R)|$ is real and non-negative. The term ``phase" in fixed phase DMC is used for the phase $\varphi(\mathcal R)$ of the wave function, and not for the phase (e.g. liquid, crystal) of the system. The variational energy of the system of interacting electrons in a magnetic field is given by $\langle \Psi(\mathcal R)|H|\Psi(\mathcal R) \rangle=\langle \Phi(\mathcal R)|H_R|\Phi(\mathcal R) \rangle$ with $H_R= \sum_{j=1}^N \left[ \vec{p}_j^2 +[\hbar\vec{\nabla}_j\varphi(\mathcal R)+(e/c)\vec{A}(\vec{r}_j)]^2\right]/2m+V_{\rm Coulomb}(\mathcal R)$. Now, keeping the phase $\varphi(\mathcal R)$ fixed and varying $\Phi(\mathcal R)$ gives us the lowest energy within the subspace of wave functions defined by the phase sector $\varphi(\mathcal R)$. This minimization can be conveniently accomplished by applying the DMC method to the imaginary time Schr\"odinger equation $-\hbar{\partial\over\partial t}\Phi(\mathcal R,t)=\left[ H_R(\mathcal R)-E_T)\right]\Phi(\mathcal R,t)$. The essence of the fixed phase DMC is to transform the fermionic problem into a bosonic one at the expense of an additional vector potential in the Hamiltonian that essentially corresponds to a fictitious magnetic field.  The fixed phase DMC produces the lowest energy in the chosen phase sector, and hence a variational upper bound for the exact ground state energy.  It would produce the exact ground state energy if we knew the phase of the exact ground state, which we do not. 

The accuracy of the energy obtained from fixed phase DMC is critically dependent on the choice of the phase $\varphi(\mathcal R)$.  G\"u\c{c}l\"u and Umrigar~\cite{Guclu05} found in exact diagonalization studies of certain small systems (maximum density droplets) that the phase of the wave function is not significantly altered by LL mixing. Following their lead, the calculations shown below use the accurate LLL wave functions of the CF theory as the trial wave functions to fix the phase $\varphi(\mathcal R)$. In cases where a comparison has been made, fixing the phase with the more accurate LLL wave function (e.g. the CF Fermi sea versus the Pfaffian wave function at $\nu=1/2$) produces lower energy for up to the largest values of $\kappa$ considered. This choice has another advantage: it keeps the system in the topological sector defined by the LLL trial wave function. Nonetheless, the results are subject to this assumption regarding the phase, the validity of which can ultimately be justified only by a detailed comparison of the numerical results with experiments. The calculations use the generalization by Melik-Alaveridan, Bonesteel and Ortiz~\cite{Melik-Alaverdian97, Melik-Alaverdian01} of the fixed-phase DMC method to the spherical geometry through a stereographic projection.

\subsection{Spin phase transitions}
\label{sec:spin}

\begin{figure}[t]
\begin{center}
\includegraphics[width=4.0in]{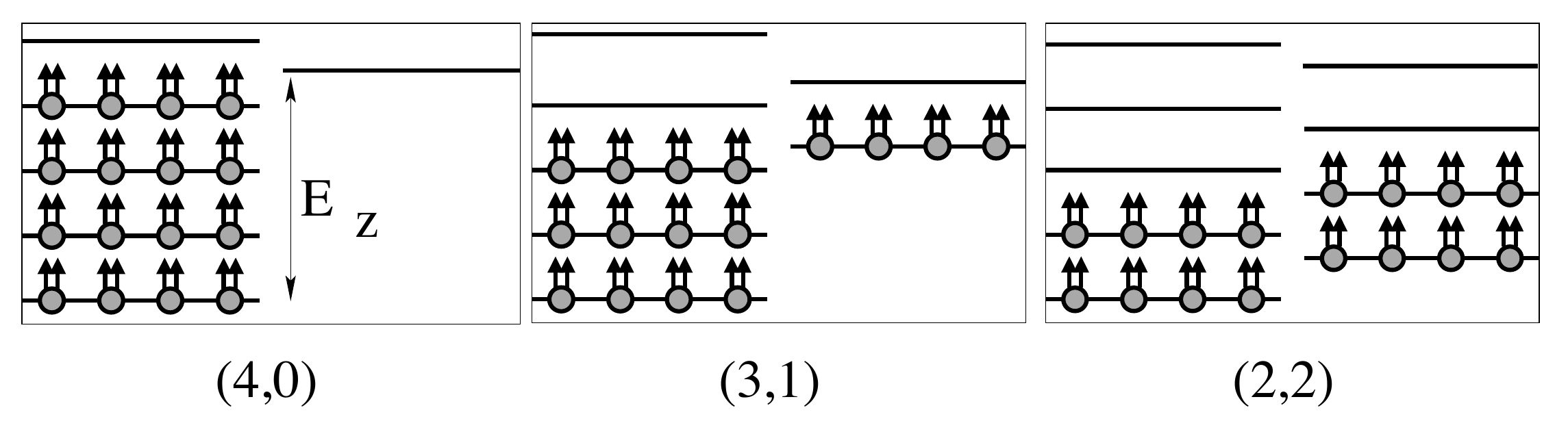}
\end{center}
\caption{Schematic view of the FQH state at $\nu=4/9$ (or $4/7$), which maps into $\nu^*=4$ filled $\Lambda$ levels, as a function of the Zeeman energy, $E_{\rm Z}$.  The three possible states are $(n_\uparrow, n_\downarrow)=(4,0)$, $(3,1)$, and $(2,2)$, which are fully polarized, partially polarized, and spin singlet, respectively. 
\label{fig:spin49}
}
\end{figure}

The explanation of FQHE as the IQHE of composite fermions also gives an understanding of the spin physics. The FQHE at $\nu=n/(2pn\pm1)$ still maps into IQHE of composite fermions at $|\nu^*|=n$ but, in general, we have $n=n_\uparrow + n_\downarrow$, where $n_\uparrow$ and $n_\downarrow$ are the number of occupied up-spin and down-spin $\Lambda$ levels. These states are labeled $(n_\uparrow, n_\downarrow)$. The allowed spin polarizations are then given by $\gamma=(n_\uparrow - n_\downarrow)/(n_\uparrow + n_\downarrow)$. Fig.~\ref{fig:spin49} depicts the situation for $\nu=4/9$ or $4/7$, where three distinct spin polarizations are possible. In particular, in the limit of zero Zeeman energy, the states at fractions with even numerators are predicted to be spin singlet, whereas those at fractions with odd numerators are predicted to be fully spin polarized for $n=1$ and partially spin polarized for $n\geq 3$.

Experimentally, transitions between differently spin polarized FQH states can be driven by tuning the Zeeman energy, which can be accomplished either by application of an additional parallel magnetic field (tilted field experiments), or by changing the density. A wealth of experimental information exists for the critical energies where such transitions occur~\cite{Eisenstein89, Eisenstein90, Engel92, Du95,Kang97, Kukushkin99, Yeh99,Kukushkin00, Melinte00,Freytag01, Tiemann12,Feldman13,Liu14}, and the number of transitions seen in experiments is generally in agreement with the prediction from the CF theory. The physical picture is that the transitions are essentially $\Lambda$L crossing transitions occurring due to a competition between the CF cyclotron energy and the Zeeman splitting. 

To obtain a more quantitative comparison, it is convenient to quote the Zeeman energy in units of the Coulomb energy, which we denote as $\alpha_{\rm Z}=E_{\rm Z}/(e^{2}/\epsilon l)$. The critical Zeeman energy for the transition between two successive states $(n_{\uparrow}, n_{\downarrow})$ and $(n_{\uparrow}-1, n_{\downarrow}+1)$ is given by
\begin{equation}
\alpha_{\rm Z}^{\rm crit}  = {E_{\rm Z}^{\rm crit}\over e^{2}/\epsilon l}=(n_{\uparrow}+n_{\downarrow})\left [ \frac{ E_{(n_{\uparrow}, n_{\downarrow})} - E_{(n_{\uparrow}-1, n_{\downarrow}+1)} }{e^{2}/\epsilon l} \right] .
\label{Ec}
\end{equation}
where $E_{(n_{\uparrow}, n_{\downarrow})}$ is the per particle Coulomb energy of the states $(n_{\uparrow}, n_{\downarrow})$. The critical Zeeman energy is thus a direct measure of the difference between the Coulomb energies of the two competing states. These energy differences are on the order of 1\% or less of the individual Coulomb energies~\cite{Park98}, and their calculation thus serves as a sensitive test of the quantitative accuracy of the theory. 

Let us first consider the ideal system with no LL mixing and no finite width corrections. In this limit, the LLL wave functions for the states $(n_{\uparrow}, n_{\downarrow})$ are accurately given by
\begin{equation}
\Psi^{\rm full}_{n\over 2pn\pm 1}  = A[\Psi_{(n_\uparrow,n_\downarrow)} u_1\cdots u_{N_\uparrow} d_{N_\uparrow+1}\cdots d_N] 
\label{CFWFfull}
\end{equation}
with
\begin{equation}
\Psi_{(n_\uparrow,n_\downarrow)}   =  \mathcal{P}_{\text{LLL}}\Phi_{\pm n_{\uparrow}}(z_1,\cdots z_{N_\uparrow}) \Phi_{\pm n_{\downarrow}}(z_{N_\uparrow+1}\cdots z_N) \prod_{j<k}(z_j-z_k)^{2p} 
\label{CFWF}
\end{equation}
Here A denotes antisymmetrization, and $u_j$ and $d_j$ are the up and down spinor wave functions. 
The LLL projection is performed using the method in Refs.~\refcite{Jain97,Jain97b,Jain07,Davenport12}. We note that these wave functions automatically satisfy Fock condition, i.e. have are eigenstates of $\vec{S}^2$ with eigenvalue $S(S+1)$ with $S=S_z$. For the calculation of the Coulomb energy, it is sufficient to work with Eq.~\ref{CFWF} rather than Eq.~\ref{CFWFfull}. 

\begin{figure}[t]
\begin{center}
\includegraphics[width=3.0in]{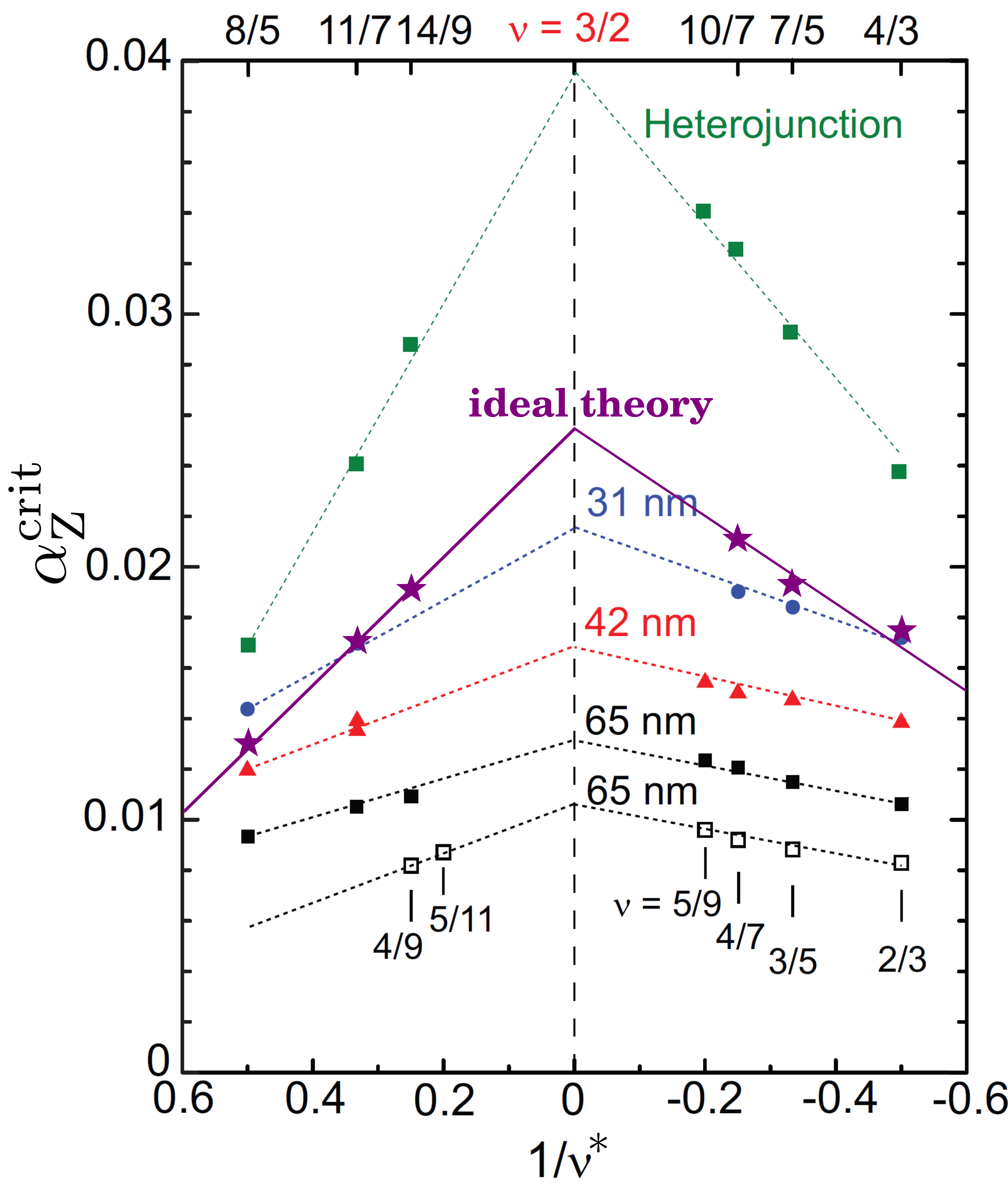}
\end{center}
\caption{Critical Zeeman energies for spin phase transitions. The purple stars marked ``ideal theory" show the theoretical prediction for the critical Zeeman energies for transitions from the fully spin polarized state into a partially spin polarized or a spin singlet state for states at $\nu=n/(2n\pm 1)$; these are obtained from exact diagonalization for a system with zero thickness and no LL mixing~\cite{Balram15a}. All other symbols are from experiments. The top green symbols are taken from Du {\em et al.}~\cite{Du95}, obtained in the heterojunction geometry. All other  results are from Liu {\em et al.}~\cite{Liu14} for experiments in quantum wells of various thicknesses shown on the figure. All experimental results are for filling factors of the form $\nu=2-n/(2n\pm 1)$, shown on the top, except for the lowest results, which are for filling factors of the form $\nu=n/(2n\pm 1)$ shown on the figure. Source: Y. Liu, S. Hasdemir, A. W\'ojs, J. K. Jain, L. N. Pfeiffer, K. W. West, K. W. Baldwin, and M. Shayegan, Phys. Rev. B {\bf 90}, 085301 (2014)~\cite{Liu14}.
\label{spin_fig}
}
\end{figure}

From these wave functions, the energies of the ground states of various spin polarizations have been calculated for many filling factors~\cite{Park98, Park99, Davenport12,Balram15a}. The thermodynamic energies are determined from an extrapolation of finite system results. The predicted critical Zeeman energies are shown in Fig.~\ref{spin_fig}, along with experimental results. [The theoretical critical energies shown in this figure are actually determined from exact diagonalization studies~\cite{Balram15a}. The CF energies for the states at $n/(2n+1)$ are very accurate with the standard projection method. For the $n/(2n-1)$ states, on the other hand, the standard projection method~\cite{Jain97,Davenport12} slightly overestimates the probability of spatial coincidence of electrons in the nonfully polarized states, and thereby overestimates their energies. The hard-core projection of Ref.~\refcite{Wu93} produces very accurate energies, but is not amenable to large scale numerical evaluations.] The theory successfully captures the energy ordering of the differently polarized states, produces critical Zeeman energies that are generally consistent with experiments, and also captures the tent-like behavior of the critical Zeeman energy around $\nu=1/2$.

Some discrepancy between theory and experiment remains, however, which is not surprising given that the theoretical calculations omit the effects of finite width and LL mixing. Here are the primary deviations: (i)  The actual numbers for $\alpha_{\rm Z}^{\rm crit}$ can be off by up to a factor of 2-3.  (ii) The spin phase transitions appear to be strongly affected by the breaking of PH symmetry due to LL mixing. For a system confined to the LLL and interacting by a two-body interaction, there is an exact PH symmetry relating filling factors $\nu$ and $2-\nu$, which implies that $\alpha_{\rm Z}^{\rm crit}$ for a fraction $\nu=2-n/(2n\pm 1)$ is identical to that for $\nu=n/(2n\pm 1)$. That is not the case in experiments, however. Spin transitions are readily observed for fractions $\nu=2-n/(2n\pm 1)$ but not for $\nu=n/(2n\pm 1)$ (even after using densities so that the fractions are seen at the same $B$). As an example, the $\nu=8/5$ was the first state were a spin transition was observed~\cite{Eisenstein89}, but a transition at 2/5 could be seen only after reducing the Land\'e g-factor substantially by application of hydrostatic pressure~\cite{Kang97}. (iii) For the heterojunction samples, the measured critical Zeeman energies at $\nu=2-n/(2n\pm 1)$ lie above the ideal theoretical values. This is surprising because both finite width and LL mixing reduce the Coulomb energies, and therefore should generally reduce the critical Zeeman energies.

\begin{figure}[t]
\begin{center}
\includegraphics[width=0.48\textwidth]{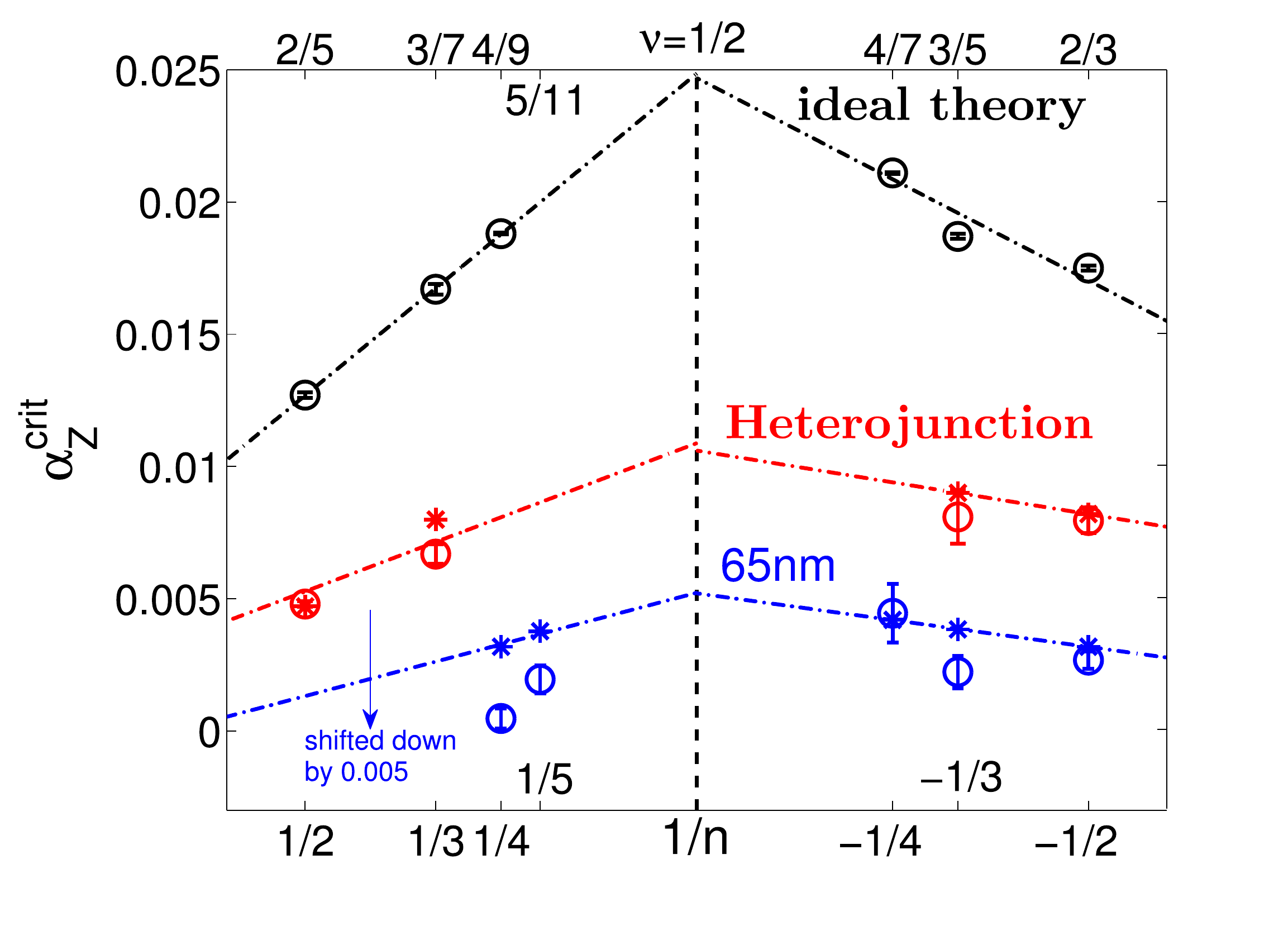} 
\includegraphics[width=0.43\textwidth]{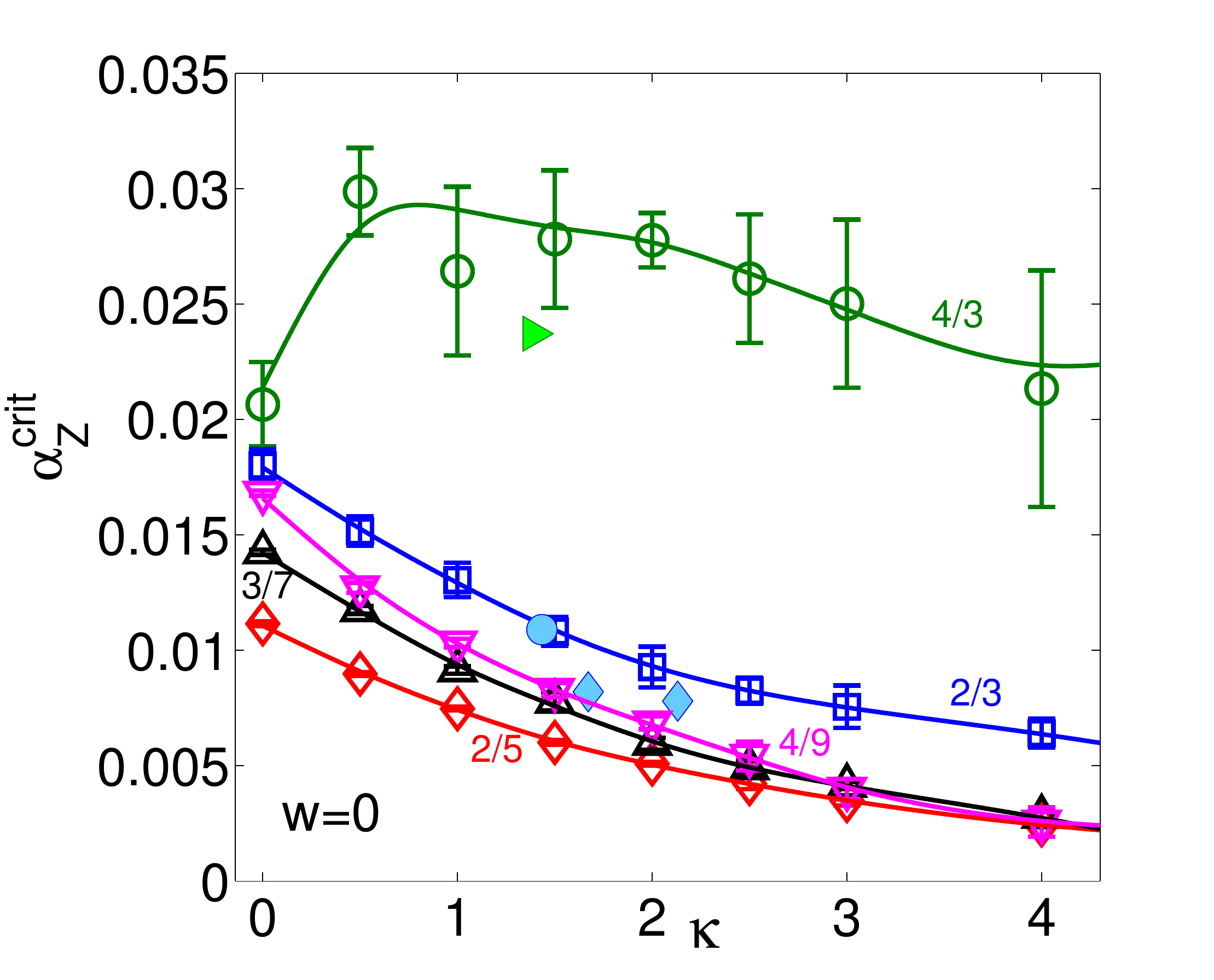}
\end{center}
\caption{
Comparison between experimental critical Zeeman energies $\alpha_{\rm Z}^{\rm crit}=E_{\rm Z}^{\rm crit}/(e^{2}/\epsilon l)$ with theoretical results from fixed phase DMC. Left panel: The blue and red stars show results from  experiments on a 65 nm wide quantum well and a heterojunction, taken from Liu {\em et al.}~\cite{Liu14}, Engel {\em et al.}~\cite{Engel92} and Kang {\em et al.}~\cite{Kang97}. (For the experiment of Kang {\em et al.}, we estimate the value of the Land\'e factor $g_0$ by assuming that it changes linearly and passes through zero at a pressure of roughly 18 Kbar~\cite{Leadley97}.) The blue and red circles show the results from fixed phase DMC calculation for corresponding widths and densities. The results for the 65 nm quantum well are shifted down by 0.005 for ease of depiction. The theoretical results without including the effects of LL mixing and finite width are also shown for comparison. The dashed lines are a guide to the eye. Right panel: Theoretical critical Zeeman energies for $w=0$ as a function of the LL mixing parameter $\kappa$ obtained from the DMC method for $\nu=4/3$ (green circle), $2/3$ (blue square), $4/9$ (magenta downward triangle), $3/7$ (black upward triangle), and $2/5$ (red diamond).  The solid lines are an approximate guide to the eye. The filled symbols indicate the experimental data from heterojunction samples at $\nu=2/3$ (light blue) and $4/3$ (green) taken from Eisenstein {\em et al.}~\cite{Eisenstein90} (circle), Engel {\em et al.}~\cite{Engel92} (diamond), and Du {\em et al.}~\cite{Du95} (rightward triangle). Source: Y. Zhang, A. W\'ojs, and J. K. Jain, Phys. Rev. Lett. {\bf 117}, 116803 (2016)~\cite{Zhang16}.
}
\label{compare}
\end{figure}

Ref.~\refcite{Zhang16} has investigated how the results are modified when we include the effects of finite width and LL mixing, by evaluating the thermodynamic limits of the Coulomb energies of the relevant states as a function of the quantum well widths and densities. Fig.~\ref{compare} (left panel) shows theoretical results for several states of the form $n/(2n\pm 1)$ obtained using the experimental parameters (width, density and LL mixing), along with the experimental results. 

How about the breaking of PH symmetry seen in experiments? It is computationally expensive to deal with the states at $\nu=2-n/(2n\pm 1)$, and therefore Ref.~\refcite{Zhang16} only compared the spin transitions at $\nu=2/3$ and $\nu=4/3$ for a zero width system. To obtain accurate results, the initial trial wave function is chosen as (i) the exact LLL ($\kappa=0$) Coulomb ground states for the spin singlet states at $2/3$ and $4/3$; (ii) Eq.~\ref{CFWF} for the fully polarized state at $\nu=2/3$; and (iii) $\Phi_{1\uparrow} \Psi_{1/3 \downarrow}$ for the partially polarized state at $\nu=4/3$. Fig.~\ref{compare} (right panel) shows that the $\alpha_{\rm Z}^{\rm crit}$ for 4/3 is substantially higher than that for 2/3 for the typical experimental value of $\kappa\approx 1-2$.  This figure also contains the experimental data from GaAs-Al$_{x}$Ga$_{1-x}$As heterojunction samples, because these have the smallest effective width. (The comparison with zero width results is meaningful, because at relatively large $\kappa$ the results are not particularly sensitive to the width.)  

The high degree of agreement between theory and experiment seen in Fig.~\ref{compare} demonstrates that the CF theory predicts the energy difference between the Coulomb energies of differently spin polarized states, which can be as small as $\sim 0.002 e^2/\epsilon l$ for the systems studied, to within a few \% accuracy. These comparisons also provide an a posteriori justification for fixing the phase using the LLL wave functions. 

FQHE has also been observed in systems with valley degeneracies, such as AlAs quantum wells~\cite{Bishop07,Padmanabhan09}, graphene~\cite{Xu09,Bolotin09,Dean11,Feldman12,Feldman13,Amet15},  and  H-terminated Si(111) surface~\cite{Kott14}. In many of these studies, transitions between differently valley polarized states have been observed. The CF theory can be generalized to treat such systems~\cite{Balram15,Balram15a}, but a careful treatment of the finite width and LL mixing corrections has not yet been performed.

\subsection{Phase diagram of the CF crystal}
\label{sec:crystal}

As noted above, Tsui, Stormer and Gossard's motivation for going to higher magnetic fields was to look for the Wigner crystal. While the crystal phase is superseded by the formation of a CF liquid for a range of filling factors, a crystal must ultimately be stabilized as the filling factor is reduced and the electrons behave more and more classically (as the distance between them measured in units of the magnetic length increases).  An insulating phase is observed at very low fillings, which is interpreted as a pinned crystal state. Extensive experimental work probing the state in transport and optical experiments~\cite{Shayegan07,Fertig07,Jiang90, Goldman90,Williams91, Paalanen92, Santos92,Santos92b,Manoharan94a, Engel97, Pan02, Li00, Ye02, Chen04, Csathy05, Pan05,Sambandamurthy06, Chen06,Liu14a,Zhang15c,Deng16,Jang17, Chen19}
has revealed a rich interplay between the crystal and FQHE. Direct evidence for a periodic lattice at very low filling factors has been obtained through commensurability oscillations in the CF Fermi sea in a nearby layer~\cite{Deng16} (see the Chapter by Shayegan). 

The experimental facts relevant to our discussion below can be summarized as follows. For n-doped GaAs samples, in the limit of zero temperature, an insulating phase is seen for $\nu<1/5$, and also for a narrow range of fillings between $1/5$ and $2/9$. These features have persisted as the sample quality has significantly improved. The fact that the an insulating state is flanked by two strongly correlated FQH liquids (1/5 and 2/9) supports the notion that the insulator is a pinned crystal rather than a state with individual carrier freeze-out. The behavior in p-doped GaAs systems is qualitatively different~\cite{Santos92,Santos92b,Pan05,Csathy05} from that in n-doped GaAs systems. In low-density p-doped GaAs systems, an insulating phase is observed for filling factors below 1/3, and even between 1/3 and 2/5. The FQH states at 1/3 and 2/5 are robust, however.  Experiments in ZnO quantum wells~\cite{Maryenko18} also show insulating phases intermingled with the FQH states at 1/3, 2/5, 3/7 etc.

Early theoretical studies~\cite{Lam84,Levesque84} suggested that the crystal should be stabilized for filling factors below approximately $\nu\approx 1/6.5$~\cite{Lam84,Levesque84}. These only considered competition between the Laughlin state and the crystal at filling factors of the form $\nu=1/m$, and thus could not account for the re-entrant crystal phase between 1/5 and 2/9 in the n-doped GaAs systems. Several authors~\cite{Zhu93, Price93, Platzman93,Ortiz93} attributed the difference between n- and p-doped GaAs to the stronger LL mixing in p-doped GaAs quantum wells due to the larger effective mass of holes. (LL mixing is also much larger in ZnO quantum wells.) They showed that LL mixing generally favors the crystal phase by studying the competition between the Laughlin liquid and the crystal state at fractions $\nu=1/3$, 1/5 and 1/7 through variational~\cite{Zhu93, Price93, Platzman93}, diffusion~\cite{Ortiz93}, and path integral Monte Carlo~\cite{He05}. These studies also considered only the $\nu=1/m$ FQH states.

More recent calculations addressing these issues have shown that a quantitative explanation of the above experimental facts requires a consideration of composite-fermion crystals~\cite{Yi98,Narevich01,Chang05,Archer13} rather than ordinary electron crystals (i.e. vortices are bound to electrons in the crystal phase as well). There are two types of CF crystals (CFCs): 

\underline{Type-I CFC}: The crystal in which all composite fermions arrange themselves on a lattice is called a type-I CFC, sometimes referred to simply as a CFC. An insulating phase is obtained when a type-I CFC is pinned by disorder. 

In the disk geometry, the wave function for a type-I CFC is given by 
\be
\Psi^{\rm CFC}=\prod_{j<k}(z_j-z_k)^{2p}\Psi^{\rm EC},
\label{CFC}
\ee 
where $\Psi^{\rm EC}=\frac{1}{\sqrt{N!}}\sum_P \epsilon_P \prod_{j=1}^N \phi_{\vec{R}_j}(\vec{r}_{Pj})$ is the Hartree-Fock electron crystal (EC) in which electrons are placed in maximally localized wave packets at $\vec{R}_j=(X_j,Y_j)$, with $\phi_{\vec{R}}(\vec{r})=\frac{1}{\sqrt{2\pi}} \exp\left(-\frac{1}{4}(\vec{r}-\vec{R})^2+\frac{i}{2}
(xY-yX)\right)$.  The filling factor $\nu$ of the CFC is related to the filling factor $\nu^*$ of the EC by the standard relation $\nu=\nu^*/(2p\nu^*+1)$. The vorticity $2p$ is a non-negative even integer, treated as a variational parameter, and it is assumed that $\nu^*<1$ (so an electron crystal may be formed within the LLL). Ref.~\refcite{Chang05} tested the CFC wave function against the exact Coulomb ground state wave function at total angular momenta $L=7N(N-1)/2$ and $L=9N(N-1)/2$, which correspond to $\nu=1/7$ and 1/9, for a system of $N=6$ particles in the disk geometry. (The CFC wave function was projected into the appropriate angular momentum $L$ for this calculation, which in effect produces a rotating crystal~\cite{Yannouleas03}.) The lowest energy CFCs were obtained for $2p=4$ at $\nu=1/7$ and $2p=6$ at $\nu=1/9$.  The overlaps of the CFC wave functions with the exact Coulomb ground states were found to be 0.997 at $\nu=1/7$ and 0.999 at $\nu=1/9$. These overlaps are significant given that the dimensions of the Hilbert spaces are  large (117,788 and 436,140), and are also much better than the overlaps of the exact ground states with the Laughlin wave functions (0.71 and 0.66). The energies of the CFCs are also very close to the exact energies: they are 0.016\% (0.006\%) higher than the exact energies at $\nu=1/7$ ($\nu=1/9$). These calculations establish the validity of the CFC wave functions at low fillings.

\underline{Type-II CFC / FQHE}: We also need a model for the FQH state as a continuous function of $\nu$. The incompressible states correspond to $\nu^*=n$ filled $\Lambda$Ls. For non-integer values of $\nu^*$ the topmost $\Lambda$L is partially occupied. What state these composite fermions will form (some possibilities being Wigner crystal, bubble crystal, stripes, Fermi sea, FQH liquid, paired state) is governed by the interaction between them and is a complex issue in itself. We note, however, that the dominant contribution to the total energy comes from the ``kinetic energy" of these composite fermions, and the interaction between them is relatively weak\footnote{The inter-CF interaction is suppressed~\cite{Lee01,Lee02} because the total charge of a CF-particle or a CF-hole, $e^*=e/(2pn\pm 1)$, is small and also spread out.}. It is therefore reasonable to expect that any configuration that builds repulsive correlations between composite fermions should be a decent first approximation. We will assume, for simplicity, that the CF-particles or CF-holes in the topmost partially-filled $\Lambda$L form a crystal. This crystal rides on a FQH state, and is called a type-II crystal by analogy to the Abrikosov flux lattice in a type-II superconductor~\cite{Archer11}. The system exhibits FQHE when a type-II crystal is pinned by disorder. For that reason, the type-II CFC will often be labeled simply as ``FQHE" in this subsection.  Following our earlier discussion (see Fig.~\ref{fig:Jeonel-CF}), the type-II CFCs are Wigner crystals of fractionally-charged quasiparticles or quasiholes of a FQH state. The idea that the quasiparticles or quasiholes of an incompressible FQH state should form a crystal, provided that their density is small and there is no disorder, has long been a part of the FQHE literature; see Halperin~\cite{Halperin84} for example. The CF theory, however, provides accurate wave functions that enable reliable estimates of their energies.

To study the interplay between FQHE and CFC states and determine the thermodynamic limits of various energies, it is convenient to employ the spherical geometry. A difficulty here is that a hexagonal lattice cannot be fitted perfectly on the surface of a sphere. We work with the ``Thomson crystal" instead, wherein we choose our crystal sites that minimize the Coulomb energy of point charges on a sphere. This is the famous Thomson problem~\cite{Thomson04} which had been proposed in 1900 as the model of an atom. The positions of electrons in a Thomson crystal have been evaluated numerically and are available in the literature~\cite{Wales06,Wales09,Thomson}. As expected, the Thomson lattice locally has a triangular structure but contains some defects. The correction to energy due to defects is negligible in large systems, and can be eliminated altogether by evaluating thermodynamic limits. A type-I CFC on the sphere is constructed by first forming a Hartree-Fock electron crystal at flux $2Q^*=2Q-2p(N-1)$, and then attaching to each electron $2p$ vortices by multiplication by an appropriate Jastrow factor. As before, $2p$ is a non-negative even integer, treated as a variational parameter, and we choose $\nu^*<1$ (i.e. $2Q^*+1>N$).

The FQH states at $\nu=n/(2pn\pm 1)$ map into $\nu^*=n$ of composite fermions, as discussed earlier. To calculate the energy  the FQH state as a continuous function of $\nu$, we assume that at non-integer values of $\nu^*>1$, the composite fermions in the topmost partially filled $\Lambda$L form type-II CFC, again modeled as a Thomson crystal on the sphere.  This is expected to be an excellent approximation when the density of composite fermions in the partially filled $\Lambda$L is small. It would be more appropriate to consider a crystal of CF-holes when a $\Lambda$L is more than half full, but the wave function for that state is technically more complicated to work with. We expect, however, that the CF-particle crystal will continue to produce a reasonable approximation for the energy even when a $\Lambda$L is more than half full, given that it properly captures the kinetic energy energy of composite fermions and that it is guaranteed to produce an accurate total energy in the limit when the $\Lambda$L becomes completely full. For $\nu^*<1$, we assume a crystal of CF-holes in the lowest $\Lambda$L, i.e. a crystal of Laughlin quasiholes.

\begin{figure}[t]
\begin{center}
\includegraphics[scale=0.35]{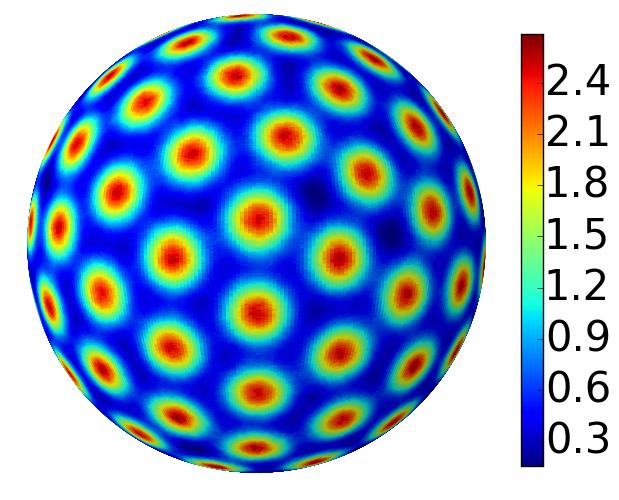}
\includegraphics[scale=0.35]{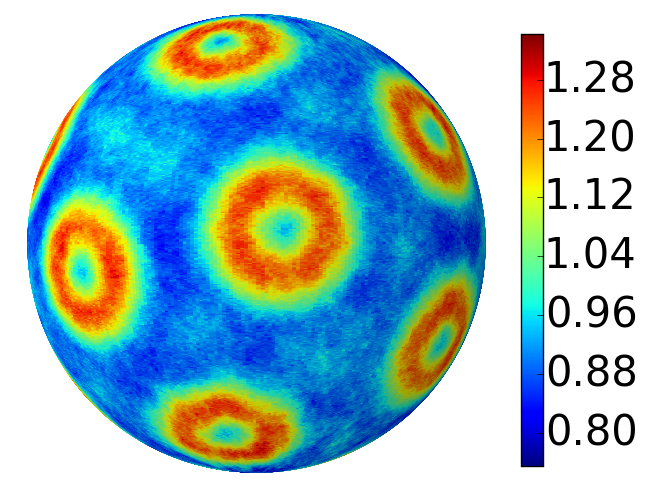}
\end{center}
\caption{
Density profiles of two crystals for a total of $N=96$ electrons at filling factors slightly higher than 1/3. Left shows a type-I electron crystal for $\nu=0.394$ ($2Q=240$). The right panel shows a type-II CF crystal for $\nu=0.351$ ($2Q=270$), where the composite fermions in the partially filled second $\Lambda$L (i.e., quasiparticles of the 1/3 state) form a crystal. 
The density is given in units of the average density. All results are for $\kappa=0$. Source: J. Zhao, Y. Zhang, and J. K. Jain, Phys. Rev. Lett. {\bf 121}, 116802 (2018)~\cite{Zhao18}.
}
\label{structure}
\end{figure}

Fig.~\ref{structure} depicts examples of two crystals on the surface of a sphere. Both panels are for a total of 96 particles at filling factors slightly above 1/3. The left panel shows a type-I electron crystal; the density profile of a type-I CFC is very similar. The right panel depicts a type-II CFC of composite fermions in the second $\Lambda$L. The density profile of an isolated composite fermion in the second $\Lambda$L resembles a smoke ring (just as that of an electron in the second LL does), producing a smoke-ring crystal when the CF density in the second $\Lambda$L is small. More intricate density patterns appear for type-II CFCs in higher $\Lambda$Ls or when the CF-particles begin to overlap.

\begin{figure}[t]
\begin{center}
\includegraphics[width=0.70\textwidth]{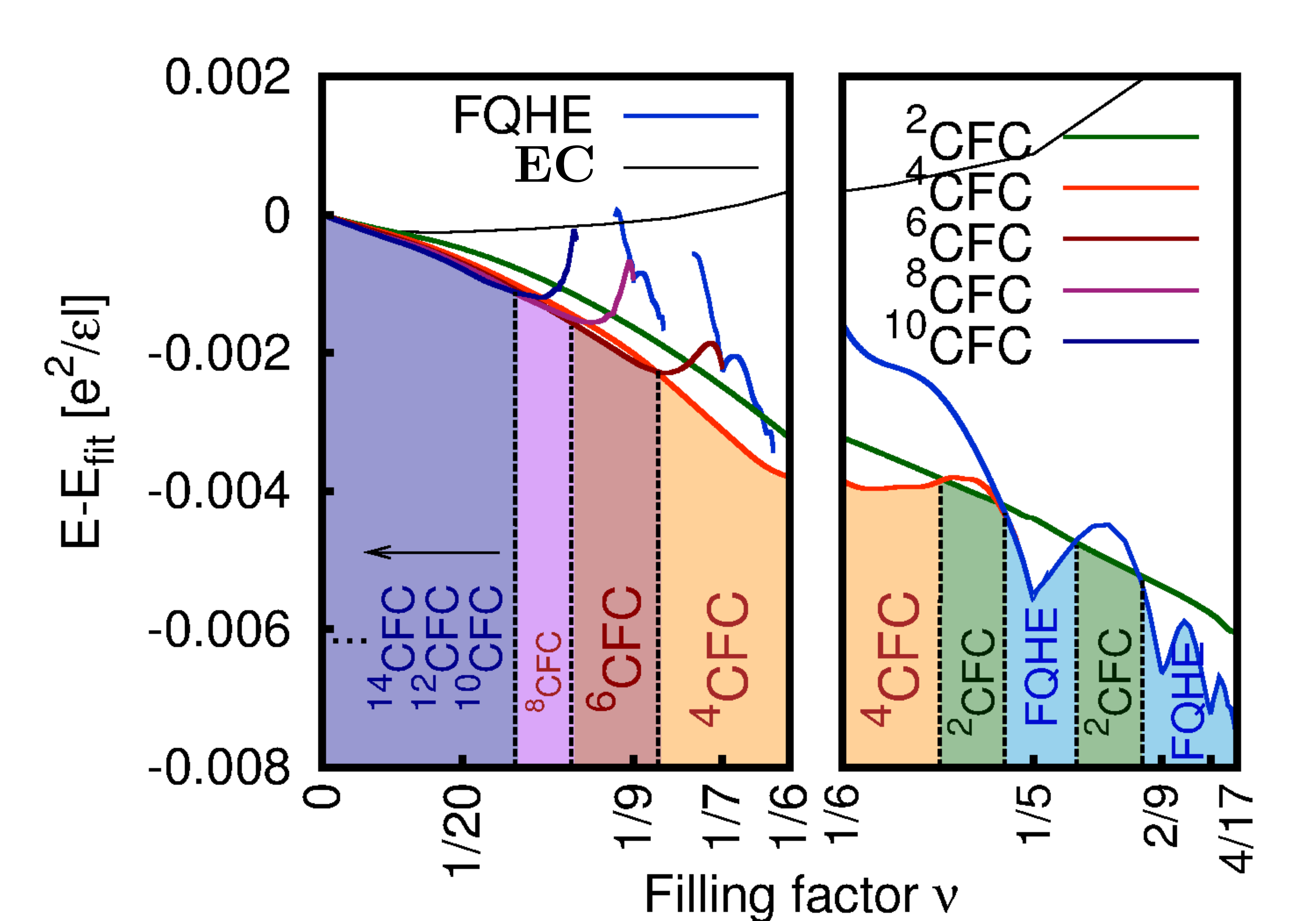}
\caption{Energy per particle as a function of the filling factor for various CF crystal and FQH states. The label $^{2p}$CFC refers to a type-I crystal of composite fermions carrying $2p$ vortices, and EC to the type-I electron crystal. The label FQHE refers to a type-II CF crystal (see text) which will show quantized Hall conductance in the presence of pinning by disorder. The energy of the FQH state as a function of $\nu$ exhibits downward cusps at the magic filling factors.  All energies are quoted relative to the reference energy $E_{\rm fit}=-0.782133\nu^{1/2}+0.2623\nu^{3/2}+0.18\nu^{5/2}-15.1e^{-2.07/\nu}$.
The regions $0<\nu<1/6$ and $1/6<\nu<4/17$ are shown in separate panels because different filling factor scales are used for them. Source: A. C. Archer, K. Park, and J. K. Jain, Phys. Rev. Lett. {\bf 111}, 146804 (2013)~\cite{Archer13}.
}
\label{fig:Etot}
\end{center}
\end{figure}

\begin{figure}[t]
\begin{center}
\includegraphics[scale=0.33]{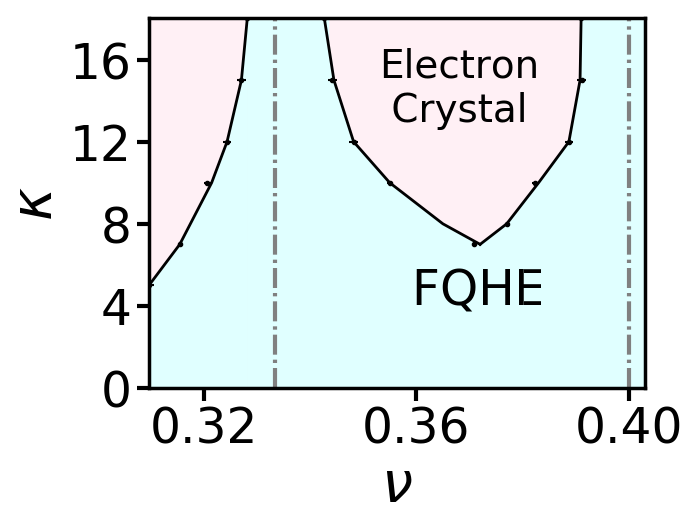}
\includegraphics[scale=0.33]{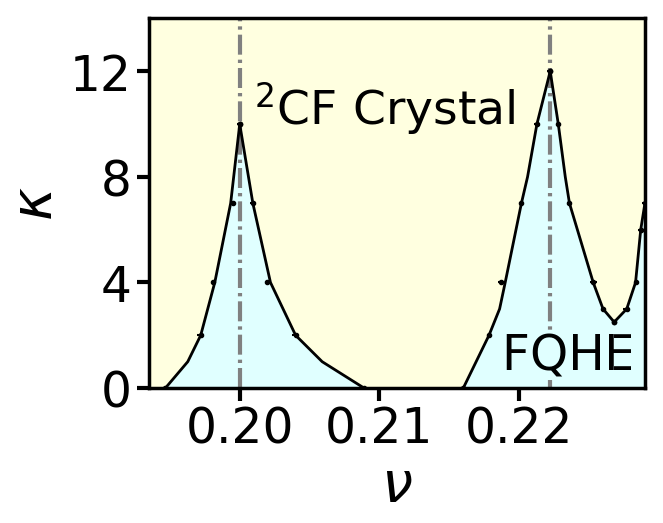}
\end{center}
\caption{
Left panel: The phase diagram of the electron crystal and the FQHE in a filling factor range including $\nu=1/3$ and $\nu=2/5$ as a function of the LL mixing parameter $\kappa$. While at 1/3 and 2/5 the FQH states are very robust to LL mixing, for intermediate fillings the type-I electron crystal appears for $\kappa\gtrsim 7$. Rght panel: The theoretical phase diagram of the type-I $^2$CF crystal and FQH state in a filling factor range including $\nu=1/5$ and $\nu=2/9$. The type-I electron crystal has substantially higher energy than the type-I $^2$CF crystal in this filling factor region. Source: J. Zhao, Y. Zhang, and J. K. Jain, Phys. Rev. Lett. {\bf 121}, 116802 (2018)~\cite{Zhao18}. 
}
\label{phase}
\end{figure}

Fig.~\ref{fig:Etot} displays the energies per particle for various type-I CFC and FQH (i.e. type-II CFC) states as a function of the filling factor~\cite{Archer13}. (The calculations presented in Figs.~\ref{fig:Etot} and \ref{phase} assume zero width.) The energy of the FQH state has cusps at the special filling factors $\nu=n/(2pn\pm 1)$. The curve for the energy of the type-I crystal of composite fermions carrying two vortices intersects the FQHE curve between 1/5 and 2/9, thus explaining the appearance of a crystal state in between these two filling factors. The CFC beats the FQH liquid in this region by a tiny energy of $\sim$0.0005 $e^2/\epsilon l$ per particle. (The simple Hartree-Fock crystal of electrons, labeled EC in Fig.~\ref{fig:Etot}, has a much higher energy and fails to produce a re-entrant transition here.) As the filling factor is further lowered, a sequence of transitions take place into type-I crystals of composite fermions with increasingly higher vorticity; this persists all the way to $\nu=0$, although the energy differences between various kinds of crystals become vanishingly small in that limit. Refs.~\refcite{Archer13,Rhim15} have determined the shear modulus of the various crystals and predicted a discontinuity at the transition points, which should reflect in various observables, for example in the magneto-phonon energy~\cite{Jang17} or the melting temperature of the crystal (although, in practice, disorder will broaden the transitions).

The type-I CFC has the same periodicity and lattice constant as the simple Hartree-Fock electron crystal and the two also have very similar density profiles. Why, then, does a type-I CFC provide a better description at low filling factors?  The reason is because, unlike the Hartree-Fock electron crystal, the CF crystal also properly accounts, through the Jastrow factor, for correlations between the zero-point fluctuations of neighboring electrons around their equilibrium positions.

To address the difference between the behaviors in p and n doped GaAs systems, Ref.~\refcite{Zhao18} has included the effect of LL mixing using fixed phase DMC, using the above wave functions to fix the phase. The resulting phase diagrams in the $\nu$-$\kappa$ plane is shown in Fig.~\ref{phase}. The most striking feature they reveal is the strong $\nu$ dependence of the phase boundary separating the FQH and the crystal phases. For example, FQHE at $\nu=1/3$ and 2/5 survives up to the largest value of $\kappa$ ($=18$) considered, but the electron crystal appears already at $\kappa\gtrsim 7$ for certain $\nu$ in between 1/3 and 2/5, and at even lower values of $\kappa$ for $\nu<1/3$.  Another notable feature is that in the vicinity of $\nu=1/5$ and 2/9, LL mixing induces a transition into the strongly correlated $^2$CF crystal rather than an electron crystal.

In n-type GaAs quantum wells, with $\epsilon=12.5$ and $m_b=0.067m_e$, the LL mixing parameter is given by $\kappa\approx 2.6/\sqrt{B[T]}\approx1.28\sqrt{\nu/(\rho/10^{11}\rm{cm}^{-2})}$. For typical densities, we have $\kappa \lesssim 1.0$ in the vicinity of $\nu=1/3$ and $\nu=1/5$. For these values we expect a crystal phase only between 1/5 and 2/9. The $\kappa$ for holes in p-doped GaAs is $\approx 5.6$ times that for electrons at the same $B$~\cite{Sodemann13}. Santos {\em et al.}~\cite{Santos92b} find that an insulating phase appears between $\nu=1/3$ and $\nu=2/5$ at $\rho\approx 7\times 10^{10}$ cm$^{-2}$, which corresponds to $\kappa\approx 5$. Given various approximations made in our calculation and our neglect of disorder (disorder should favor a crystal, because a crystal can more readily adjust to it than an incompressible liquid), we regard the level of agreement to be  satisfactory. Similar considerations apply to ZnO quantum wells~\cite{Maryenko18} for which $\kappa$ is $\sim 6.4$ times larger than that for n-doped GaAs systems~\cite{Sodemann13}.

Why does LL mixing favor the crystal phase? Both the liquid and the crystal states lower their energies by taking advantage of LL mixing, but one can expect that the crystal has more flexibility, because LL mixing allows the wave packet at each site to become more localized. The competition is subtle and complicated, however, and only a detailed calculation can tell if and where a transition into a crystal takes place.

There is additional experimental support for CF nature of the crystal. 
Jang {\em et al.}~\cite{Jang17} have measured vibrations of the crystal phase using electron tunneling spectroscopy and found that the stiffening of the resonance is consistent with the shear modulus evaluated in Refs.~\refcite{Archer13,Rhim15} for the CF crystal. Evidence for type-II crystals has been seen by Zhu {\em et al.}\cite{Zhu10} in optical experiments through observation of collective pinning modes in the vicinity of $\nu=1/3$.  In a theoretical work, Shi and Ji~\cite{Shi18} have predicted that CF nature of the type-I crystal results in a magnetoroton-like phonon. 

\section{Kohn-Sham density functional theory of the FQHE}
\label{sec:KS*}

This section is a minimally modified reproduction of an article by Yayun Hu and the author~\cite{Hu19}.

The Kohn-Sham (KS) density-functional theory (DFT) uses the electron density to construct a single particle formalism that incorporates the complex effects of many-particle interactions through a universal exchange correlation function~\cite{Giuliani08}.  It is an invaluable tool for treating systems of interacting electrons spanning the disciplines of physics, chemistry, materials science and biology, but very little work has been done~\cite{Ferconi95,Heinonen95,Zhao17} toward applying this method to the FQHE. The reasons are evident.  To begin with, even though the KS-DFT is in principle exact, its accuracy, in practice, is dictated by the availability of exchange correlation (xc) potentials, and it works best when the xc contribution is small compared to the kinetic energy. In the FQHE problem, the kinetic energy is altogether absent (at least in the convenient limit of very high magnetic fields) and the physics is governed entirely by the xc energy. A more fundamental impediment is that, by construction, the KS-DFT eventually obtains a single Slater determinant solution, whereas the ground state for the FQHE problem is an extremely complex, filling factor-dependent wave function that is not adiabatically connected to a single Slater determinant. In particular, a mapping into a problem of non-interacting electrons in a KS potential will produce a ground state that locally has integer fillings, whereas nature displays preference for certain fractional fillings. Finally, a mapping into a system of weakly interacting electrons will also fail to capture topological features of the FQHE, such as fractional charge and fractional braid statistics for the quasiparticles. At a fundamental level, these difficulties can be traced back to the fact that the space of ground states in the LLL is highly degenerate for non-interacting electrons, and the interaction causes a non-perturbative reorganization to produce the FQHE.

\subsection{KS equations for composite fermions} 

To make progress, we exploit the fact that the strongly interacting electrons in the FQH regime turn into weakly interacting composite fermions, which suggests using an auxiliary system of non-interacting composite fermions to construct a KS-DFT formulation of the FQHE. This is the approach taken here. A crucial aspect of our KS Theory is that it properly incorporates the physics of long range gauge interaction between composite fermions induced by the Berry phases due to the quantum mechanical vortices attached to them, which is responsible for the topological properties of the FQHE, such as fractional charge and statistics~\cite{Jeon04,Jain07,Zhang14}. That effectively amounts to using a non-local exchange-correlation potential. Certain previous DFT formulations of the FQHE~\cite{Ferconi95,Heinonen95,Zhao17} employ a local exchange-correlation potential and thus do not capture the topological features of the FQHE.

We consider the Hamiltonian for fully spin polarized electrons confined to the LLL:
\begin{equation}
\hat{\mathcal H}=\hat{H}_{\rm ee}+\int d\vec{r} V_{\rm ext}(\vec{r})\hat{\rho}(\vec{r})\;.
\end{equation}
Within the so-called magnetic-field DFT~\cite{Grayce94,kohn04,Tellgren12,Tellgren18b}, the Hohenberg-Kohn (HK) theorem also applies to interacting electrons in the FQH regime and  implies that
the ground state density and energy can be obtained by minimizing the energy functional
\begin{equation}
E[\rho]=F[\rho]+\int d\vec{r} V_{\rm ext}(\vec{r})\rho(\vec{r}),
\end{equation}
where the HK functional is given by~\cite{Levy79,Lieb83}
\begin{equation}
F[\rho]=\min_{\Psi_{\rm LLL}\rightarrow \rho(\vec{r})} \langle\Psi_{\rm LLL}|\hat{H}_{\rm ee}|\Psi_{\rm LLL}\rangle \equiv E_{\rm xc}[\rho]+E_{\rm H}[\rho].
\end{equation}
(The $B$ dependence of the energy functional has been suppressed for notational convenience). Here $E_{\rm xc}[\rho]$ and $E_{\rm H}[\rho]$ are the xc and Hartree energy functionals of electrons and $\Psi_{\rm LLL}$ represents a LLL wave function.  
The conventional KS mapping into non-interacting electrons is problematic due to the absence of kinetic energy.

We instead map the FQHE into the auxiliary problem of ``non-interacting" composite fermions. Even though we use the term non-interacting, the Berry phases associated with the bound vortices induce a long range gauge interaction between composite fermions, as a result of which they  experience a density dependent magnetic field $B^*(\vec{r})=B- 2\rho(\vec{r})\phi_0$, where $\phi_0=hc/e$ is a flux quantum. We therefore write
\begin{equation}
\left[\frac{1}{2m^*}\left(\vec{p}+\frac{e}{c}\vec{A}^*(\vec{r};[\rho])\right)^2+V_{\rm KS}^*(\vec{r})\right] \psi_{\alpha}(\vec{r})=
\epsilon_{\alpha} \psi_{\alpha}(\vec{r}),
\label{singleCFKS}
\end{equation}
where
$V_{\rm KS}^*(\vec{r})$ is the KS potential for composite fermions, $m^*$ is the CF mass (taken to be $m^*=0.079 \sqrt{B[T]} \; m_e$; see Sec.~\ref{sec:verifications}), and $\nabla \times \textbf{A}^*(\vec{r};[\rho])=B^*(\vec{r})$. 
As a result of the gauge interaction, the solution for any given orbital depends, through the $\rho(\vec{r})$ dependence of the vector potential, on the occupation of all other orbitals. Eq.~\ref{singleCFKS} must therefore be solved self-consistently, i.e., the single-CF orbitals $\psi_\alpha(\vec{r})$ must satisfy the condition that the ground state density $\rho(\vec{r})=\sum_{\alpha} c_\alpha |\psi_{\alpha}(\vec{r})|^2$, where $c_\alpha=1$ (0) for the lowest energy occupied (higher energy unoccupied) single-CF orbitals, is equal to the density that appears in the kinetic energy of the Hamiltonian. The energy levels of Eq.~\ref{singleCFKS} are the self consistent $\Lambda$Ls.  For the special case of a spatially uniform density and constant $V_{\rm KS}^*$, Eq.~\ref{singleCFKS} reduces to the problem of non-interacting particles in a uniform $B^*$. 
Importantly, once a self-consistent solution is found for a given $V^*_{\rm KS}(\vec{r})$, for the corresponding density in the Hamiltonian in Eq.~\ref{singleCFKS}, the ground state satisfies, by definition, the self-consistency condition and also the variational theorem, and the standard proof for the HK theorem follows~\cite{Hu19}. We define the CF kinetic energy functional as
\be
T_{\rm s}^*[\rho]=\min_{\Psi\rightarrow \rho}\langle\Psi |  \frac{1}{2m^*}\sum_{j=1}^N\left(\vec{p}_j+\frac{e}{c}\vec{A}^*(\vec{r}_j;[\rho])\right)^2 | \Psi\rangle,
\ee
where we perform a constrained search over all single Slater determinant wave functions $\Psi$ that correspond to the density $\rho(\vec{r})$, following the strategy of the generalized KS scheme~\cite{Seidl96,Hu19}.

The next key step is to write $E_{\rm xc}[\rho]=T_{\rm s}^*[\rho]+E^*_{\rm xc}[\rho]$, or $F[\rho]=T_{\rm s}^*[\rho]+E_{\rm H}[\rho]+E^*_{\rm xc}[\rho]$. (Note that $T_{\rm s}^*[\rho]$ and thus $E_{\rm xc}[\rho]$ is a non-local functional of the density.)  Such a partitioning of $F[\rho]$ can, in principle, always be made given our assumptions, but is practically useful only if the $T_{\rm s}^*[\rho]$ and $E_{\rm H}[\rho]$ capture the significant part of $F[\rho]$, and the remainder $E^*_{\rm xc}[\rho]$, called the exchange-correlation energy of composite fermions, makes a relatively small contribution. This appears plausible given that the CF kinetic energy term captures the topological aspects of the FQHE, and also because the model of weakly interacting composite fermions has been known to be rather successful in describing a large class of experiments.

Minimization of the energy
$
E[\rho]=T_{\rm s}^*[\rho]+E_{\rm H}[\rho]+E^*_{\rm xc}[\rho]+\int d\vec{r} V_{\rm ext}(\vec{r})\rho(\vec{r})
$
with respect to $\rho(\vec{r})=\sum_{\alpha} c_\alpha |\psi_{\alpha}(\vec{r})|^2$, subject to the constraint $\int d \vec{r} \psi_{\alpha}^*(\vec{r})\psi_{\beta}(\vec{r})=\delta_{\alpha\beta}$, yields~\cite{Hu19} Eq.~\ref{singleCFKS} with
\begin{equation}
V^*_{\rm KS}[\rho,\{\psi_\alpha\}]=V_{\rm H}(\vec{r})+V_{\rm xc}^{\rm  *}(\vec{r})+V_{\rm ext}(\vec{r})+V^*_{\rm T}(\vec{r}),
\label{VKS*}
\end{equation}
where $V_{\rm H}(\vec{r})= \delta E_{\rm H}/\delta \rho(\vec{r})$ and $V_{\rm xc}^{\rm  *}(\vec{r})=\delta E_{\rm xc}^*/\delta \rho(\vec{r})$ are the Hartree and CF-xc potentials. The non-standard potential
$V^*_{\rm T}(\vec{r})=\sum_{\alpha} c_\alpha \langle\psi_{\alpha}|{\delta T^*}/{\delta \rho(\vec{r})}|\psi_{\alpha}\rangle$
with $T^*=\frac{1}{2m^*}\left(\vec{p}+\frac{e}{c}\vec{A}^*(\vec{r};[\rho])\right)^2$ arises due to the density-dependence of the CF kinetic energy.  $V^*_{\rm T}$ describes the change in $T_{\rm s}^*$ to a local disturbance in density for a fixed choice of the KS orbitals.  Because $V^*_{\rm T}(\vec{r})$ depends not only on the density but also on the occupied orbitals, we are actually working with what is known as the ``orbital dependent DFT"~\cite{Kummel08}.

Having formulated the CF-DFT equations, we now proceed to obtain solutions for some representative cases. The primary advantage of our approach is evident without any calculations. Take the example of a uniform density FQH state at $\nu=n/(2pn\pm 1)$. It is an enormously complicated state in terms of electrons, but maps into the CF state at filling factor $\nu^*=n$ with a spatially uniform magnetic field, thereby producing the correct density.  For non-uniform densities, the state of non-interacting composite fermions will produce configurations where composite fermions {\it locally} have $\nu^*\approx n$, which corresponds to an electronic state where the local filling factor is $\nu\approx n/(2pn\pm 1)$, which is a reasonable description, and certainly a far superior representation of the reality than any state of non-interacting electrons.

For a more quantitative treatment we need a model for the xc energy. To this end, we assume the LDA form $E_{\rm xc}^*[\rho]=\int d\vec{r} \epsilon^*_{\rm xc}[\rho(\vec{r})]\rho(\vec{r})$, where $\epsilon^*_{\rm xc}[\rho]$ is the xc energy per CF. We express all lengths in units of the magnetic length and energies in units of $e^2/\epsilon l$. The density is related to the local filling factor as $\nu(\vec{r})=\rho(\vec{r}) 2\pi l^2$.  
 We take the model $\epsilon^*_{\rm xc}[\rho]=a\nu^{1/2}+(b-f/2) \nu+ g$,  with $a=-0.78213$, $b=0.2774$, $f=0.33$, $g=-0.04981$. The form is chosen empirically so that the sum of $\epsilon^*_{\rm xc}$ and the CF kinetic energy accurately reproduces the known electronic xc energies at $\nu=n/(2n+1)$. (The term $a\nu^{1/2}$ is chosen to match with the known classical value of energy of the Wigner crystal in the limit $\nu\rightarrow 0$~\cite{Bonsall77}.) Although optimized for $\nu=n/(2n+1)$, we shall uncritically assume this form of $\epsilon^*_{\rm xc}(\nu)$ for all $\nu$.   Our aim here is to establish the proof-of-principle validity and the applicability of our approach and its ability to capture topological features, which are largely robust against the precise form of the xc energy. The xc potential is given by $V^*_{\rm xc}=\delta E_{\rm xc}^*/\delta \rho(\vec{r})=\frac{3}{2}a\nu^{1/2}+(2b-f)\nu+g$. We note that while the CF xc potential $V^*_{\rm xc}$ is a continuous functions of density, the {\it electron} xc potential $V_{\rm xc}$ has derivative discontinuities at $\nu=n/(2n\pm 1)$, arising from the kinetic energy of the composite fermions.

In our applications below, we will consider $N$ electrons in a potential $V_{\rm ext}(\vec{r})=-\int d^2 \vec{r}' {\rho_{\rm b}(\vec{r}') \over \sqrt{|\vec{r}-\vec{r}'|^2+d^2}}$ generated by a two-dimensional uniform background charge density $\rho_b=\nu_0/2\pi l^2$ distributed on a disk of radius $R_{\rm b}$ satisfying $\pi R_{\rm b}^2 \rho_b=N$ at a separation of $d$ from the plane of the electron liquid. This produces an electron system at filling factor $\nu=\nu_0$ in the interior of the disk. We use $\nu_0=1/3$ and $d/l\rightarrow 0$ in our calculations below.  For the vector potential, we assume circular symmetry and choose the gauge $\textbf{A}^*(\vec{r})=\frac{r\mathcal{B}(r)}{2}\mathbf{e}_{\phi}$, with $\mathcal{B}(r)=\frac{1}{\pi r^2}\int_0^r 2\pi r'B^*(r') dr'$.  We obtain self-consistent solutions of Eqs.~\ref{singleCFKS} and \ref{VKS*} by an iterative process.

\subsection{Density profile of the FQH droplet}

\begin{figure}[t]
\hspace{1.5cm}\includegraphics[width=3.5in]{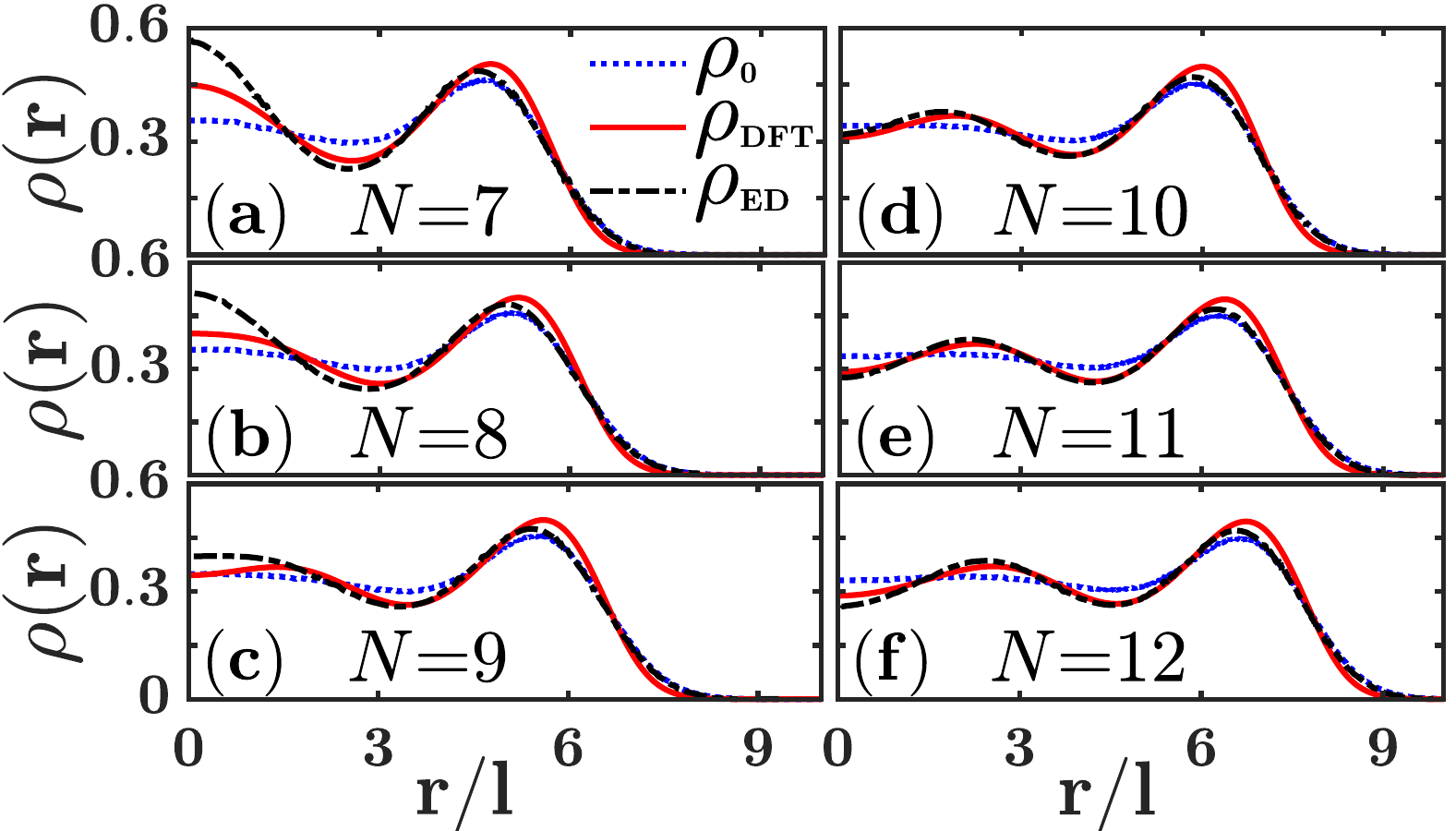}
\caption{Density profile for 1/3 droplets. This figure shows the density of a system of $N$ composite fermions. $\rho_0$ is the density for Laughlin's 1/3 wave function~\cite{Laughlin83}, and $\rho_{\rm ED}$ is obtained from exact diagonalization (ED) of the Coulomb interaction at total angular momentum $L_{\rm total}=3N(N-1)/2$~\cite{Tsiper01}. The density $\rho_{\rm DFT}$ is calculated from the solution of the KS equations for composite fermions in an external potential produced by a uniform positively charged disk of radius $R$ so that $\pi R^2\rho_b=N$. The total angular momentum of the CF state is $L^*_{\rm tot}$, which is related to the total angular momentum of the electron state by $L_{\rm tot}=L^*_{\rm tot}+N(N-1)$~\cite{Jain95}. The CF-DFT solution produces $L^*_{\rm tot}=N(N-1)/2$, which is consistent with $L_{\rm tot}=3N(N-1)/2$. 
All densities are quoted in units of $(2\pi l^2)^{-1}$, the density at $\nu=1$. We take $\rho_b=1/3$. Source: Y. Hu and J. K. Jain, Phys. Rev. Lett. {\bf 123}, 176802 (2019)~\cite{Hu19}.}\label{MCDFTcompare}
\end{figure}

As a first application, we consider the density profile of the $\nu_0=1/3$ droplet. Fig.~\ref{MCDFTcompare} shows the density profiles calculated from Laughlin's trial wave function as well as that obtained from exact diagonalization at total angular momentum $L=3N(N-1)/2$~\cite{Tsiper01}.  Also shown are the density profiles obtained from the above KS equations. The  density profile from our CF-DFT captures that obtained in exact diagonalization well, especially for $N\geq 10$. Remarkably, it reproduces the characteristic shape near the edge where the density exhibits oscillations and overshoots the bulk value before descending to zero. This qualitative behavior is fairly insensitive to the choice of $V^*_{\rm xc}$, and is largely a result of the self-consistency requirement in Eq.~\ref{singleCFKS}.

\subsection{Screening by the FQH state}

\begin{figure}[t]
\centering\includegraphics[width=3in]{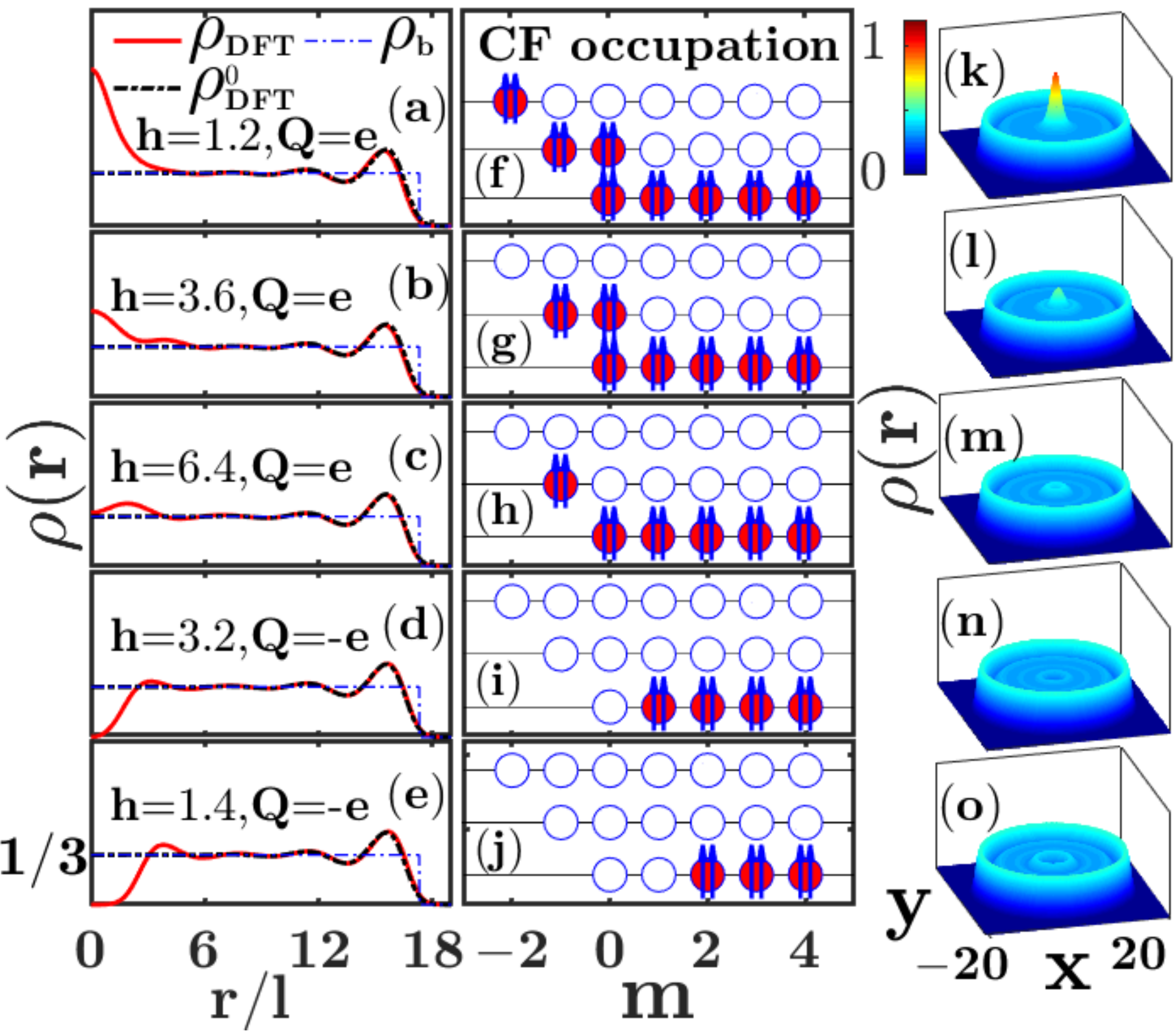}
\centering\includegraphics[width=3in]{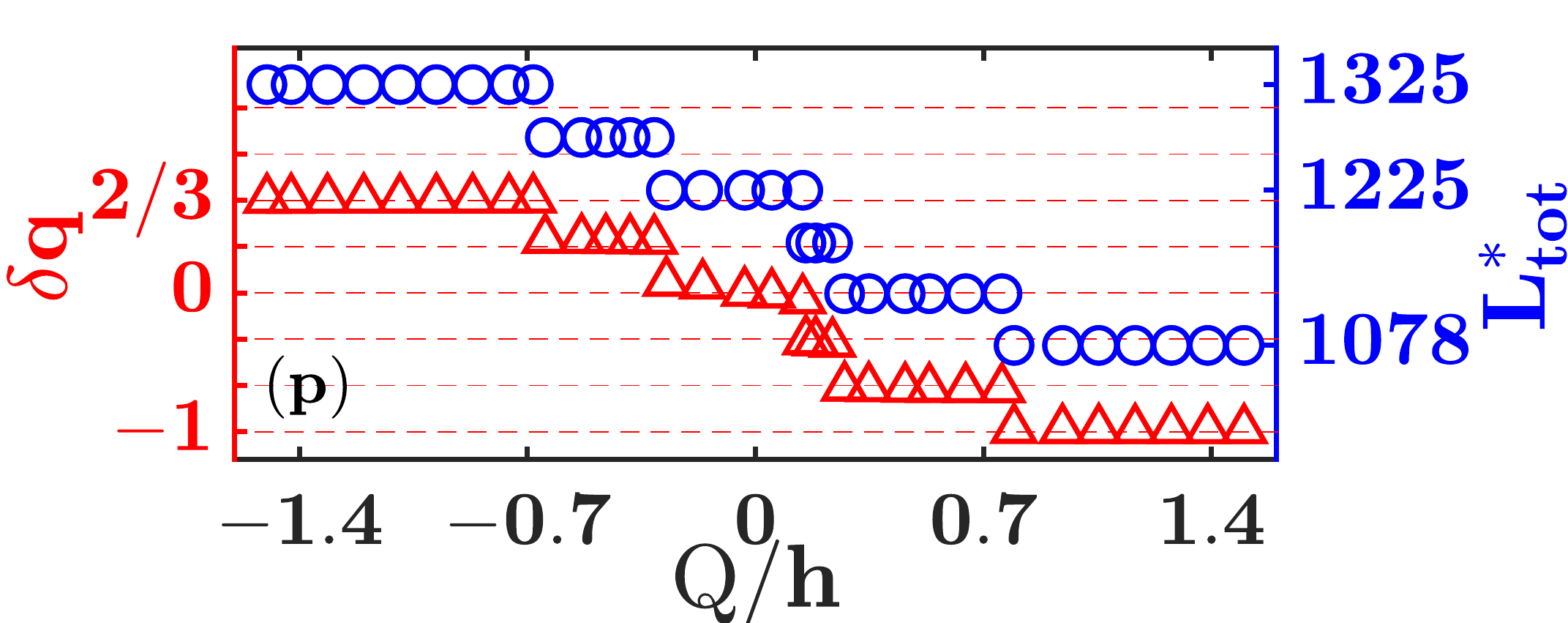}
\caption{Screening and fractional charge. This figure shows how the 1/3 state screens a charged impurity of strength $Q=\pm e$ located at a perpendicular distance $h$ from the origin. The panels {(a)-(e)} and {(k)-(o)} show the self-consistent density $\rho_{\rm DFT}(\vec{r})$. Also shown are $\rho^0_{\rm DFT}(\vec{r})$, the ``unperturbed" density (for $Q=0$), and $\rho_b$, which is the density of the positively charged background. Panels {(f-j)} show the occupation of renormalized $\Lambda {\rm L}$s in the vicinity of the origin; each composite fermion is depicted as an electron with two arrows, which represent quantized vortices. (The single particle angular momentum is given by $m=-n, -n+1, \cdots$ in the $n^{\rm th}$ $\Lambda$L.) The panel {\bf(p)} shows the evolution of the excess charge $\delta q$ and the total CF angular momentum $L^*_{\rm tot}$ as a function of the impurity potential strength at the origin $V_{\textrm{imp}}(r =0)=Q/h$. Change in the charge at the origin is associated with a change in $L^*_{\rm tot}$. The system contains a total of $N=50$ composite fermions. For $h=\infty$, we have $L^*_{\rm tot}=1225$ and $\delta q=0$. For one and two quasiholes, we have $L^*_{\rm tot}=1225$ and 1275, whereas for one, two and three quasiparticles we have $L^*_{\rm tot}=1175$, 1127 and 1078, precisely as expected from the configurations in panels {(f)-(j)}~\cite{Jain95}. Source: Y. Hu and J. K. Jain, 
 Phys. Rev. Lett. {\bf 123}, 176802 (2019)~\cite{Hu19}.}\label{kBT1N50IM}
\end{figure}

We next consider screening of an impurity with charge $Q=\pm e$ at a height $h$ directly above the center of the FQH droplet. The strength of its potential
\begin{equation}\label{Vr}
V_{\rm imp}(\vec{r})=\frac{Q}{\sqrt{|\vec{r}|^2+h^2}}
\end{equation}
can be tuned by varying $h$.
Panels (a)-(e) in Fig.~\ref{kBT1N50IM} show the density $\rho$ for certain representative values of $h$.
It is important to note that the CF orbitals in the self-consistent solution form strongly renormalized $\Lambda {\rm L}s$ (i.e. include the effect of mixing between the unperturbed $\Lambda$Ls). Panels (f)-(j) show the occupation of the $\Lambda {\rm L}s$. The presence of the impurity either empties some CF orbitals from the lowest $\Lambda$L or fills those in higher $\Lambda$Ls. Each empty orbital in the lowest $\Lambda$L corresponds to a charge 1/3 quasihole, whereas each filled orbital in an excited $\Lambda$L to a charge $-1/3$ quasiparticle~\cite{Jain07}. The excess charge is defined as $\delta q=\int_{|r|<r_0} d^2\vec{r} [\rho_0-\rho(\vec{r})]$ in a circular area of radius $r_0=10 l$ around the origin.  Panel (p) shows how $\delta q$ and $L^*_{\rm tot}$ change as a function of the potential at the origin $V_{\rm imp}(\vec{r}=0)=-Q/h$.  The excess charge $\delta q$ is seen to be quantized at an integer multiple of $\pm 1/3$.

\subsection{Fractional braid statistics}

We finally come to fractional braid statistics. Particles obeying such statistics, called anyons, are characterized by the property that the phase associated with a closed loop of a particle depends on whether the loop encloses other particles. In particular, for abelian anyons, each enclosed particle contributes a phase factor of $e^{i 2\pi \alpha}$, where $\alpha$ is called the statistics parameter. [For non-interacting bosons (fermions), $\alpha$ is an even (odd) integer.] In the FQHE, the quasiparticles are excited composite fermions and quasiholes are missing composite fermions. Let us consider quasiholes of the 1/3 state for illustration. A convenient way to ascertain the statistics parameter within our KS-DFT is to ask how the location of a quasihole in angular momentum $m$ orbital changes when another quasihole is inserted at the origin in the $m=0$ orbital. Let us first recall the expected behavior arising from fractional braid statistics. In an effective description, the wave function of a single quasihole in angular momentum $m$ orbital is given by $z^m e^{-|z|^2/4l^{*2}}$ ($z\equiv x-iy$), which is maximally localized at $r_{\rm ex}=(2m)^{1/2}l^{*}=(6m)^{1/2}l$, with $l^*=\sqrt{3}l$ (as appropriate for $\nu_0=1/3$). When another quasihole is present at the origin, it induces an additional statistical phase factor $e^{i2\pi\alpha}$, where $\alpha$ is the statistics parameter. This changes the wave function of the outer quasihole to $z^{m-\alpha} e^{-|z|^2/4l^{*2}}$, which is now localized at $r_{\rm ex}'=[6(m-\alpha)]^{1/2}l$. We now determine $\alpha$ from our KS-DFT formalism.

\begin{figure}[t]
\hspace{1.2cm}\includegraphics[width=4in]{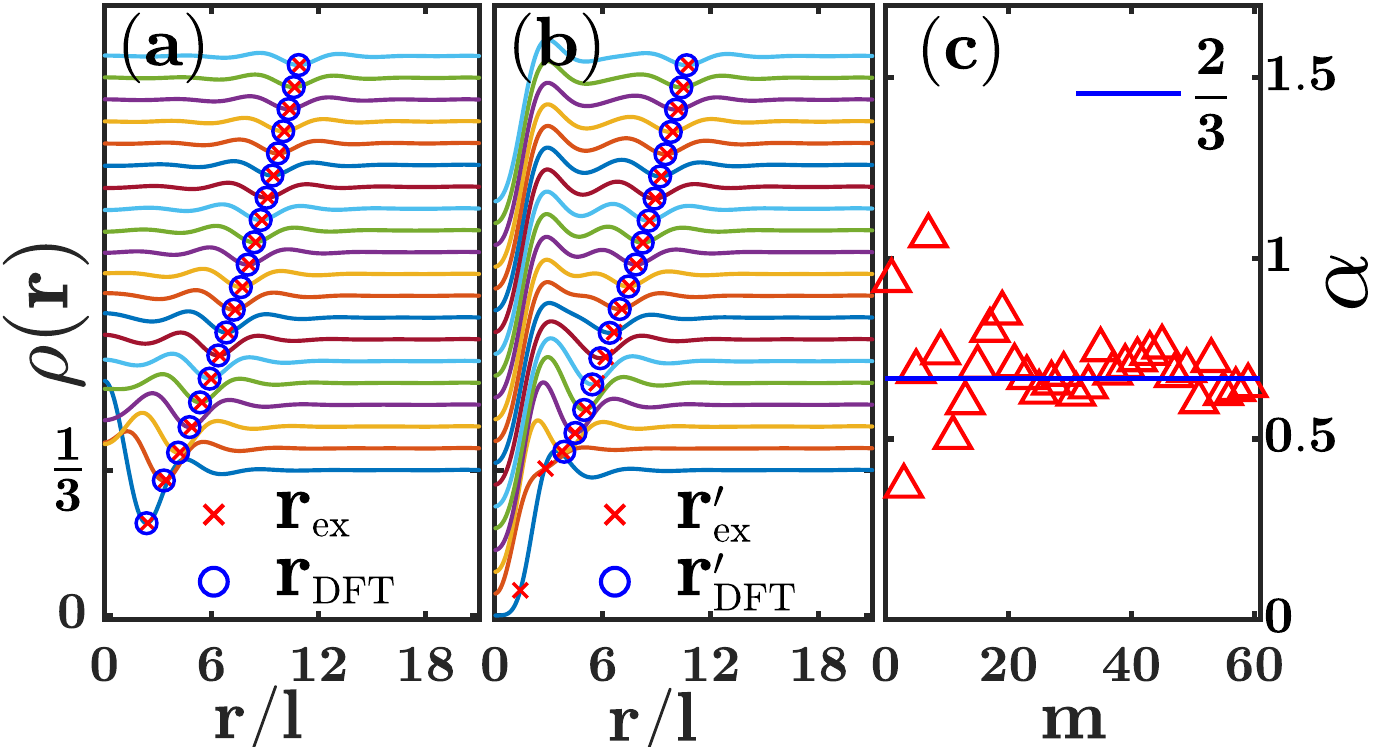}
\caption{Fractional braid statistics. Panel {\bf(a)} shows the electron density for a system with a quasihole in angular momentum $m$ orbital, with $m$ changing from 1 to 20 for the curves from the bottom to the top. (Each successive curve has been shifted up vertically for clarity.) Panel {\bf(b)} shows the same in the presence of another quasihole at the origin. For each $m$, we indicate the expected position of the outer quasihole (red cross) as well as the position obtained from the DFT density determined by locating the local minimum (blue circle). Panel {\bf(c)} shows the calculated statistics parameter $\alpha\equiv (r^2_{\rm{\rm DFT}}-r^{\prime 2}_{\rm{\rm DFT}})/6l^2 $. The calculation has been performed for $N=200$ composite fermions at $\nu_0=1/3$. Source: Y. Hu and J. K. Jain, 
 Phys. Rev. Lett. {\bf 123}, 176802 (2019)~\cite{Hu19}.}\label{Braiding}
\end{figure}

A quasihole can be treated in a constrained DFT~\cite{Kaduk12} wherein we leave a certain angular momentum orbital unoccupied. The panels (a) and (b) of Fig.~\ref{Braiding} show the self-consistent KS density profiles of the state with a quasihole in angular momentum $m$, without and with another quasihole in the $m=0$ orbital. The locations of the outer quasihole, $r_{\rm DFT}$ and $r'_{\rm DFT}$, are determined from the minimum in the density. These are in reasonable agreement with the expected positions $r_{\rm ex}$ and $r_{\rm ex}'$ (provided $m> 3$).  More importantly, the calculated statistics parameter $\alpha\equiv (r^2_{\rm{\rm DFT}}-r^{\prime 2}_{\rm{\rm DFT}})/6l^2$ is in excellent agreement with the expected fractional value of $\alpha=2/3$~\cite{Halperin84,Jain07} provided that the two quasiparticles are not close to one another, indicating that our method properly captures the physics of fractional braid statistics. The small deviation from 2/3 for large $m$ arises from the fact that the density of the unperturbed system itself has slight oscillations due to the finite system size, which causes a slight shift in the position of the local minimum due to an additional quasihole. Correcting for that effect produces a value much closer to $\alpha=2/3$~\cite{Hu19}.

These studies demonstrate that the Kohn-Sham DFT faithfully captures the topological characteristics of the FQH state. This opens a new strategy for exploring a variety of problems of interest.

\section{Looking beyond composite fermions: The parton paradigm}
\label{sec:parton}

Soon following the CF theory, a generalization was introduced in 1989 known as the parton construction~\cite{Jain89b,Jain90}, which further exploits the connection between the FQHE and the IQHE. While composite fermions are the building blocks of the CF theory, IQH states are the building blocks of the parton theory. The parton construction produces candidate FQH states that are products of IQH states. These include all of the states of the CF theory but also states beyond the CF theory. The states in the latter category are interesting in their own right, but also because, as shown by Wen~\cite{Wen91}, they include non-Abelian states. All of the states of the parton theory are in principle valid, and one can attempt to construct a model interaction whose ground state is well represented by a given parton state. The important question, however, is whether the new (beyond-CF) states are realized in some known systems. One of the simplest candidates beyond the CF theory, namely the 221 state (defined below), is a non-Abelian state at $\nu=1/2$. It was considered in early 1990s by the author and his collaborators as a candidate for the 5/2 state~\cite{Willett87} (i.e., 1/2 in the second LL) but was not found to be stabilized by the second LL Coulomb interaction. In 1991, a Pfaffian wave function was introduced by Moore and Read~\cite{Moore91}, which is also a non-Abelian state at $\nu=1/2$ (although distinct from the 221 state). The Pfaffian state was seen in numerical diagonalization studies~\cite{Morf98} to provide a reasonable description for the 5/2 FQHE.  As a result of these developments, interest in the parton construction subsided. However, a recent work by Balram, Barkeshli and Rudner (BBR)~\cite{Balram18} has breathed a new life into the parton theory. These authors have demonstrated that a different non-Abelian state from the parton construction, labeled $\bar{2}\bar{2}111$ (see below), does provide a good account of the 5/2 FQHE. This work has inspired further studies that have indicated possible realizations of certain other beyond-CF states as well. 

It is a remarkable fact that the physics of strong correlations in the FQHE can be captured by wave functions that are products of Slater determinants of IQH states.  One wonders if all experimentally realized FQH states conform to this paradigm. That, in the author's view, would be very satisfying, and also appears to be the case so far, as discussed below. 

The subsection~\ref{sec:parton1} outlines the parton construction, and gives a brief account of the topological properties of these states, appearance of non-Abelian statistics, and connection of some of these states to topological superconductivity of composite fermions. Rest of the section discusses several states that are of possible experimental relevance.

\subsection{The parton construction: Abelian and non-Abelian states}
\label{sec:parton1}

\begin{figure}[t]
\begin{center}
\includegraphics[width=0.6\textwidth]{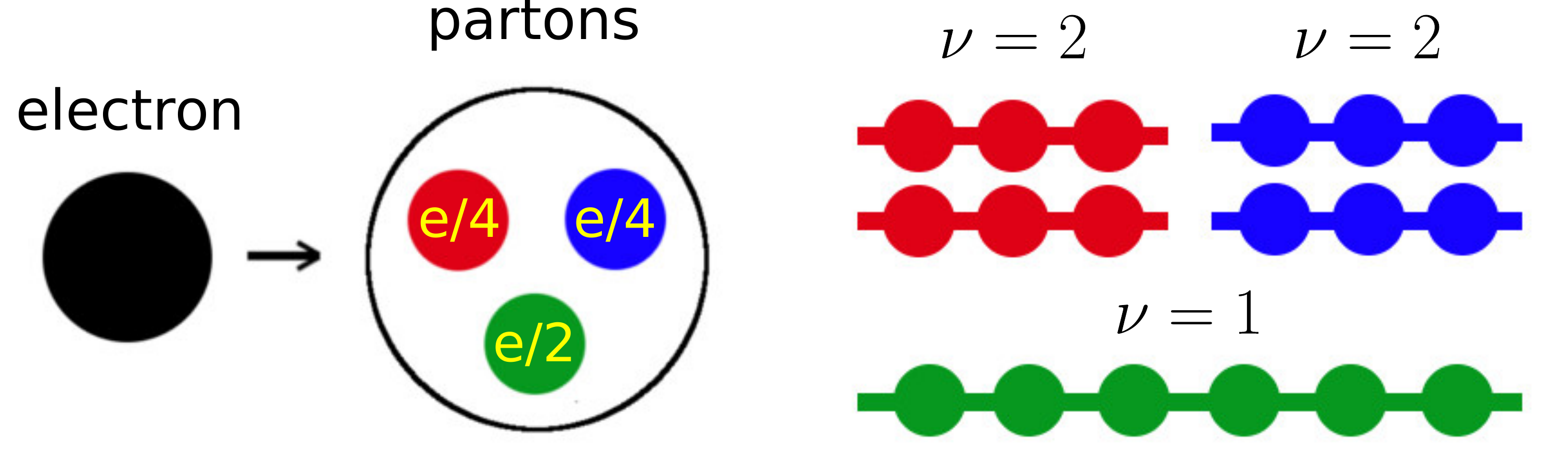}
\caption{Parton construction of the 221 state at $\nu=1/2$. Figure taken from Y. Wu, T. Shi, and J. K. Jain, Nano Letters. {\bf 17}, 4643-4647 (2017)~\cite{Wu17b}.
}
\label{fig:parton}
\end{center}
\end{figure}

We begin by asking if it is possible to construct new incompressible states from known incompressible states, such as the IQH states. The parton construction seeks to accomplish this goal in the following manner (Fig.~\ref{fig:parton}). We first decompose each electron into $m$ fictitious particles called partons, which, in the simplest implementation, are taken to be fermionic. We then place each species of partons into an IQH state with filling factor $n_\lambda$, where $\lambda=1,\cdots, m$ labels different parton species. Finally, we glue the partons back together to recover the physical electrons. It is intuitively sensible that the resulting state will be incompressible. 

The density of each species of partons must be the same as the density of the physical electrons, which implies, recalling $\rho= \nu e B / hc$, that each species must satisfy $n_\lambda q_\lambda=\nu e$, where $q_\lambda$ is the charge of the $\lambda$-parton. Substituting $q_\lambda=e \nu/n_\lambda$ into $\sum_{\lambda}q_\lambda=e$ gives the relation $\nu=( \sum_\lambda n_\lambda^{-1} )^{-1}$. The wave function for the partons is given by 
$\prod_{\lambda=1}^m \Phi_{n_\lambda}(\{z^{\lambda}_j\})$. We identify $z_j^{\lambda}=z_j$ and project into the LLL to obtain ``the $n_1\cdots n_m$ state"\footnote{These have been called Jain (parton) states in the literature.} 
\be
\Psi^{n_1\cdots n_m}_{\nu}= \mathcal{P}_{\rm LLL} \prod_{\lambda=1}^m \Phi_{n_\lambda}(\{z_j\}),\;\; \nu=\left( \sum_{\lambda=1}^m {1\over n_\lambda} \right)^{-1}, \;\; q_\lambda={\nu\over n_\lambda}.
\label{jainparton}
\ee
Negative values of $q_\lambda$ produce negative filling factors $n_{\lambda}$, which correspond to partons in a negative magnetic field.  For notational ease, it is customary to write $-n=\bar{n}$, with  $\Phi_{-n}=\Phi_{\bar{n}}=[\Phi_n]^*$. The LLL projection, as before, is not expected to alter the topological character of the unprojected product state. Interestingly, even though the partons are unphysical, they leave their footprints in the physical world: an excitation  in the factor $\Phi_{n_\lambda}$ has a charge $q_\lambda=e \nu/n_\lambda$ associated with it. In general, it is expected that the lowest energy quasiparticles correspond to the excitations in the factor $\Phi_{n_\lambda}$ with the largest $|n_{\lambda}|$, as they have the smallest charge. The $n_1\cdots n_m$ state occurs at shift ${\cal S}=\sum_{\lambda=1}^m n_m$.

The standard states of Eq.~\ref{jainwf2} are a part of the parton construction because 
$\Phi_1\sim \prod_{j<k}(z_j-z_k)$. Specifically, we have 
\be
\Psi^{n11\cdots}_{\nu={n\over 2pn+1}}=\mathcal{P}_{\rm LLL} \Phi_n\Phi_1^{2p} = \mathcal{P}_{\rm LLL} \Phi_n\prod_{j<k}(z_j-z_k)^{2p},
\ee
\be
\Psi^{\bar{n}11\cdots}_{\nu={n\over 2pn-1}}=\mathcal{P}_{\rm LLL} \Phi_{\bar{n}}\Phi_1^{2p} = \mathcal{P}_{\rm LLL} [\Phi_n]^*\prod_{j<k}(z_j-z_k)^{2p}.
\ee
The fact that the states in Eq.~\ref{jainwf2} can be written as products of IQH states was the original motivation for the parton construction.  The standard CF theory corresponds to $n_1\cdots n_m$ states with no more than one integer different from 1.

A field theoretical description\footnote{The discussion here owes greatly to insights from Ajit Balram and Maissam Barkeshli.}  of the wave functions in Eq.~\ref{jainparton} was developed by Wen and others~\cite{Blok90,Blok90b,Wen91,Wen92b,Balram18,Balram18a}. Let us outline the derivation of the CS theory for the standard $\nu=n/(2pn\pm 1)$ FQHE starting from the parton construction. We begin by noting that the CS Lagrangian for the $\nu=\pm n$ state of charge $q$ fermions is given by~\cite{Zee10,Tong16} 
\be
{\cal L}=\mp {1\over 4\pi} \sum_{j=1}^n a^j \partial a^j + {1\over  2\pi}   \sum_{j=1}^n t^j A\partial a^j,
\ee 
where $j$ is the LL index, $A$ is the physical vector potential (treated as a non-dynamical background field), $a^j$ is an emergent gauge field associated with the $j$th LL, $t=(q, q, \cdots, q)^T$ is the charge vector (same charge in each LL), and we have used the notation $a\partial b = \epsilon^{\mu \nu \delta}a_\mu \partial_\nu b_\delta$ and also set $e=\hbar=c=1$. The particle current density is given by $J^\mu=\delta S/\delta A_\mu = (1/2\pi)  \sum_{j=1}^n t^j \epsilon^{\mu\nu\lambda}\partial_\nu a_\lambda^j$.

The CS theory for the state at $n/(2n\pm 1)$ is constructed as follows. Let us represent the electron operator as $c=f_1  f_2f_3$, where the parton $f_1$ is in the $\nu=\pm n$ state and the partons $f_2$ and $f_3$ are in $\nu=1$ state. Of course, the partons are unphysical and the final theory must glue the partons into physical electrons. The redundancy of the electron operator implies an internal local gauge symmetry in which the local U(1) transformation $f_1\rightarrow e^{i\theta_1} f_1$, $f_2\rightarrow e^{i\theta_2}f_2$ and $f_3\rightarrow e^{-i\theta_1-i\theta_2}f_3$ leaves the theory invariant. This constraint is imposed by introducing two local gauge fields, denoted $b_1$ and $b_2$ below. 
The Lagrangian for the $\pm n 11$ state is given by
\begin{eqnarray}
{\cal L}&=& \left[\mp {1\over 4\pi} \sum_{j=1}^n a_1^j \partial a_1^j + {1\over 2\pi} \sum_{j=1}^n t_1^j A \partial a_1^j \right] + \left[ -{1\over 4\pi} a_2\partial a_2 + {1\over 2\pi} t_2 A \partial a_2\right] \nonumber \\
& & +\left[- {1\over 4\pi} a_3\partial a_3 + {1\over 2\pi} t_3 A \partial a_3 \right] 
+ \left[ b_1 \sum_{j=1}^n \partial a_1^j + b_2 \partial a_2 - (b_1+b_2)\partial a_3\right].
\label{standardL}
\end{eqnarray}
Here, $t_1=(q_1, q_1, \cdots, q_1)^T$, $t_2=q_2$ and $t_3=q_3$, where $q_1=\pm 1/(2n\pm 1)$, $q_2=q_3=n/(2n\pm 1)$ are the charges of the three partons in units of $e$. The terms in the first three square brackets on the right come from the individual factors, and the last square brackets contain the constraints. The constraints yield $\partial a_2=\partial a_3 = \sum_{j=1}^n \partial a_1^j$. These are equivalent to $a_2=a_3+c$ and $a_2= \sum_{j=1}^n a_1^j + d$, with $\epsilon^{\mu\nu\lambda}\partial_\nu c_\lambda=0$ and $\epsilon^{\mu\nu\lambda}\partial_\nu d_\lambda=0$. Substituting into Eq.~\ref{standardL} and noting that the terms containing $c$ and $d$ vanish~\cite{Balram18a}, the final form of the CS Lagrangian is obtained:
\be
{\cal L}=  -{1\over 4\pi} \sum_{i,j=1}^n a^i K^{ij} \partial a^j + {1\over 2\pi} \sum_{j=1}^n t^j  A \partial a^j, \;\; K^{ij}=\pm \delta_{ij}+2p, \;\; t=(1, 1, \cdots, 1)^T,
\label{Kmatrix}
\ee
where we have generalized to $\nu=n/(2pn\pm 1)$.

The CS theory of Abelian FQH states is in general given by a Lagrangian of the type shown in Eq.~\ref{Kmatrix}, which is defined by a symmetric integer valued $K$ matrix and a charge vector $t$. 
The $n_1n_2\cdots n_m$ state where no integer other than 1 is repeated is an Abelian state (as explained below).  In this case, there are $\sum_\lambda |n_\lambda|$ gauge fields prior to projection into the physical space, but the $m-1$ U$(1)$-constraints gluing the partons reduce the number of physical gauge fields to $\sum_\lambda |n_\lambda|-(m-1)$, which gives the dimension of the $K$ matrix. The $K$ matrix and the $t$ vector can be determined in the manner outlined above, and encode information about many topological properties of the state~\cite{Zee10,Tong16,Wen91b,Wen92b,Moore98}. The Hall conductance is given by $\sigma_{xy}=\sum_{i,j}t^i (K^{-1})^{ij} t^j$; the charge of the quasihole coupled to the field $a^i$ is $q^i=\sum_j(K^{-1})^{ij}t^j$; the relative braid statistics of the quasiholes is $\alpha^{ij}=(K^{-1})^{ij}$; and the ground state degeneracy on genus-$g$ surface is $|\det K|^g$. The charge and statistics are obtained by adding to the Lagrangian a term $\sum_j l^j a^j_\mu J'^\mu$, where $J'^\mu$ is the quasiparticle current and $l^j$ are positive (negative) integers representing the number of quasiparticles (quasiholes) coupled to the gauge field $a^j$. The K-matrix also contains information about edge states. Assuming that there is no edge reconstruction, the dimension of the matrix gives the number of independent edge modes and the number of positive (negative) eigenvalues gives the number of downstream (upstream) modes. The central charge $c$ is equal to the number of downstream minus the number of upstream modes, which is not affected by edge reconstruction. The central charge can be experimentally ascertained by a measurement of the thermal Hall conductance~\cite{Kane97}, which is given by $\kappa_{xy} =c \pi^2 k_{\rm B}^2T /3h$.

Wen showed~\cite{Wen91} that the parton construction also produces non-Abelian states, which are $n_1n_2\cdots n_m$ states where an integer $\geq 2$ is repeated. In particular, he considered states of the form $\Psi_{n/m}=[\Phi_n]^m$, i.e. $nn\cdots n$ states, for which all partons have charge $q_\lambda=e/m$. Because all parton species are indistinguishable, only those states are physical that are invariant under a local SU$(m)$ transformation within the parton space. This local SU$(m)$ symmetry is implemented through a non-Abelian SU$(m)$ gauge field coupled to the partons.  Integrating out the parton fields yields an SU$(m)_n$ CS theory. The quasiparticles in this theory are particle or hole excitations in $\Phi_n$, dressed by an SU$(m)_n$ CS gauge field. The non-Abelian braid properties of the excitations are determined from the properties of the SU$(m)_n$ CS theory.  Wen also considered the edge theory of the $\Psi_{n/m}=[\Phi_n]^m$ state. Before imposing the constraint that combines the unphysical partons into physical electrons, $mn$ chiral edge states arise from $n$ LLs of each of the $m$ partons, producing a central charge of $c=mn$. One must now project the theory into the physical space\footnote{This projection, which glues the partons back to produce the physical electrons, is not to be confused with the LLL projection.} by eliminating all of the fluctuations that transform non-trivially under SU$(m)$ transformation. The projection can be carried out by using the level-rank duality of Wess-Zumino-Witten models in conformal field theory as follows. Bosonization of $mn$ chiral fermions (assuming the same velocity for them) gives a U$(mn)_1$ algebra, the Hilbert space of which can be represented as a direct product of a U$(1)$ Kac-Moody algebra, an SU$(n)_m$ Kac-Moody algebra, and an SU$(m)_n$ Kac-Moody algebra. The central charges of these three algebras add to $mn$: $1+m(n^2-1)/(m+n)+n(m^2-1)/(m+n)=mn$. Projection is equivalent to removing the Hilbert space of SU$(m)_n$ Kac-Moody algebra, which leaves the central charge $c=1+m(n^2-1)/(m+n)=n(mn+1)/(m+n)$. The central charge for the complex conjugate state $[\Psi_{n/m}]^*=[\Phi^*_n]^m=[\Phi_{\bar{n}}]^m$ is given by $c=-n(mn+1)/(m+n)$.

Let us take some examples. For the Laughlin $1/m$ state, we have $n=1$, which gives $c=1$. Furthermore, the SU$(m)_1$ CS theory is abelian, implying Abelian statistics for the excitations (which is why repeated 1's do not yield non-Abelian statistics). For the 22 state $\Phi_2^2$, we get  $c=5/2$ and the fusion rules for quasiparticles correspond to the SU$(2)_2$ Ising topological quantum field theory. The $\bar{2}\bar{2}$ state $[\Phi_2]^{*2}$ has $c=-5/2$.

In addition to $[\Phi_n]^m$, states containing factors of $[\Phi_n]^m$, with $n\geq 2$ and $m\geq 2$, are also non-Abelian. An  interesting state is the $221$ state $\Phi_2^2\Phi_1$ at $\nu=1/2$. At the mean field level (before fusing partons into physical electrons) it has 5 chiral edge states, i.e. $c=5$. Gauge constraint must project out SU$(2)_2 \times$U$(1)$, which has central charge $3/2+1=5/2$, producing the central charge of $c=5-5/2=5/2$ for the $221$ state. The same remains true for the  $2211111$ states at $\nu=1/4$, because both $1$ and $111$ have central charge $c=1$. The lowest energy quasiparticles, which are excitations in the factors $\Phi_2$, have Ising fusion rules. 

The states containing factors of $\Phi_2^2$ can be interpreted as topological f-wave superconductors of composite fermions~\cite{Balram18}. To see this, note that the central charge of an $s$ wave superconductor is zero, whereas the $(p_x\pm ip_y)^l$ superconductor has central charge $c=\pm l/2$. Now consider the wave function $\Psi^{\rm paired}_{l}\Phi_1$, where $\Psi^{\rm paired}_{l}$ is the wave function of a paired state with relative angular momentum $l$ pairing. This wave function has filling factor $\nu=1$, Ising fusion rules for the quasiparticles (vortices in the superconductor), and central charge $c=1+l/2$. Furthermore, there is a unique topological quantum field theory for each central charge satisfying these properties. It therefore follows that $\Psi^{\rm paired}_{l=3}\Phi_1$ and $\Phi_2^2$ are topologically equivalent, i.e. belong to the same universality class. The 221 wave function $\Phi_2^2\Phi_1$ at $\nu=1/2$ and 22111 wave function $\Phi_2^2\Phi^3_1$ at $\nu=1/4$ are similarly topologically equivalent to the f-wave superconductor of composite fermions, $\Psi^{\rm paired}_{l=3}\Phi_1^{2p}$, with $c=5/2$.  Other topological superconductors of composite fermions have been considered in the past. The Pfaffian wave function $\Psi^{\rm paired}_{l=1}\Phi_1^2$, which is a $p_x+ ip_y$ superconductor of composite fermions, has central charge $c=1+1/2=3/2$, with the factor $\Phi_1^2$ contributing 1 and the Pfaffian factor $1/2$. Its hole partner, the anti-Pfaffian wave function, has central charge $c=1-3/2=-1/2$, because it is the Pfaffian of holes (contributing $-3/2$) in the background of $\nu=1$ state (contributing $+1$). Finally, the so-called PH Pfaffian~\cite{Son15} $\Psi^{\rm paired}_{l=-1}\Phi_1^2$ has $c=1/2$. When occurring in the second LL, the central charge for these states has additional contribution of $+2$ from the lowest filled Landau level.

The topological properties can, in principle, be derived directly from the wave functions. For the $\Psi_{n/m}=[\Phi_n]^m$ state with $n\geq 2$ and $m\geq 2$, specifying the positions of the quasiholes, in general, does not fully specify the wave function, because the different distributions of quasiholes in different factors do not necessarily produce identical wave functions. This lies at the root of non-Abelian statistics. The state $\Psi_{1/m}=[\Phi_1]^m$ does not produce non-Abelian statistics because here a hole in $\Phi_1$ at position $\eta$ corresponds to multiplication by the factor $\prod_j(z_j-\eta)$, and thus the wave function for several quasiholes simply produces an overall multiplicative factor $\prod_{j,\alpha}(z_j-\eta_\alpha)$ independent of which factors of $\Phi_1$ the holes were created in originally, thus defining the wave function uniquely. One can, in principle, obtain the braid statistics of the quasiparticles from the explicit wave functions, and the properties of the edge states and the central charge by studying the entanglement spectra.

All of the states of Eq.~\ref{jainparton} are mathematically well defined and presumably occur for some specially designed model interactions. The LLL is known to stabilize composite fermions. Can states beyond the CF theory be realized in experiments? For that one must look to higher LLs, to monolayer or bilayer graphene, or to LLL systems in wide quantum wells, all of which have different Coulomb matrix elements than purely two-dimensional electrons in the LLL. We review recent work that has found certain states of Eq.~\ref{jainparton} to be promising candidates for experimentally observed FQH states.

We do not discuss here a further generalization of the parton construction, called the projective construction, where the electron is represented as $c=f_0(f_1f_2+ \cdots + f_{2k-1}f_{2k})$, where $f_i$ represent fermion species. This can produce many other non-Abelian states, such as the Pfaffian state, and has been used to determine the bulk and edge field theories and other topological properties of these states~\cite{Wen99,Barkeshli10,Barkeshli14,Repellin15,Goldman19}.

\subsection{$\bar{2}\bar{2}111$ at $\nu=5/2$}

The 221 state $\mathcal{P}_{\rm LLL} \Phi_2^2\Phi_1$ is the simplest state beyond the standard CF theory~\cite{Jain89b,Jain90,Wu17b,Bandyopadhyay18}.  It is also interesting because it occurs at an even denominator fraction and is thought to support non-Abelian quasiparticles. It was considered as a candidate for the 5/2 FQHE~\cite{Willett87} but deemed unsatisfactory, because exact diagonalization in the second LL does not produce an incompressible state at the corresponding ``shift" on the sphere (see, for example, Ref.~\refcite{Wojs09}).  

BBR have considered~\cite{Balram18} the $\bar{2}\bar{2}111$ state as a candidate for the half filled second LL of the 5/2 FQH state. This state can be constructed conveniently by evaluating the LLL projection as\footnote{The interaction in the $n$th LL is fully defined by the Haldane pseudopotentials $V^{(n)}_m$, which are the energies of two electrons in relative angular momentum $m$. The problem of interacting electrons in the $n$th LL is formally equivalent to that of electrons in the LLL interacting with an effective interaction $V^{\rm eff}(r)$ that produces pseudopotentials $V^{(n)}_m$. This mapping allows us to conveniently work within the LLL even for the higher LL states, so long as LL mixing is disallowed.}
\be
\Psi^{\bar{2}\bar{2}111}_{\nu=1/2}={\cal P}_{\rm LLL} [\Phi_2^*]^2\Phi_1^3\sim [{\cal P}_{\rm LLL} \Phi_2^*\Phi_1^2]^2/\Phi_1=[\Psi_{2/3}]^2/\Phi_1.
\ee
BBR showed that this state has a reasonably high overlap with the exact ground state. The $\bar{2}\bar{2}111$ state belongs in the same universality class as the anti-Pfaffian. The two states occur at the same shift ($S=-1$), have decent overlaps, and produce very similar entanglement spectra~\cite{Balram18}. Furthermore they have the same central charge. To see this, we note that $[\Phi_2]^{*2}\Phi_1^3\sim \Psi^{\rm paired}_{l=-3}[\Phi_1]^*\Phi_1^3 \sim \Psi^{\rm paired}_{l=-3}\Phi_1^2$, where $\sim$ refers to topological equivalence and we have made use of the fact that multiplication by $[\Phi_1]^*\Phi_1$ does not alter the topological structure. The $\bar{2}\bar{2}111$ state thus has $c=1-3/2=-1/2$, which is the same as that for the anti-Pfaffian state. 

As seen in the context of $\nu=1/2$, many different states can be constructed for a given fraction. How does one decide which of these is plausible? The answer to this question must ultimately come from detailed calculations, and will depend on the form of the interaction. (As seen below, the 221 state may also be realized under different conditions.) The rule of thumb is that the more 1's, the better, and the fewer non-1's, the better.

\subsection{$\bar{3}\bar{2}111$ at $\nu=2+6/13$}

Kumar \emph{et al.}~\cite{Kumar10} reported the formation of a FQH state at $2+6/13$. Many facts suggest that this is unlikely to be analogous to the 6/13 state in the LLL, which is understood as six filled $\Lambda$Ls of composite fermions, or, alternatively, as the $611$ state. While the path to the 6/13 state in the LLL passes through 1/3, 2/5, 3/7, 4/9 and 5/11, the last three are not observed in the second LL~\cite{Shingla18}, and even 2+2/5 is believed to be distinct from the standard 211 state of the LLL~\cite{Read99,Rezayi09,Sreejith13,Zhu15,Mong15,Pakrouski16}. Furthermore, 6/13 is close to half filling, where the lowest and the second LLs exhibit qualitatively distinct behaviors. The observation of $2+6/13$ thus gives a clue into the different organizing principle in the second LL. (The $\nu=7/13$ FQH state observed in the $n=1$ LL of bilayer graphene~\cite{Zibrov16} is likely the hole partner of the $2+6/13$ state in GaAs quantum wells.)

Following the BBR insight, Balram {\em et al.}~\cite{Balram18a} considered the sequence $\bar{n}\bar{2}111$, which corresponds to the wave functions:
\be
\Psi^{\bar{n}\bar{2}111}_{\nu=2n/(5n-2)}\sim {[\mathcal{P}_{\rm LLL}\Phi_{\bar{n}}\Phi_1^2] 
[\mathcal{P}_{\rm LLL}\Phi_{\bar{2}}\Phi_1^2]\over \Phi_1}=
\frac{\Psi_{n/(2n-1)}\Psi_{2/3}}{\Phi_{1}}.
\ee
The first member of this sequence is $\bar{1}\bar{2}111$ at $\nu=2/3$, which is essentially identical to the standard $\bar{2}11$ state. The second member is the 1/2 state discussed in the previous subsection. Encouragingly, the third member $\bar{3}\bar{2}111$ occurs at 6/13. In the second LL, its energy ($-0.366$) is lower than that of the 611 state (-0.355), in contrast to the lowest LL where the energy of the $\bar{3}\bar{2}111$ and 611 states are -0.438 and -0.453, respectively (all energies are thermodynamic limits, quoted in units of $e^2/\epsilon l$). Additionally, $\bar{3}\bar{2}111$ has a reasonably high overlap of 0.754 with the exact 12 particle state in the second LL. These results make the $\bar{3}\bar{2}111$ state plausible.

The $\bar{3}\bar{2}111$ state has quasiparticles with charges $\mp 3/13$ and $\mp 2/13$ which correspond to particles and holes in the $\Phi_{\bar{3}}$ and $\Phi_{\bar{2}}$ factors; these can be combined to produce an excitation with charge $\pm 1/13$, but that is a composite object. $\bar{3}\bar{2}111$ occurs at a shift $S=-2$ on the sphere.  To obtain other topological properties, we consider the low-energy effective theory of the edge. Before we glue the partons to recover electrons, there are a total of eight edge states: three from $\Phi_{\bar{3}}$, two from $\Phi_{\bar{2}}$, and one from each $\Phi_1$. Gluing the partons gives four constraints, reducing the number of independent edge modes to four. Following the methods outlined above, one obtains the $K$ matrix~\cite{Balram18a} 
\be
K_{\bar{3}\bar{2}111} =    \begin{pmatrix} 
      -2 & -1 & 0 & 1 \\
      -1 & -2 & 0 & 1 \\
       0 & 0 & -2 & 1 \\
       1 & 1 &  1 & 1 \\
   \end{pmatrix},\quad
\ee
and the charge vector $t=(0, 0, 0, 1)^T$.  The ground state degeneracy on a manifold with genus $g$ is $|{\rm det}(K)|^{g}=13^{g}$.  The $K$ matrix above has one positive and three negative eigenvalues, giving central charge $c=-2$. In contrast, the 611 state occurs at shift $S=8$, and has central charge $c=6$ with all edge modes moving downstream (assuming absence of edge reconstruction). The $\bar{3}\bar{2}111$ and $611$ states may, in principle, be distinguished by shot noise experiments, which have been used to measure the presence of upstream modes~\cite{Bid10,Dolev11,Gross12,Inoue14}, or by a measurement of the thermal Hall conductance~\cite{Banerjee17,Banerjee17b}. (For $\nu=2+6/13$, we must also add $c=2$ coming from the two edge states of the lowest filled LL.)

Levin and Halperin~\cite{Levin09a} have proposed to obtain a FQHE at $\nu=2+6/13$ in a hierarchy starting from the anti-Pfaffian. Although this construction does not produce a microscopic wave function, it is possibly topologically equivalent to the $\bar{3}\bar{2}111$ state.

\subsection{$221$ at $\nu=1/2$ in single and multi-layer graphene}

A FQHE at $\nu=1/2$ has been seen in the $n=3$ LL of graphene~\cite{Kim18}. Exact diagonalization studies using the interaction pseudopotentials of the $n=3$ graphene LL do not support any of the known single or two component candidate incompressible states. However, a slight change of the interaction stabilizes the 221 state~\cite{Kim18}, which makes it a plausible candidate for the observed FQHE (given that the actual interaction is modified, for example, due to LL mixing or screening by a nearby conducting layer). A definitive identification will require further investigation.

A model Hamiltonian can be constructed for which the 221 state $\Phi^2_2\Phi_1$ (without the LLL projection) is the exact and unique zero energy ground state. In this model, one takes the lowest three LLs with orbital index $n=0, 1, 2$ to be degenerate and considers the Trugman-Kivelson interaction~\cite{Trugman85} $V_{\rm TK} = 4\pi  \nabla^2 \delta^{(2)}(\mathbf{r})$ between electrons\footnote{If the lowest {\em two} LLs with $n=0,\,1$ are taken to be degenerate, this model produces the unprojected $2/5$ state $\Phi_2\Phi_1^2$ as the unique zero energy ground state\cite{Jain90}. In this case, the state is seen, in numerical studies,\cite{Rezayi91} to evolve continuously into the LLL 2/5 state, without any gap closing, as the splitting between the two levels is increased to infinity.}. The kinetic energy is zero because $\Phi^2_2\Phi_1$ it involves only the lowest three LLs, and the interaction energy is zero because the wave function vanishes as $r^3$ when two electrons approach one another. One can further show that $\Phi^2_2\Phi_1$ is the unique state with these properties~\cite{Wu17b,Bandyopadhyay18}, as also confirmed in exact diagonalization studies in the spherical geometry for $N=6$ and 8 particles~\cite{Wu17b}.

The model where the lowest three LLs are degenerate but well separated from other LLs appears unphysical, but it turns out that precisely this situation occurs in multilayer graphene.  The low-energy Hamiltonian of Bernel stacked bilayer graphene (BLG) and ABC stacked trilayer graphene (TLG) can be approximately described, for each of the two valleys, by~\cite{McCann06,Barlas12} 
\begin{eqnarray}
H = T_J \left[
\begin{array}{cc}
0 & (\pi_x+i\pi_y)^J \\
(\pi_x-i\pi_y)^J & 0
\end{array}
\right].
\label{SingleHamilton}
\end{eqnarray}
Here 
$\vec{\pi}=\vec{p}+(e/c)\vec{A}$ is the canonical momentum operator,  $J=2$ ($3$) for BLG (TLG) is the chirality, and $T_J$ is a constant depending on microscopic details. The zeroth LL of Eq.~(\ref{SingleHamilton}) contains $J$-fold degenerate states, the wave functions for which, in the simplest approximation, are the wave functions of the lowest $J$ LLs of non-relativistic fermions. The degeneracy of the LLs is split by various features left out in Eq.~\ref{SingleHamilton}, and the splitting can be tuned by applying a transverse electric field~\cite{Kim15,Cote10,Apalkov11,Snizhko12}. For a proper choice of parameters, it appears possible to obtain situations where two or three orbital levels are approximately degenerate, producing the ideal condition for the realization of the 221 state. 

\begin{figure}[t]
\begin{center}
\includegraphics[width=0.5\textwidth]{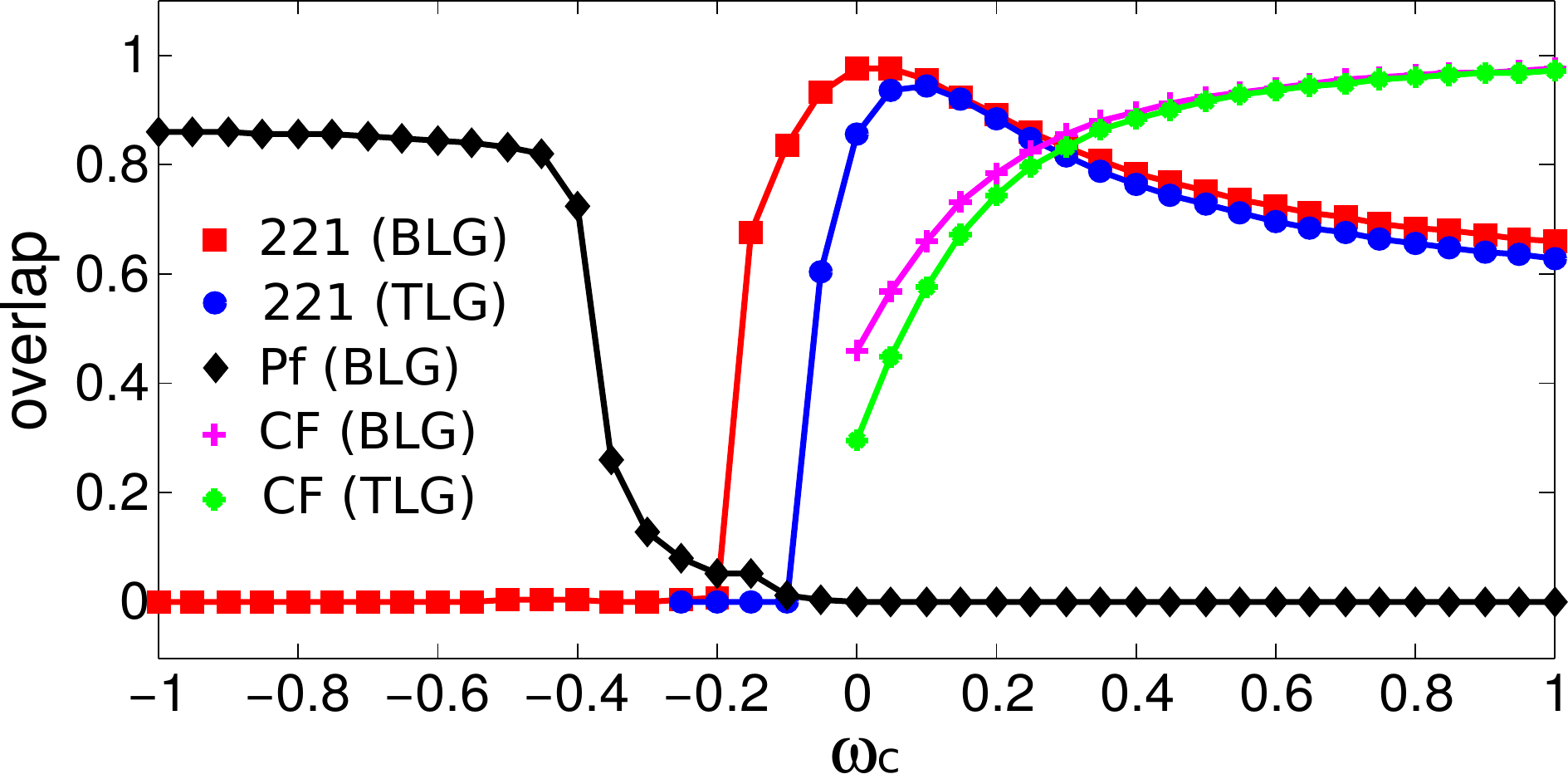}
\caption{
Overlap between the exact Coulomb ground states of model BLG and TLG Hamiltonians with various trial wave functions: the $\Phi_2^2\Phi_1$ state (221), the CF Fermi liquid state (CF), and the Pfaffian state (Pf). The BLG (TLG) Hamiltonian is defined by a model in which the lowest two (three) LLs are considered, with a variable LL splitting of $\omega_c$.  Results are shown for $(N,2Q)=(8,11)$ for which the $L=0$ subspace contains $418$ ($18212$) independent states in BLG (TLG). Y. Wu, T. Shi, and J. K. Jain, Nano Letters. {\bf 17}, 4643-4647, (2017)~\cite{Wu17b}.}
\label{Figure2-Yinghai}
\end{center}
\end{figure}

Wu {\em et al.}~\cite{Wu17b} have investigated if the 221 state can be realized in these systems for the Coulomb interaction. Fig.~\ref{Figure2-Yinghai} shows the overlap of the exact Coulomb ground state for 8 particles with the 221 state as a function of the LL splitting $\omega_c$ (quoted in units of $e^2/\epsilon l$). Overlaps are also shown for the Pfaffian and the CF Fermi sea states. For a range of splittings near $\omega_c=0$ the 221 state  has a large overlaps with the exact Coulomb state for both BLG and TLG. For large positive $\omega_c$ the CF Fermi sea is obtained, as expected, in both BLG and TLG. For large negative $\omega_c$ in a BLG, ordering of the lowest two LLs is inverted, stabilizing the Pfaffian 1/2 state in the $N=1$ orbital. These results suggest that the 221 state should occur for both BLG and TLG in the vicinity of $\omega_c=0$.  Should this state be observed, it would be the first example where LL mixing is fundamentally responsible for creating a new FQH state. The 221 state may be relevant to the 1/2 FQH state reported in BLG,~\cite{Kim15} and possibly in TLG~\cite{Bao10}. It ought to be noted that the actual wave functions for the BLG and TLG graphene LLs are more complicated than the model considered above (see the chapter by Dean, Kim, Li and Young), which will need to be incorporated in a more realistic calculation.

\begin{figure}[t]
\begin{center}
\includegraphics[width=0.5\textwidth]{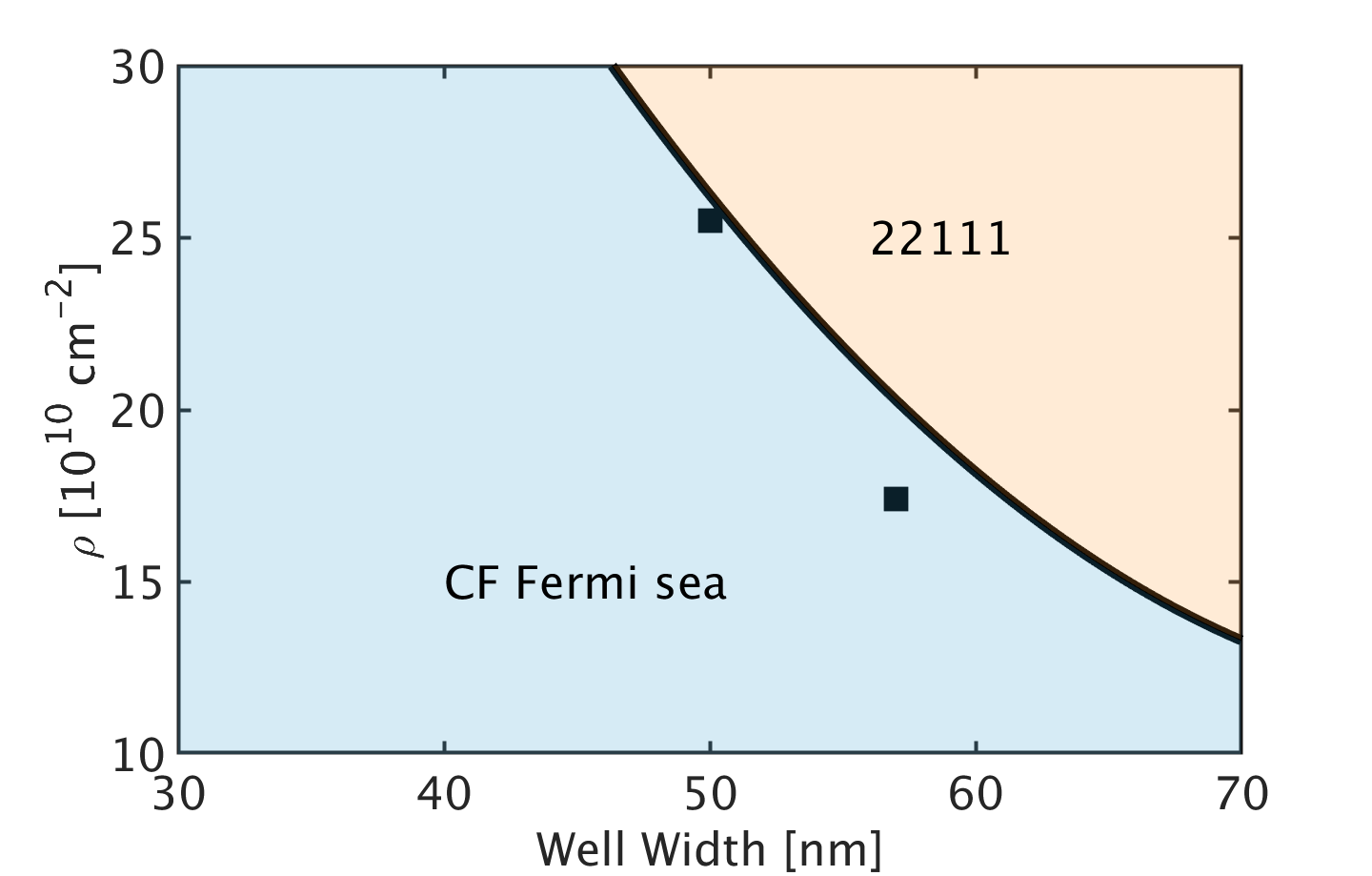}
\caption{\label{fig:faugno}
The calculated phase diagram at $\nu=1/4$ as a function of the quantum well width and density considering single-component candidate states. Only the CFFS and 22111 states are realized. Black squares, taken from Luhman {\em et al.}\cite{Luhman08} and Shabani {\em et al.}\cite{Shabani09a}, indicate the experimental phase boundary. Source: W. N. Faugno, A. C. Balram, M. Barkeshli, and J. K. Jain, Phys. Rev. Lett. {\bf 123}, 016802 (2019)~\cite{Faugno19}.
} 
\end{center}
\end{figure}

\subsection{$22111$ at $\nu=1/4$ in wide quantum wells}

There exists evidence for FQHE at filling factor $\nu=1/4$ in wide quantum wells~\cite{Luhman08,Shabani09a,Shabani09b,Shabani13}. This FQHE is induced by a change in the form of the interaction due to finite width, as the $\nu=1/4$ state is known to be a CFFS for small widths. Faugno {\em et al.}~\cite{Faugno19} have determined the variational energies of many single component states: the CF Fermi sea, 22111, $\bar{2}\bar{2}11111$, ${\rm Pf}[(z_j-z_k)^{-1}]\Phi_1^4$, and ${\rm Pf}[(z_j-z_k)^{-3}]\Phi_1^4$, as a function of the quantum well width and the density, with the finite width effect treated in LDA (see Section~\ref{sec:laboratory}). The CF Fermi sea is seen to become unstable to the 22111 state as the density and / or the quantum well width is increased. The calculated phase diagram shown in Fig.~\ref{fig:faugno} is in good agreement with the onset of the 1/4 FQHE in experiments.  Wide quantum wells can behave like bilayer systems and one may ask if the observed state might be an incompressible bilayer state.  An energetic comparison between the one- and two-component states is complicated by the fact that the energy separation between the symmetric and antisymmetric subbands ($\Delta_{\rm SAS}$) is known much less precisely than the Coulomb energy differences between the various candidates states. Faugno {\em et al.}~\cite{Faugno19} have also considered a large number of two-component candidate states at $\nu=1/4$ for an ideal bilayer system consisting of two two-dimensional planes, and found that no incompressible state is stabilized for any value of the interlayer separation. This result, combined with the agreement between theory and experiment in Fig.~\ref{fig:faugno}, supports the view that the observed $\nu=1/4$ FQHE in wide quantum wells has a single component origin.

\subsection{$\bar{2}\bar{2}\bar{2}1111$ for $\nu=2+2/5$}

Balram {\em et al.}~\cite{Balram19} have considered states of the form $\bar{2}^k 1^{k+1}$ (in obvious notation) at $\nu=2/(k+2)$, described by the wave function ${\cal P}_{\rm LLL} [\Phi_2^*]^k \Phi_1^{k+1}\sim [\Psi_{2/3}]^k/\Phi_1^{k-1}$. They have shown that these states are in the same universality class as the hole conjugates of the so-called parafermion states~\cite{Read99} at $\nu=k/(k+2)$. There is theoretical evidence that for $k=3$ this state is relevant for the FQHE at $\nu=2+3/5$ (and, via particle hole symmetry, also for $\nu=2+2/5$).

{\bf Acknowledgment:} 
This chapter features results from fruitful collaboration with many wonderful students and colleagues, including Alexander Archer, Ajit Balram, Maissam Barkeshli,  William Faugno, Mikael Fremling, Yayun Hu, Manish Jain, Yang Liu, Sudhansu Mandal, Sutirtha Mukherjee, Kwon Park, Loren Pfeiffer, Songyang Pu, Mark Rudner, Diptiman Sen, Mansour Shayegan, Jurgen Smet, G. J. Sreejith, Manisha Thakurathi, Csaba T\"oke, Arkadiusz W\'ojs, Yinghai Wu, Yuhe Zhang, and Jianyun Zhao. The author is deeply grateful to all of them. Thanks are also due to Ajit Balram, Yayun Hu, Dwipesh Majumder, Songyang Pu and Yinghai Wu for help with figures, to the US Department of Energy for financial support under Grant No. DE-SC0005042, and to the Infosys Foundation for enabling a visit to the Indian Institute of Science, Bangalore, where part of this article was written. The author  expresses gratitude to Bert Halperin for many valuable suggestions on the manuscript.

\begin{appendix}

\section{Landau levels in the symmetric gauge}
\label{sec:symmetric}

This Appendix is reproduced from Ref.~\refcite{Jain15}. The Hamiltonian for a non-relativistic electron moving in two-dimensions in a perpendicular magnetic field is given by
\begin{equation}
H = {1\over 2 m_b}\left(\vec{p}+{e\over c}\vec{A}\right)^2,
\end{equation}
where $m_b$ is the band mass of the electron. Choosing the symmetric gauge 
$\vec{A} = \frac{\vec{B}\times\vec{r}}{2}=\frac{B}{2} (-y, x, 0)$, and taking 
the units of length as the magnetic length $l=\sqrt{\hbar c/eB}=1$, the Hamiltonian becomes 
\begin{equation}
H=\hbar\omega_c\left(a^{\dagger}a+\frac{1}{2}\right),
\end{equation}
where $\hbar\omega_c=\hbar eB/m_bc$ is the cyclotron energy and the ladder operators are defined as $a^\dagger = \frac{1}{\sqrt{2}}\left(\frac{\bar{z}}{2} - 2\frac{\partial}{\partial z}\right)$ 
and 
$a= \frac{1}{\sqrt{2}} \left(\frac{z}{2} +2\frac{\partial}{\partial \bar{z}}\right)$ 
in terms of the complex coordinates 
$z= x-iy=r{\rm e}^{-i\theta}, \;\; \bar{z}= x+iy=r{\rm e}^{i\theta}$. 
Further defining 
$b= \frac{1}{\sqrt{2}} \left(\frac{\bar{z}}{2} + 2\frac{\partial}{\partial z}\right)$ 
and 
$b^\dagger = \frac{1}{\sqrt{2}} \left(\frac{z}{2} -2\frac{\partial}{\partial \bar{z}}\right)$, 
one can check that 
$[a,a^{\dagger}]=1$, $[b,b^{\dagger}]=1$, 
and all other commutators vanish. 
The LL index $n$ is the eigenvalue of $a^{\dagger}a$, and the 
$z$ component of the angular momentum operator is defined as 
$L = -i\hbar \frac{\partial}{\partial \theta} =-\hbar (b^\dagger b-a^\dagger a)\equiv -\hbar m$, with 
$m=-n, -n+1, \cdots 0, 1, \cdots$ in the $n^{\rm th}$ LL. The single particle eigenstates are obtained in the standard manner by successive applications of ladder operators 
\begin{equation}
|n,m\rangle =\frac{(b^{\dagger})^{m+n}}{\sqrt{(m+n)!}}
\frac{(a^{\dagger})^{n}}{\sqrt{n!}}\;|0,0\rangle \;,
\end{equation}
with eigenenergies $E_{n}=\hbar\omega_c\left(n+\frac{1}{2}\right)$, and the bottom state $\langle\vec{r}|0,0\rangle \equiv \eta_{0,0}(\vec{r})=\frac{1}{\sqrt{2\pi}}\;{\rm e}^{-\frac{1}{4}z\bar{z}}$ is annihilated by $a$ and $b$. 
The single-particle states are especially simple in the LLL ($n=0$):
\begin{equation}
\eta_{0,m} =\langle \vec{r} | 0,m \rangle = \frac{\left(b^\dagger\right)^m}
{\sqrt{m!}}\eta_{0,0}=\frac{ z^m
{\rm e}^{-\frac{1}{4}z\bar{z}}}{\sqrt{2\pi 2^m m!}}.
\end{equation}
Aside from the ubiquitous Gaussian factor, a general single particle state in the lowest LL is an analytic function of $z$, i.e. it does not involve any $\bar z$.   A general many-particle wave function confined to the LLL therefore has the form $\Psi=F[\{  z_j\}] \exp\left[-\frac{1}{4}\sum_{i}|z_{i}|^{2}\right]$ where $F[\{  z_j\}]$ is an antisymmetric function of the $z_j$.

The LL degeneracy can be obtained by considering 
a region of radius $R$ centered at the origin, and asking
how many single particle states lie inside it. 
For the LLL, the
eigenstate $|0,m\rangle $ has its weight located at the circle of radius
$r=\sqrt{2m}\cdot l$.
The largest value of $m$ for which the single particle state falls inside our circular region
is given by $M=R^{2}/2l ^{2}$, 
which is also the total number of single particle eigenstates in the
LLL that fall inside the disk (neglecting order one corrections).
Thus, the degeneracy per unit area is 
$M/\pi R^2= 1/(2\pi l^2)=B/ \phi_0$
which is the number of flux quanta (with a single flux quantum defined as $\phi_0=hc/e$) penetrating the sample through a unit area.
The filling factor, which is the nominal number of filled LLs, is equal to the number of electrons per flux quantum, given by
\begin{equation}
\nu= {\rho\over B/\phi_0}=2\pi l^{2} \rho,
\end{equation}
where $\rho$ is the 2D density of electrons.

The wave function $\Phi_n$ of the state with $n$ fully filled LL (in which all states inside a disk of some radius are filled) is precisely known; it is the Slater determinant formed from the occupied single particle orbitals.  The wave function of the lowest filled LL, $\Phi_1$, has a particularly simple form (apart from a normalization factor):
\be
\Phi_{1}=\left|\begin{array}{ccccc}
1&1&1&.&.\\
z_{1}&z_{2}&z_{3}&.&.\\
z_{1}^{2}&z_{2}^{2}&z_{3}^{2}&.&.\\
.&.&.&.&.\\
.&.&.&.&.
\end{array}
\right|\;\;
\exp\left[-\frac{1}{4}\sum_{i}|z_{i}|^{2}\right]=\prod_{j<k}(z_{j}-z_{k})\;
\exp\left[-\frac{1}{4}\sum_{i}|z_{i}|^{2}\right].
\label{phi1}
\ee

\end{appendix}

\bibliography{../../Latex-Revtex-etc./biblio_fqhe.bib}

\begin{thebibliography}{247}
\providecommand{\natexlab}[1]{#1}
\providecommand{\url}[1]{\texttt{#1}}
\expandafter\ifx\csname urlstyle\endcsname\relax
  \providecommand{\doi}[1]{doi: #1}\else
  \providecommand{\doi}{doi: \begingroup \urlstyle{rm}\Url}\fi

\bibitem{Stormer07}
H.~L. Stormer and D.~C. Tsui.
\newblock Composite fermions in the fractional quantum {Hall} effect.
\newblock In \emph{Perspectives in Quantum Hall Effects}, pp. 385--421.
  Wiley-VCH Verlag GmbH,  (2007).
\newblock ISBN 9783527617258.
\newblock \doi{10.1002/9783527617258.ch10}.
\newblock URL \url{http://dx.doi.org/10.1002/9783527617258.ch10}.

\bibitem{Haldane83}
F.~D.~M. Haldane, Fractional quantization of the {Hall} effect: A hierarchy of
  incompressible quantum fluid states, \emph{Phys. Rev. Lett.} {\bf 51},
  \penalty0 605--608 (Aug, 1983).
\newblock \doi{10.1103/PhysRevLett.51.605}.
\newblock URL \url{http://link.aps.org/doi/10.1103/PhysRevLett.51.605}.

\bibitem{Klitzing80}
K.~v. Klitzing, G.~Dorda, and M.~Pepper, New method for high-accuracy
  determination of the fine-structure constant based on quantized {Hall}
  resistance, \emph{Phys. Rev. Lett.} {\bf 45}, \penalty0 494--497 (Aug, 1980).
\newblock \doi{10.1103/PhysRevLett.45.494}.
\newblock URL \url{http://link.aps.org/doi/10.1103/PhysRevLett.45.494}.

\bibitem{Laughlin81}
R.~B. Laughlin, Quantized {Hall} conductivity in two dimensions, \emph{Phys.
  Rev. B}. {\bf 23}, \penalty0 5632--5633 (May, 1981).
\newblock \doi{10.1103/PhysRevB.23.5632}.
\newblock URL \url{http://link.aps.org/doi/10.1103/PhysRevB.23.5632}.

\bibitem{Thouless82}
D.~J. Thouless, M.~Kohmoto, M.~P. Nightingale, and M.~den Nijs, Quantized
  {Hall} conductance in a two-dimensional periodic potential, \emph{Phys. Rev.
  Lett.} {\bf 49}, \penalty0 405--408 (Aug, 1982).
\newblock \doi{10.1103/PhysRevLett.49.405}.
\newblock URL \url{http://link.aps.org/doi/10.1103/PhysRevLett.49.405}.

\bibitem{Haldane88}
F.~D.~M. Haldane, Model for a quantum {Hall} effect without {Landau} levels:
  Condensed-matter realization of the "parity anomaly", \emph{Phys. Rev. Lett.}
  {\bf 61}, \penalty0 2015--2018 (Oct, 1988).
\newblock \doi{10.1103/PhysRevLett.61.2015}.
\newblock URL \url{https://link.aps.org/doi/10.1103/PhysRevLett.61.2015}.

\bibitem{Tsui82}
D.~C. Tsui, H.~L. Stormer, and A.~C. Gossard, Two-dimensional magnetotransport
  in the extreme quantum limit, \emph{Phys. Rev. Lett.} {\bf 48}, \penalty0
  1559--1562 (May, 1982).
\newblock \doi{10.1103/PhysRevLett.48.1559}.
\newblock URL \url{http://link.aps.org/doi/10.1103/PhysRevLett.48.1559}.

\bibitem{Wigner34}
E.~Wigner, On the interaction of electrons in metals., \emph{Phys. Rev.} {\bf
  46}, \penalty0 1002,  (1934).

\bibitem{Lozovik75}
Y.~E. Lozovik and V.~I. Yudson, Feasibility of superfluidity of paired
  spatially separated electrons and holes; a new superconductivity mechanism.,
  \emph{JETP Lett.} {\bf 22}, \penalty0 11,  (1975).

\bibitem{Laughlin83}
R.~B. Laughlin, Anomalous quantum {Hall} effect: An incompressible quantum
  fluid with fractionally charged excitations, \emph{Phys. Rev. Lett.} {\bf
  50}, \penalty0 1395--1398 (May, 1983).
\newblock \doi{10.1103/PhysRevLett.50.1395}.
\newblock URL \url{http://link.aps.org/doi/10.1103/PhysRevLett.50.1395}.

\bibitem{Leinaas77}
J.~Leinaas and J.~Myrheim, On the theory of identical particles, \emph{Il Nuovo
  Cimento B Series 11}. {\bf 37}\penalty0 (1), \penalty0 1--23,  (1977).
\newblock ISSN 0369-3554.
\newblock \doi{10.1007/BF02727953}.
\newblock URL \url{http://dx.doi.org/10.1007/BF02727953}.

\bibitem{Wilczek82}
F.~Wilczek, Quantum mechanics of fractional-spin particles, \emph{Phys. Rev.
  Lett.} {\bf 49}, \penalty0 957--959 (Oct, 1982).
\newblock \doi{10.1103/PhysRevLett.49.957}.
\newblock URL \url{http://link.aps.org/doi/10.1103/PhysRevLett.49.957}.

\bibitem{Halperin84}
B.~I. Halperin, Statistics of quasiparticles and the hierarchy of fractional
  quantized {Hall} states, \emph{Phys. Rev. Lett.} {\bf 52}, \penalty0
  1583--1586 (Apr, 1984).
\newblock \doi{10.1103/PhysRevLett.52.1583}.
\newblock URL \url{http://link.aps.org/doi/10.1103/PhysRevLett.52.1583}.

\bibitem{Arovas84}
D.~Arovas, J.~R. Schrieffer, and F.~Wilczek, Fractional statistics and the
  quantum {Hall} effect, \emph{Phys. Rev. Lett.} {\bf 53}, \penalty0 722--723
  (Aug, 1984).
\newblock \doi{10.1103/PhysRevLett.53.722}.
\newblock URL \url{http://link.aps.org/doi/10.1103/PhysRevLett.53.722}.

\bibitem{Girvin87}
S.~M. Girvin and A.~H. MacDonald, Off-diagonal long-range order, oblique
  confinement, and the fractional quantum {Hall} effect, \emph{Phys. Rev.
  Lett.} {\bf 58}, \penalty0 1252--1255 (Mar, 1987).
\newblock \doi{10.1103/PhysRevLett.58.1252}.
\newblock URL \url{http://link.aps.org/doi/10.1103/PhysRevLett.58.1252}.

\bibitem{Zhang89}
S.~C. Zhang, T.~H. Hansson, and S.~Kivelson, Effective-field-theory model for
  the fractional quantum {Hall} effect, \emph{Phys. Rev. Lett.} {\bf 62},
  \penalty0 82--85 (Jan, 1989).
\newblock \doi{10.1103/PhysRevLett.62.82}.
\newblock URL \url{http://link.aps.org/doi/10.1103/PhysRevLett.62.82}.

\bibitem{Jain89}
J.~K. Jain, Composite-fermion approach for the fractional quantum {Hall}
  effect, \emph{Phys. Rev. Lett.} {\bf 63}, \penalty0 199--202 (Jul, 1989).
\newblock \doi{10.1103/PhysRevLett.63.199}.
\newblock URL \url{http://link.aps.org/doi/10.1103/PhysRevLett.63.199}.

\bibitem{Jain07}
J.~K. Jain, \emph{Composite Fermions}. (Cambridge University Press, New York,
  US, 2007).

\bibitem{Trivedi91}
N.~Trivedi and J.~K. Jain, Numerical study of {Jastrow}-{Slater} trial states
  for the fractional quantum {Hall} effect, \emph{Mod. Phys. Lett. B}. {\bf
  05}\penalty0 (07), \penalty0 503--510,  (1991).
\newblock \doi{10.1142/S0217984991000599}.

\bibitem{Kamilla97b}
R.~K. Kamilla and J.~K. Jain, Variational study of the vortex structure of
  composite fermions, \emph{Phys. Rev. B}. {\bf 55}, \penalty0 9824--9827 (Apr,
  1997).
\newblock \doi{10.1103/PhysRevB.55.9824}.
\newblock URL \url{http://link.aps.org/doi/10.1103/PhysRevB.55.9824}.

\bibitem{Jain97}
J.~K. Jain and R.~K. Kamilla, Composite fermions in the {Hilbert} space of the
  lowest electronic {Landau} level, \emph{Int. J. Mod. Phys. B}. {\bf
  11}\penalty0 (22), \penalty0 2621--2660,  (1997).
\newblock \doi{10.1142/S0217979297001301}.

\bibitem{Jain97b}
J.~K. Jain and R.~K. Kamilla, Quantitative study of large composite-fermion
  systems, \emph{Phys. Rev. B}. {\bf 55}, \penalty0 R4895--R4898 (Feb, 1997).
\newblock \doi{10.1103/PhysRevB.55.R4895}.
\newblock URL \url{http://link.aps.org/doi/10.1103/PhysRevB.55.R4895}.

\bibitem{Mandal02}
S.~S. Mandal and J.~K. Jain, Theoretical search for the nested quantum {Hall}
  effect of composite fermions, \emph{Phys. Rev. B}. {\bf 66}, \penalty0 155302
  (Oct, 2002).
\newblock \doi{10.1103/PhysRevB.66.155302}.
\newblock URL \url{http://link.aps.org/doi/10.1103/PhysRevB.66.155302}.

\bibitem{Haldane85}
F.~D.~M. Haldane and E.~H. Rezayi, Periodic {Laughlin}-{Jastrow} wave functions
  for the fractional quantized {Hall} effect, \emph{Phys. Rev. B}. {\bf 31},
  \penalty0 2529--2531 (Feb, 1985).
\newblock \doi{10.1103/PhysRevB.31.2529}.
\newblock URL \url{http://link.aps.org/doi/10.1103/PhysRevB.31.2529}.

\bibitem{Hermanns13}
M.~Hermanns, Composite fermion states on the torus, \emph{Phys. Rev. B}. {\bf
  87}, \penalty0 235128 (Jun, 2013).
\newblock \doi{10.1103/PhysRevB.87.235128}.
\newblock URL \url{http://link.aps.org/doi/10.1103/PhysRevB.87.235128}.

\bibitem{Pu17}
S.~Pu, Y.-H. Wu, and J.~K. Jain, Composite fermions on a torus, \emph{Phys.
  Rev. B}. {\bf 96}, \penalty0 195302 (Nov, 2017).
\newblock \doi{10.1103/PhysRevB.96.195302}.
\newblock URL \url{https://link.aps.org/doi/10.1103/PhysRevB.96.195302}.

\bibitem{Lopez91}
A.~Lopez and E.~Fradkin, Fractional quantum {Hall} effect and {Chern}-{Simons}
  gauge theories, \emph{Phys. Rev. B}. {\bf 44}, \penalty0 5246--5262 (Sep,
  1991).
\newblock \doi{10.1103/PhysRevB.44.5246}.
\newblock URL \url{http://link.aps.org/doi/10.1103/PhysRevB.44.5246}.

\bibitem{Halperin93}
B.~I. Halperin, P.~A. Lee, and N.~Read, Theory of the half-filled {Landau}
  level, \emph{Phys. Rev. B}. {\bf 47}, \penalty0 7312--7343 (Mar, 1993).
\newblock \doi{10.1103/PhysRevB.47.7312}.
\newblock URL \url{http://link.aps.org/doi/10.1103/PhysRevB.47.7312}.

\bibitem{Hansson17}
T.~H. Hansson, M.~Hermanns, S.~H. Simon, and S.~F. Viefers, Quantum {Hall}
  physics: Hierarchies and conformal field theory techniques, \emph{Rev. Mod.
  Phys.} {\bf 89}, \penalty0 025005 (May, 2017).
\newblock \doi{10.1103/RevModPhys.89.025005}.
\newblock URL \url{https://link.aps.org/doi/10.1103/RevModPhys.89.025005}.

\bibitem{Kang93}
W.~Kang, H.~L. Stormer, L.~N. Pfeiffer, K.~W. Baldwin, and K.~W. West, How real
  are composite fermions?, \emph{Phys. Rev. Lett.} {\bf 71}, \penalty0
  3850--3853 (Dec, 1993).
\newblock \doi{10.1103/PhysRevLett.71.3850}.
\newblock URL \url{http://link.aps.org/doi/10.1103/PhysRevLett.71.3850}.

\bibitem{Stormer}
{H.L. Stormer}.
\newblock Private communication.

\bibitem{Pan03}
W.~Pan, H.~L. Stormer, D.~C. Tsui, L.~N. Pfeiffer, K.~W. Baldwin, and K.~W.
  West, Fractional quantum {Hall} effect of composite fermions, \emph{Phys.
  Rev. Lett.} {\bf 90}, \penalty0 016801 (Jan, 2003).
\newblock \doi{10.1103/PhysRevLett.90.016801}.
\newblock URL \url{http://link.aps.org/doi/10.1103/PhysRevLett.90.016801}.

\bibitem{Pan02}
W.~Pan, H.~L. Stormer, D.~C. Tsui, L.~N. Pfeiffer, K.~W. Baldwin, and K.~W.
  West, Transition from an electron solid to the sequence of fractional quantum
  {Hall} states at very low {Landau} level filling factor, \emph{Phys. Rev.
  Lett.} {\bf 88}, \penalty0 176802 (Apr, 2002).
\newblock \doi{10.1103/PhysRevLett.88.176802}.

\bibitem{Jain14}
J.~K. Jain, A note contrasting two microscopic theories of the fractional
  quantum {Hall} effect, \emph{Indian Journal of Physics}. {\bf 88}, \penalty0
  915--929 (Sep, 2014).
\newblock ISSN 0974-9845.
\newblock \doi{10.1007/s12648-014-0491-9}.
\newblock URL \url{http://dx.doi.org/10.1007/s12648-014-0491-9}.

\bibitem{Scarola02}
V.~W. Scarola, S.-Y. Lee, and J.~K. Jain, Excitation gaps of incompressible
  composite fermion states: Approach to the {Fermi} sea, \emph{Phys. Rev. B}.
  {\bf 66}, \penalty0 155320 (Oct, 2002).
\newblock \doi{10.1103/PhysRevB.66.155320}.
\newblock URL \url{http://link.aps.org/doi/10.1103/PhysRevB.66.155320}.

\bibitem{Du93}
R.~R. Du, H.~L. Stormer, D.~C. Tsui, L.~N. Pfeiffer, and K.~W. West,
  Experimental evidence for new particles in the fractional quantum {Hall}
  effect, \emph{Phys. Rev. Lett.} {\bf 70}, \penalty0 2944--2947 (May, 1993).
\newblock \doi{10.1103/PhysRevLett.70.2944}.
\newblock URL \url{http://link.aps.org/doi/10.1103/PhysRevLett.70.2944}.

\bibitem{Manoharan94}
H.~C. Manoharan, M.~Shayegan, and S.~J. Klepper, Signatures of a novel {Fermi}
  liquid in a two-dimensional composite particle metal, \emph{Phys. Rev. Lett.}
  {\bf 73}, \penalty0 3270--3273 (Dec, 1994).
\newblock \doi{10.1103/PhysRevLett.73.3270}.
\newblock URL \url{http://link.aps.org/doi/10.1103/PhysRevLett.73.3270}.

\bibitem{Pinczuk93}
A.~Pinczuk, B.~S. Dennis, L.~N. Pfeiffer, and K.~West, Observation of
  collective excitations in the fractional quantum {Hall} effect, \emph{Phys.
  Rev. Lett.} {\bf 70}, \penalty0 3983--3986 (Jun, 1993).
\newblock \doi{10.1103/PhysRevLett.70.3983}.
\newblock URL \url{http://link.aps.org/doi/10.1103/PhysRevLett.70.3983}.

\bibitem{Kang00}
M.~Kang, A.~Pinczuk, B.~S. Dennis, M.~A. Eriksson, L.~N. Pfeiffer, and K.~W.
  West, Inelastic light scattering by gap excitations of fractional quantum
  {Hall} states at $1/3\ge \nu \le 2/3$, \emph{Phys. Rev. Lett.} {\bf 84},
  \penalty0 546--549 (Jan, 2000).
\newblock \doi{10.1103/PhysRevLett.84.546}.
\newblock URL \url{http://link.aps.org/doi/10.1103/PhysRevLett.84.546}.

\bibitem{Kukushkin00}
I.~V. Kukushkin, J.~H. Smet, K.~von Klitzing, and K.~Eberl, Optical
  investigation of spin-wave excitations in fractional quantum {Hall} states
  and of interaction between composite fermions, \emph{Phys. Rev. Lett.} {\bf
  85}, \penalty0 3688--3691 (Oct, 2000).
\newblock \doi{10.1103/PhysRevLett.85.3688}.
\newblock URL \url{http://link.aps.org/doi/10.1103/PhysRevLett.85.3688}.

\bibitem{Kang01}
M.~Kang, A.~Pinczuk, B.~S. Dennis, L.~N. Pfeiffer, and K.~W. West, Observation
  of multiple magnetorotons in the fractional quantum {Hall} effect,
  \emph{Phys. Rev. Lett.} {\bf 86}, \penalty0 2637--2640 (Mar, 2001).
\newblock \doi{10.1103/PhysRevLett.86.2637}.
\newblock URL \url{http://link.aps.org/doi/10.1103/PhysRevLett.86.2637}.

\bibitem{Dujovne03}
I.~Dujovne, A.~Pinczuk, M.~Kang, B.~S. Dennis, L.~N. Pfeiffer, and K.~W. West,
  Evidence of {Landau} levels and interactions in low-lying excitations of
  composite fermions at
  $1/3\ensuremath{\le}\ensuremath{\nu}\ensuremath{\le}2/5$, \emph{Phys. Rev.
  Lett.} {\bf 90}, \penalty0 036803 (Jan, 2003).
\newblock \doi{10.1103/PhysRevLett.90.036803}.
\newblock URL \url{http://link.aps.org/doi/10.1103/PhysRevLett.90.036803}.

\bibitem{Dujovne05}
I.~Dujovne, A.~Pinczuk, M.~Kang, B.~S. Dennis, L.~N. Pfeiffer, and K.~W. West,
  Composite-fermion spin excitations as $\ensuremath{\nu}$ approaches $1/2$:
  Interactions in the {Fermi} sea, \emph{Phys. Rev. Lett.} {\bf 95}, \penalty0
  056808 (Jul, 2005).
\newblock \doi{10.1103/PhysRevLett.95.056808}.
\newblock URL \url{http://link.aps.org/doi/10.1103/PhysRevLett.95.056808}.

\bibitem{Kukushkin09}
I.~V. Kukushkin, J.~H. Smet, V.~W. Scarola, V.~Umansky, and K.~von Klitzing,
  Dispersion of the excitations of fractional quantum {Hall} states,
  \emph{Science}. {\bf 324}\penalty0 (5930), \penalty0 1044--1047,  (2009).
\newblock \doi{10.1126/science.1171472}.
\newblock URL \url{http://www.sciencemag.org/content/324/5930/1044.abstract}.

\bibitem{Rhone11}
T.~D. Rhone, D.~Majumder, B.~S. Dennis, C.~Hirjibehedin, I.~Dujovne, J.~G.
  Groshaus, Y.~Gallais, J.~K. Jain, S.~S. Mandal, A.~Pinczuk, L.~Pfeiffer, and
  K.~West, Higher-energy composite fermion levels in the fractional quantum
  {Hall} effect, \emph{Phys. Rev. Lett.} {\bf 106}, \penalty0 096803 (Mar,
  2011).
\newblock \doi{10.1103/PhysRevLett.106.096803}.
\newblock URL \url{http://link.aps.org/doi/10.1103/PhysRevLett.106.096803}.

\bibitem{Wurstbauer11}
U.~Wurstbauer, D.~Majumder, S.~S. Mandal, I.~Dujovne, T.~D. Rhone, B.~S.
  Dennis, A.~F. Rigosi, J.~K. Jain, A.~Pinczuk, K.~W. West, and L.~N. Pfeiffer,
  Observation of nonconventional spin waves in composite-fermion ferromagnets,
  \emph{Phys. Rev. Lett.} {\bf 107}, \penalty0 066804 (Aug, 2011).
\newblock \doi{10.1103/PhysRevLett.107.066804}.
\newblock URL \url{http://link.aps.org/doi/10.1103/PhysRevLett.107.066804}.

\bibitem{Wu93}
X.~G. Wu, G.~Dev, and J.~K. Jain, Mixed-spin incompressible states in the
  fractional quantum {Hall} effect, \emph{Phys. Rev. Lett.} {\bf 71}, \penalty0
  153--156 (Jul, 1993).
\newblock \doi{10.1103/PhysRevLett.71.153}.
\newblock URL \url{http://link.aps.org/doi/10.1103/PhysRevLett.71.153}.

\bibitem{Park98}
K.~Park and J.~K. Jain, Phase diagram of the spin polarization of composite
  fermions and a new effective mass, \emph{Phys. Rev. Lett.} {\bf 80},
  \penalty0 4237--4240 (May, 1998).
\newblock \doi{10.1103/PhysRevLett.80.4237}.
\newblock URL \url{http://link.aps.org/doi/10.1103/PhysRevLett.80.4237}.

\bibitem{Du95}
R.~R. Du, A.~S. Yeh, H.~L. Stormer, D.~C. Tsui, L.~N. Pfeiffer, and K.~W. West,
  Fractional quantum {Hall} effect around $\nu=3/2$: Composite fermions with a
  spin, \emph{Phys. Rev. Lett.} {\bf 75}, \penalty0 3926--3929 (Nov, 1995).
\newblock \doi{10.1103/PhysRevLett.75.3926}.
\newblock URL \url{http://link.aps.org/doi/10.1103/PhysRevLett.75.3926}.

\bibitem{Du97}
R.~R. Du, A.~S. Yeh, H.~L. Stormer, D.~C. Tsui, L.~N. Pfeiffer, and K.~W. West,
  g factor of composite fermions around $\nu=3/2$ from angular-dependent
  activation-energy measurements, \emph{Phys. Rev. B}. {\bf 55}, \penalty0
  R7351--R7354 (Mar, 1997).
\newblock \doi{10.1103/PhysRevB.55.R7351}.
\newblock URL \url{http://link.aps.org/doi/10.1103/PhysRevB.55.R7351}.

\bibitem{Kukushkin99}
I.~V. Kukushkin, K.~v.~Klitzing, and K.~Eberl, Spin polarization of composite
  fermions: Measurements of the {Fermi} energy, \emph{Phys. Rev. Lett.} {\bf
  82}, \penalty0 3665--3668 (May, 1999).
\newblock \doi{10.1103/PhysRevLett.82.3665}.
\newblock URL \url{http://link.aps.org/doi/10.1103/PhysRevLett.82.3665}.

\bibitem{Bishop07}
N.~C. Bishop, M.~Padmanabhan, K.~Vakili, Y.~P. Shkolnikov, E.~P. De~Poortere,
  and M.~Shayegan, Valley polarization and susceptibility of composite fermions
  around a filling factor $\nu=3/2$, \emph{Phys. Rev. Lett.} {\bf 98},
  \penalty0 266404 (Jun, 2007).
\newblock \doi{10.1103/PhysRevLett.98.266404}.
\newblock URL \url{http://link.aps.org/doi/10.1103/PhysRevLett.98.266404}.

\bibitem{Padmanabhan09}
M.~Padmanabhan, T.~Gokmen, and M.~Shayegan, Density dependence of valley
  polarization energy for composite fermions, \emph{Phys. Rev. B}. {\bf 80},
  \penalty0 035423 (Jul, 2009).
\newblock \doi{10.1103/PhysRevB.80.035423}.
\newblock URL \url{http://link.aps.org/doi/10.1103/PhysRevB.80.035423}.

\bibitem{Feldman13}
B.~E. Feldman, A.~J. Levin, B.~Krauss, D.~A. Abanin, B.~I. Halperin, J.~H.
  Smet, and A.~Yacoby, Fractional quantum {Hall} phase transitions and
  four-flux states in graphene, \emph{Phys. Rev. Lett.} {\bf 111}, \penalty0
  076802 (Aug, 2013).
\newblock \doi{10.1103/PhysRevLett.111.076802}.
\newblock URL \url{http://link.aps.org/doi/10.1103/PhysRevLett.111.076802}.

\bibitem{Willett93}
R.~L. Willett, R.~R. Ruel, K.~W. West, and L.~N. Pfeiffer, Experimental
  demonstration of a {Fermi} surface at one-half filling of the lowest {Landau}
  level, \emph{Phys. Rev. Lett.} {\bf 71}, \penalty0 3846--3849 (Dec, 1993).
\newblock \doi{10.1103/PhysRevLett.71.3846}.
\newblock URL \url{http://link.aps.org/doi/10.1103/PhysRevLett.71.3846}.

\bibitem{Goldman94}
V.~J. Goldman, B.~Su, and J.~K. Jain, Detection of composite fermions by
  magnetic focusing, \emph{Phys. Rev. Lett.} {\bf 72}, \penalty0 2065--2068
  (Mar, 1994).
\newblock \doi{10.1103/PhysRevLett.72.2065}.
\newblock URL \url{http://link.aps.org/doi/10.1103/PhysRevLett.72.2065}.

\bibitem{Smet96}
J.~H. Smet, D.~Weiss, R.~H. Blick, G.~L\"utjering, K.~von Klitzing,
  R.~Fleischmann, R.~Ketzmerick, T.~Geisel, and G.~Weimann, Magnetic focusing
  of composite fermions through arrays of cavities, \emph{Phys. Rev. Lett.}
  {\bf 77}, \penalty0 2272--2275 (Sep, 1996).
\newblock \doi{10.1103/PhysRevLett.77.2272}.
\newblock URL \url{http://link.aps.org/doi/10.1103/PhysRevLett.77.2272}.

\bibitem{Smet98}
J.~H. Smet, \emph{Ballistic transport of composite fermions in semiconductor
  nanostructures}, In \emph{Composite Fermions}, chapter~7, pp. 443--491.
\newblock World Scientific Pub Co Inc,  (1998).
\newblock \doi{10.1142/9789812815989\textunderscore0007}.
\newblock URL
  \url{http://www.worldscientific.com/doi/abs/10.1142/9789812815989_0007}.

\bibitem{Willett99}
R.~L. Willett, K.~W. West, and L.~N. Pfeiffer, Geometric resonance of composite
  fermion cyclotron orbits with a fictitious magnetic field modulation,
  \emph{Phys. Rev. Lett.} {\bf 83}, \penalty0 2624--2627 (Sep, 1999).
\newblock \doi{10.1103/PhysRevLett.83.2624}.
\newblock URL \url{http://link.aps.org/doi/10.1103/PhysRevLett.83.2624}.

\bibitem{Smet99}
J.~H. Smet, S.~Jobst, K.~von Klitzing, D.~Weiss, W.~Wegscheider, and
  V.~Umansky, Commensurate composite fermions in weak periodic electrostatic
  potentials: Direct evidence of a periodic effective magnetic field,
  \emph{Phys. Rev. Lett.} {\bf 83}, \penalty0 2620--2623 (Sep, 1999).
\newblock \doi{10.1103/PhysRevLett.83.2620}.
\newblock URL \url{http://link.aps.org/doi/10.1103/PhysRevLett.83.2620}.

\bibitem{Kamburov12}
D.~Kamburov, M.~Shayegan, L.~N. Pfeiffer, K.~W. West, and K.~W. Baldwin,
  Commensurability oscillations of hole-flux composite fermions, \emph{Phys.
  Rev. Lett.} {\bf 109}, \penalty0 236401 (Dec, 2012).
\newblock \doi{10.1103/PhysRevLett.109.236401}.
\newblock URL \url{http://link.aps.org/doi/10.1103/PhysRevLett.109.236401}.

\bibitem{Gokmen10}
T.~Gokmen, M.~Padmanabhan, and M.~Shayegan, Transference of transport
  anisotropy to composite fermions, \emph{Nature Physics}. {\bf 6}, \penalty0
  621--624,  (2010).

\bibitem{Kamburov13}
D.~Kamburov, Y.~Liu, M.~Shayegan, L.~N. Pfeiffer, K.~W. West, and K.~W.
  Baldwin, Composite fermions with tunable {Fermi} contour anisotropy,
  \emph{Phys. Rev. Lett.} {\bf 110}, \penalty0 206801 (May, 2013).
\newblock \doi{10.1103/PhysRevLett.110.206801}.
\newblock URL \url{http://link.aps.org/doi/10.1103/PhysRevLett.110.206801}.

\bibitem{Melinte00}
S.~Melinte, N.~Freytag, M.~Horvatic, C.~Berthier, L.~P. L\'evy, V.~Bayot, and
  M.~Shayegan, {N}{M}{R} determination of 2d electron spin polarization at
  $\nu=1/2$, \emph{Phys. Rev. Lett.} {\bf 84}, \penalty0 354--357 (Jan, 2000).
\newblock \doi{10.1103/PhysRevLett.84.354}.
\newblock URL \url{http://link.aps.org/doi/10.1103/PhysRevLett.84.354}.

\bibitem{Du94}
R.~Du, H.~Stormer, D.~Tsui, L.~Pfeiffer, and K.~West, Shubnikov-dehaas
  oscillations around $\nu=1/2$ {Landau} level filling factor, \emph{Solid
  State Communications}. {\bf 90}\penalty0 (2), \penalty0 71 -- 75,  (1994).
\newblock ISSN 0038-1098.
\newblock \doi{http://dx.doi.org/10.1016/0038-1098(94)90934-2}.
\newblock URL
  \url{http://www.sciencedirect.com/science/article/pii/0038109894909342}.

\bibitem{Leadley94}
D.~R. Leadley, R.~J. Nicholas, C.~T. Foxon, and J.~J. Harris, Measurements of
  the effective mass and scattering times of composite fermions from
  magnetotransport analysis, \emph{Phys. Rev. Lett.} {\bf 72}, \penalty0
  1906--1909 (Mar, 1994).
\newblock \doi{10.1103/PhysRevLett.72.1906}.
\newblock URL \url{http://link.aps.org/doi/10.1103/PhysRevLett.72.1906}.

\bibitem{Kukushkin02}
I.~V. Kukushkin, J.~H. Smet, K.~von Klitzing, and W.~Wegscheider, Cyclotron
  resonance of composite fermions, \emph{Nature}. {\bf 415}, \penalty0
  409--412,  (2002).

\bibitem{Kukushkin07}
I.~V. Kukushkin, J.~H. Smet, D.~Schuh, W.~Wegscheider, and K.~von Klitzing,
  Dispersion of the composite-fermion cyclotron-resonance mode, \emph{Phys.
  Rev. Lett.} {\bf 98}, \penalty0 066403 (Feb, 2007).
\newblock \doi{10.1103/PhysRevLett.98.066403}.
\newblock URL \url{http://link.aps.org/doi/10.1103/PhysRevLett.98.066403}.

\bibitem{Dev92}
G.~Dev and J.~K. Jain, Band structure of the fractional quantum {Hall} effect,
  \emph{Phys. Rev. Lett.} {\bf 69}, \penalty0 2843--2846 (Nov, 1992).
\newblock \doi{10.1103/PhysRevLett.69.2843}.
\newblock URL \url{http://link.aps.org/doi/10.1103/PhysRevLett.69.2843}.

\bibitem{Balram13}
A.~C. Balram, A.~W\'ojs, and J.~K. Jain, State counting for excited bands of
  the fractional quantum {Hall} effect: Exclusion rules for bound excitons,
  \emph{Phys. Rev. B}. {\bf 88}, \penalty0 205312 (Nov, 2013).
\newblock \doi{10.1103/PhysRevB.88.205312}.
\newblock URL \url{http://link.aps.org/doi/10.1103/PhysRevB.88.205312}.

\bibitem{Jain15}
J.~K. Jain, Composite fermion theory of exotic fractional quantum {Hall}
  effect, \emph{Annu. Rev. Condens. Matter Phys.} {\bf 6}, \penalty0 39--62,
  (2015).
\newblock \doi{10.1146/annurev-conmatphys-031214-014606}.

\bibitem{Mukherjee12}
S.~Mukherjee, S.~S. Mandal, A.~W\'ojs, and J.~K. Jain, Possible anti-{Pfaffian}
  pairing of composite fermions at $\nu=3/8$, \emph{Phys. Rev. Lett.} {\bf
  109}, \penalty0 256801 (Dec, 2012).
\newblock \doi{10.1103/PhysRevLett.109.256801}.
\newblock URL \url{http://link.aps.org/doi/10.1103/PhysRevLett.109.256801}.

\bibitem{Rezayi94}
E.~Rezayi and N.~Read, Fermi-liquid-like state in a half-filled {Landau} level,
  \emph{Phys. Rev. Lett.} {\bf 72}, \penalty0 900--903 (Feb, 1994).
\newblock \doi{10.1103/PhysRevLett.72.900}.
\newblock URL \url{http://link.aps.org/doi/10.1103/PhysRevLett.72.900}.

\bibitem{Rezayi00}
E.~H. Rezayi and F.~D.~M. Haldane, Incompressible paired {Hall} state, stripe
  order, and the composite fermion liquid phase in half-filled {Landau} levels,
  \emph{Phys. Rev. Lett.} {\bf 84}, \penalty0 4685--4688 (May, 2000).
\newblock \doi{10.1103/PhysRevLett.84.4685}.
\newblock URL \url{http://link.aps.org/doi/10.1103/PhysRevLett.84.4685}.

\bibitem{Shao15}
J.~Shao, E.-A. Kim, F.~D.~M. Haldane, and E.~H. Rezayi, Entanglement entropy of
  the $\ensuremath{\nu}=1/2$ composite fermion non-{Fermi} liquid state,
  \emph{Phys. Rev. Lett.} {\bf 114}, \penalty0 206402 (May, 2015).
\newblock \doi{10.1103/PhysRevLett.114.206402}.
\newblock URL \url{http://link.aps.org/doi/10.1103/PhysRevLett.114.206402}.

\bibitem{Wang19}
J.~Wang, S.~D. Geraedts, E.~H. Rezayi, and F.~D.~M. Haldane, Lattice {Monte}
  {Carlo} for quantum {Hall} states on a torus, \emph{Phys. Rev. B}. {\bf 99},
  \penalty0 125123 (Mar, 2019).
\newblock \doi{10.1103/PhysRevB.99.125123}.
\newblock URL \url{https://link.aps.org/doi/10.1103/PhysRevB.99.125123}.

\bibitem{Geraedts18}
S.~D. Geraedts, J.~Wang, E.~H. Rezayi, and F.~D.~M. Haldane, Berry phase and
  model wave function in the half-filled {Landau} level, \emph{Phys. Rev.
  Lett.} {\bf 121}, \penalty0 147202 (Oct, 2018).
\newblock \doi{10.1103/PhysRevLett.121.147202}.
\newblock URL \url{https://link.aps.org/doi/10.1103/PhysRevLett.121.147202}.

\bibitem{Fremling18}
M.~Fremling, N.~Moran, J.~K. Slingerland, and S.~H. Simon, Trial wave functions
  for a composite {Fermi} liquid on a torus, \emph{Phys. Rev. B}. {\bf 97},
  \penalty0 035149 (Jan, 2018).
\newblock \doi{10.1103/PhysRevB.97.035149}.
\newblock URL \url{https://link.aps.org/doi/10.1103/PhysRevB.97.035149}.

\bibitem{Pu18}
S.~Pu, M.~Fremling, and J.~K. Jain, Berry phase of the composite-fermion
  {Fermi} sea: Effect of {Landau}-level mixing, \emph{Phys. Rev. B}. {\bf 98},
  \penalty0 075304 (Aug, 2018).
\newblock \doi{10.1103/PhysRevB.98.075304}.
\newblock URL \url{https://link.aps.org/doi/10.1103/PhysRevB.98.075304}.

\bibitem{Wu95}
X.~G. Wu and J.~K. Jain, Excitation spectrum and collective modes of composite
  fermions, \emph{Phys. Rev. B}. {\bf 51}, \penalty0 1752--1761 (Jan, 1995).
\newblock \doi{10.1103/PhysRevB.51.1752}.
\newblock URL \url{http://link.aps.org/doi/10.1103/PhysRevB.51.1752}.

\bibitem{Meyer16}
M.~L. Meyer, O.~Liab{\o}tr{\o}, and S.~Viefers, Linear dependencies between
  composite fermion states, \emph{Journal of Physics A: Mathematical and
  Theoretical}. {\bf 49}\penalty0 (39), \penalty0 395201 (sep, 2016).
\newblock \doi{10.1088/1751-8113/49/39/395201}.
\newblock URL \url{https://doi.org/10.1088%2F1751-8113%2F49%2F39%2F395201}.

\bibitem{Son15}
D.~T. Son, Is the composite fermion a {Dirac} particle?, \emph{Phys. Rev. X}.
  {\bf 5}, \penalty0 031027 (Sep, 2015).
\newblock \doi{10.1103/PhysRevX.5.031027}.
\newblock URL \url{http://link.aps.org/doi/10.1103/PhysRevX.5.031027}.

\bibitem{Balram15b}
A.~C. Balram, C.~T\ifmmode~\mbox{\H{o}}\else \H{o}\fi{}ke, and J.~K. Jain,
  Luttinger theorem for the strongly correlated {Fermi} liquid of composite
  fermions, \emph{Phys. Rev. Lett.} {\bf 115}, \penalty0 186805 (Oct, 2015).
\newblock \doi{10.1103/PhysRevLett.115.186805}.
\newblock URL \url{http://link.aps.org/doi/10.1103/PhysRevLett.115.186805}.

\bibitem{Davenport12}
S.~C. Davenport and S.~H. Simon, Spinful composite fermions in a negative
  effective field, \emph{Phys. Rev. B}. {\bf 85}, \penalty0 245303 (Jun, 2012).
\newblock \doi{10.1103/PhysRevB.85.245303}.
\newblock URL \url{http://link.aps.org/doi/10.1103/PhysRevB.85.245303}.

\bibitem{Balram16b}
A.~C. Balram and J.~K. Jain, Nature of composite fermions and the role of
  particle-hole symmetry: A microscopic account, \emph{Phys. Rev. B}. {\bf 93},
  \penalty0 235152 (Jun, 2016).
\newblock \doi{10.1103/PhysRevB.93.235152}.
\newblock URL \url{http://link.aps.org/doi/10.1103/PhysRevB.93.235152}.

\bibitem{Lee02}
S.-Y. Lee, V.~W. Scarola, and J.~K. Jain, Structures for interacting composite
  fermions: Stripes, bubbles, and fractional quantum {Hall} effect, \emph{Phys.
  Rev. B}. {\bf 66}, \penalty0 085336 (Aug, 2002).
\newblock \doi{10.1103/PhysRevB.66.085336}.
\newblock URL \url{http://link.aps.org/doi/10.1103/PhysRevB.66.085336}.

\bibitem{Samkharadze15b}
N.~Samkharadze, I.~Arnold, L.~N. Pfeiffer, K.~W. West, and G.~A. Cs\'athy,
  Observation of incompressibility at $\nu=4/11$ and $\nu=5/13$, \emph{Phys.
  Rev. B}. {\bf 91}, \penalty0 081109 (Feb, 2015).
\newblock \doi{10.1103/PhysRevB.91.081109}.
\newblock URL \url{http://link.aps.org/doi/10.1103/PhysRevB.91.081109}.

\bibitem{Pan15}
W.~Pan, K.~W. Baldwin, K.~W. West, L.~N. Pfeiffer, and D.~C. Tsui, Fractional
  quantum {Hall} effect at {Landau} level filling $\ensuremath{\nu}=4/11$,
  \emph{Phys. Rev. B}. {\bf 91}, \penalty0 041301,  (2015).
\newblock \doi{10.1103/PhysRevB.91.041301}.
\newblock URL \url{http://link.aps.org/doi/10.1103/PhysRevB.91.041301}.

\bibitem{Moore91}
G.~Moore and N.~Read, Nonabelions in the fractional quantum {Hall} effect,
  \emph{Nucl. Phys. B}. {\bf 360}, \penalty0 362 -- 396,  (1991).
\newblock ISSN 0550-3213.
\newblock \doi{10.1016/0550-3213(91)90407-O}.
\newblock URL
  \url{http://www.sciencedirect.com/science/article/pii/055032139190407O}.

\bibitem{Read00}
N.~Read and D.~Green, Paired states of fermions in two dimensions with breaking
  of parity and time-reversal symmetries and the fractional quantum {Hall}
  effect, \emph{Phys. Rev. B}. {\bf 61}, \penalty0 10267--10297 (Apr, 2000).
\newblock \doi{10.1103/PhysRevB.61.10267}.
\newblock URL \url{http://link.aps.org/doi/10.1103/PhysRevB.61.10267}.

\bibitem{Scarola00}
V.~W. Scarola, K.~Park, and J.~K. Jain, Rotons of composite fermions:
  Comparison between theory and experiment, \emph{Phys. Rev. B}. {\bf 61},
  \penalty0 13064--13072 (May, 2000).
\newblock \doi{10.1103/PhysRevB.61.13064}.
\newblock URL \url{http://link.aps.org/doi/10.1103/PhysRevB.61.13064}.

\bibitem{Zee10}
A.~Zee, \emph{Quantum Field Theory in a Nutshell}. (Cambridge University Press,
  New York, US, 2010).

\bibitem{Tong16}
D.~Tong, Lectures on the quantum {Hall} effect, \emph{arXiv e-prints}. art.
  arXiv:1606.06687 (Jun, 2016).

\bibitem{Su86}
W.~P. Su, Statistics of the fractionally charged excitations in the quantum
  {Hall} effect, \emph{Phys. Rev. B}. {\bf 34}, \penalty0 1031--1033 (Jul,
  1986).
\newblock \doi{10.1103/PhysRevB.34.1031}.
\newblock URL \url{http://link.aps.org/doi/10.1103/PhysRevB.34.1031}.

\bibitem{Jeon04}
G.~S. Jeon, K.~L. Graham, and J.~K. Jain, Berry phases for composite fermions:
  Effective magnetic field and fractional statistics, \emph{Phys. Rev. B}. {\bf
  70}, \penalty0 125316 (Sep, 2004).
\newblock \doi{10.1103/PhysRevB.70.125316}.
\newblock URL \url{http://link.aps.org/doi/10.1103/PhysRevB.70.125316}.

\bibitem{Zhang16}
Y.~Zhang, A.~W\'ojs, and J.~K. Jain, Landau-level mixing and particle-hole
  symmetry breaking for spin transitions in the fractional quantum {Hall}
  effect, \emph{Phys. Rev. Lett.} {\bf 117}, \penalty0 116803 (Sep, 2016).
\newblock \doi{10.1103/PhysRevLett.117.116803}.
\newblock URL \url{http://link.aps.org/doi/10.1103/PhysRevLett.117.116803}.

\bibitem{Zhao18}
J.~Zhao, Y.~Zhang, and J.~K. Jain, Crystallization in the fractional quantum
  {Hall} regime induced by {Landau}-level mixing, \emph{Phys. Rev. Lett.} {\bf
  121}, \penalty0 116802 (Sep, 2018).
\newblock \doi{10.1103/PhysRevLett.121.116802}.
\newblock URL \url{https://link.aps.org/doi/10.1103/PhysRevLett.121.116802}.

\bibitem{Ortalano97}
M.~W. Ortalano, S.~He, and S.~Das~Sarma, Realistic calculations of correlated
  incompressible electronic states in
  {Ga}{As}-${\mathrm{al}}_{\mathrm{x}}$${\mathrm{ga}}_{1\mathrm{-}\mathrm{x}}$as
  heterostructures and quantum wells, \emph{Phys. Rev. B}. {\bf 55}, \penalty0
  7702--7714 (Mar, 1997).
\newblock \doi{10.1103/PhysRevB.55.7702}.
\newblock URL \url{http://link.aps.org/doi/10.1103/PhysRevB.55.7702}.

\bibitem{Zhang86}
F.~C. Zhang and S.~Das~Sarma, Excitation gap in the fractional quantum {Hall}
  effect: Finite layer thickness corrections, \emph{Phys. Rev. B}. {\bf 33},
  \penalty0 2903--2905 (Feb, 1986).
\newblock \doi{10.1103/PhysRevB.33.2903}.
\newblock URL \url{https://link.aps.org/doi/10.1103/PhysRevB.33.2903}.

\bibitem{MacDonald84}
A.~H. MacDonald, Influence of {Landau}-level mixing on the charge-density-wave
  state of a two-dimensional electron gas in a strong magnetic field,
  \emph{Phys. Rev. B}. {\bf 30}, \penalty0 4392--4398 (Oct, 1984).
\newblock \doi{10.1103/PhysRevB.30.4392}.
\newblock URL \url{http://link.aps.org/doi/10.1103/PhysRevB.30.4392}.

\bibitem{Melik-Alaverdian95}
V.~Melik-Alaverdian and N.~E. Bonesteel, Composite fermions and {Landau}-level
  mixing in the fractional quantum {Hall} effect, \emph{Phys. Rev. B}. {\bf
  52}, \penalty0 R17032--R17035 (Dec, 1995).
\newblock \doi{10.1103/PhysRevB.52.R17032}.
\newblock URL \url{http://link.aps.org/doi/10.1103/PhysRevB.52.R17032}.

\bibitem{Murthy02}
G.~Murthy and R.~Shankar, Hamiltonian theory of the fractional quantum {Hall}
  effect: Effect of {Landau} level mixing, \emph{Phys. Rev. B}. {\bf 65},
  \penalty0 245309 (Jun, 2002).
\newblock \doi{10.1103/PhysRevB.65.245309}.
\newblock URL \url{http://link.aps.org/doi/10.1103/PhysRevB.65.245309}.

\bibitem{Bishara09}
W.~Bishara and C.~Nayak, Effect of {Landau} level mixing on the effective
  interaction between electrons in the fractional quantum {Hall} regime,
  \emph{Phys. Rev. B}. {\bf 80}, \penalty0 121302 (Sep, 2009).
\newblock \doi{10.1103/PhysRevB.80.121302}.
\newblock URL \url{http://link.aps.org/doi/10.1103/PhysRevB.80.121302}.

\bibitem{Wojs10}
A.~W\'ojs, C.~T\ifmmode~\mbox{\H{o}}\else \H{o}\fi{}ke, and J.~K. Jain,
  Landau-level mixing and the emergence of {Pfaffian} excitations for the
  $\nu=5/2$ fractional quantum {Hall} effect, \emph{Phys. Rev. Lett.} {\bf
  105}, \penalty0 096802 (Aug, 2010).
\newblock \doi{10.1103/PhysRevLett.105.096802}.
\newblock URL \url{http://link.aps.org/doi/10.1103/PhysRevLett.105.096802}.

\bibitem{Sodemann13}
I.~Sodemann and A.~H. MacDonald, {Landau} level mixing and the fractional
  quantum {Hall} effect, \emph{Phys. Rev. B}. {\bf 87}, \penalty0 245425 (Jun,
  2013).
\newblock \doi{10.1103/PhysRevB.87.245425}.
\newblock URL \url{http://link.aps.org/doi/10.1103/PhysRevB.87.245425}.

\bibitem{Simon13}
S.~H. Simon and E.~H. Rezayi, {Landau} level mixing in the perturbative limit,
  \emph{Phys. Rev. B}. {\bf 87}, \penalty0 155426 (Apr, 2013).
\newblock \doi{10.1103/PhysRevB.87.155426}.
\newblock URL \url{http://link.aps.org/doi/10.1103/PhysRevB.87.155426}.

\bibitem{Peterson13}
M.~R. Peterson and C.~Nayak, More realistic {Hamiltonians} for the fractional
  quantum {Hall} regime in {Ga}{As} and graphene, \emph{Phys. Rev. B}. {\bf
  87}, \penalty0 245129 (Jun, 2013).
\newblock \doi{10.1103/PhysRevB.87.245129}.
\newblock URL \url{http://link.aps.org/doi/10.1103/PhysRevB.87.245129}.

\bibitem{Peterson14}
M.~R. Peterson and C.~Nayak, Effects of {Landau} level mixing on the fractional
  quantum {Hall} effect in monolayer graphene, \emph{Phys. Rev. Lett.} {\bf
  113}, \penalty0 086401 (Aug, 2014).
\newblock \doi{10.1103/PhysRevLett.113.086401}.
\newblock URL \url{http://link.aps.org/doi/10.1103/PhysRevLett.113.086401}.

\bibitem{Ortiz93}
G.~Ortiz, D.~M. Ceperley, and R.~M. Martin, New stochastic method for systems
  with broken time-reversal symmetry: 2d fermions in a magnetic field,
  \emph{Phys. Rev. Lett.} {\bf 71}, \penalty0 2777--2780 (Oct, 1993).
\newblock \doi{10.1103/PhysRevLett.71.2777}.
\newblock URL \url{http://link.aps.org/doi/10.1103/PhysRevLett.71.2777}.

\bibitem{Melik-Alaverdian97}
V.~Melik-Alaverdian, N.~E. Bonesteel, and G.~Ortiz, Quantum {Hall} fluids on
  the {Haldane} sphere: A diffusion {Monte} {Carlo} study, \emph{Phys. Rev.
  Lett.} {\bf 79}, \penalty0 5286--5289 (Dec, 1997).
\newblock \doi{10.1103/PhysRevLett.79.5286}.
\newblock URL \url{http://link.aps.org/doi/10.1103/PhysRevLett.79.5286}.

\bibitem{Melik-Alaverdian01}
V.~Melik-Alaverdian, G.~Ortiz, and N.~Bonesteel, Quantum projector method on
  curved manifolds, \emph{Journal of Statistical Physics}. {\bf 104}\penalty0
  (1-2), \penalty0 449--470,  (2001).
\newblock ISSN 0022-4715.
\newblock \doi{10.1023/A:1010326231389}.
\newblock URL \url{http://dx.doi.org/10.1023/A%3A1010326231389}.

\bibitem{Reynolds82}
P.~J. Reynolds, D.~M. Ceperley, B.~J. Alder, and W.~A. Lester~Jr., Fixed‐node
  quantum {Monte} {Carlo} for molecules, \emph{J. Chem. Phys.} {\bf
  77}\penalty0 (11), \penalty0 5593--5603,  (1982).
\newblock \doi{http://dx.doi.org/10.1063/1.443766}.
\newblock URL
  \url{http://scitation.aip.org/content/aip/journal/jcp/77/11/10.1063/1.443766}.

\bibitem{Foulkes01}
W.~M.~C. Foulkes, L.~Mitas, R.~J. Needs, and G.~Rajagopal, Quantum {Monte}
  {Carlo} simulations of solids, \emph{Rev. Mod. Phys.} {\bf 73}, \penalty0
  33--83 (Jan, 2001).
\newblock \doi{10.1103/RevModPhys.73.33}.
\newblock URL \url{http://link.aps.org/doi/10.1103/RevModPhys.73.33}.

\bibitem{Melton17}
C.~A. Melton and L.~Mitas, Quantum {Monte} {Carlo} with variable spins:
  Fixed-phase and fixed-node approximations, \emph{Phys. Rev. E}. {\bf 96},
  \penalty0 043305 (Oct, 2017).
\newblock \doi{10.1103/PhysRevE.96.043305}.
\newblock URL \url{https://link.aps.org/doi/10.1103/PhysRevE.96.043305}.

\bibitem{Guclu05}
A.~D. G\"u\c{c}l\"u and C.~J. Umrigar, Maximum-density droplet to lower-density
  droplet transition in quantum dots, \emph{Phys. Rev. B}. {\bf 72}, \penalty0
  045309 (Jul, 2005).
\newblock \doi{10.1103/PhysRevB.72.045309}.
\newblock URL \url{http://link.aps.org/doi/10.1103/PhysRevB.72.045309}.

\bibitem{Eisenstein89}
J.~P. Eisenstein, H.~L. Stormer, L.~Pfeiffer, and K.~W. West, Evidence for a
  phase transition in the fractional quantum {Hall} effect, \emph{Phys. Rev.
  Lett.} {\bf 62}, \penalty0 1540--1543 (Mar, 1989).
\newblock \doi{10.1103/PhysRevLett.62.1540}.
\newblock URL \url{http://link.aps.org/doi/10.1103/PhysRevLett.62.1540}.

\bibitem{Eisenstein90}
J.~P. Eisenstein, H.~L. Stormer, L.~N. Pfeiffer, and K.~W. West, Evidence for a
  spin transition in the $\nu=2/3$ fractional quantum {Hall} effect,
  \emph{Phys. Rev. B}. {\bf 41}, \penalty0 7910--7913 (Apr, 1990).
\newblock \doi{10.1103/PhysRevB.41.7910}.
\newblock URL \url{http://link.aps.org/doi/10.1103/PhysRevB.41.7910}.

\bibitem{Engel92}
L.~W. Engel, S.~W. Hwang, T.~Sajoto, D.~C. Tsui, and M.~Shayegan, Fractional
  quantum {Hall} effect at $\nu$=2/3 and 3/5 in tilted magnetic fields,
  \emph{Phys. Rev. B}. {\bf 45}, \penalty0 3418--3425 (Feb, 1992).
\newblock \doi{10.1103/PhysRevB.45.3418}.
\newblock URL \url{http://link.aps.org/doi/10.1103/PhysRevB.45.3418}.

\bibitem{Kang97}
W.~Kang, J.~B. Young, S.~T. Hannahs, E.~Palm, K.~L. Campman, and A.~C. Gossard,
  Evidence for a spin transition in the $\nu=2/5$ fractional quantum {Hall}
  effect, \emph{Phys. Rev. B}. {\bf 56}, \penalty0 R12776--R12779 (Nov, 1997).
\newblock \doi{10.1103/PhysRevB.56.R12776}.
\newblock URL \url{http://link.aps.org/doi/10.1103/PhysRevB.56.R12776}.

\bibitem{Yeh99}
A.~S. Yeh, H.~L. Stormer, D.~C. Tsui, L.~N. Pfeiffer, K.~W. Baldwin, and K.~W.
  West, Effective mass and $\mathit{g}$ factor of four-flux-quanta composite
  fermions, \emph{Phys. Rev. Lett.} {\bf 82}, \penalty0 592--595 (Jan, 1999).
\newblock \doi{10.1103/PhysRevLett.82.592}.
\newblock URL \url{http://link.aps.org/doi/10.1103/PhysRevLett.82.592}.

\bibitem{Freytag01}
N.~Freytag, Y.~Tokunaga, M.~Horvati\'{c}, C.~Berthier, M.~Shayegan, and L.~P.
  L\'evy, New phase transition between partially and fully polarized quantum
  {Hall} states with charge and spin gaps at $\nu=\frac{2}{3}$, \emph{Phys.
  Rev. Lett.} {\bf 87}, \penalty0 136801 (Sep, 2001).
\newblock \doi{10.1103/PhysRevLett.87.136801}.
\newblock URL \url{http://link.aps.org/doi/10.1103/PhysRevLett.87.136801}.

\bibitem{Tiemann12}
L.~Tiemann, G.~Gamez, N.~Kumada, and K.~Muraki, Unraveling the spin
  polarization of the $\nu= 5/2$ fractional quantum {Hall} state,
  \emph{Science}. {\bf 335}\penalty0 (6070), \penalty0 828--831,  (2012).
\newblock \doi{10.1126/science.1216697}.
\newblock URL \url{http://www.sciencemag.org/content/335/6070/828.abstract}.

\bibitem{Liu14}
Y.~Liu, S.~Hasdemir, A.~W\'ojs, J.~K. Jain, L.~N. Pfeiffer, K.~W. West, K.~W.
  Baldwin, and M.~Shayegan, Spin polarization of composite fermions and
  particle-hole symmetry breaking, \emph{Phys. Rev. B}. {\bf 90}, \penalty0
  085301 (Aug, 2014).
\newblock \doi{10.1103/PhysRevB.90.085301}.
\newblock URL \url{http://link.aps.org/doi/10.1103/PhysRevB.90.085301}.

\bibitem{Park99}
K.~Park and J.~K. Jain, Spontaneous magnetization of composite fermions,
  \emph{Phys. Rev. Lett.} {\bf 83}, \penalty0 5543--5546 (Dec, 1999).
\newblock \doi{10.1103/PhysRevLett.83.5543}.
\newblock URL \url{http://link.aps.org/doi/10.1103/PhysRevLett.83.5543}.

\bibitem{Balram15a}
A.~C. Balram, C.~T\"oke, A.~W\'ojs, and J.~K. Jain, Fractional quantum {Hall}
  effect in graphene: Quantitative comparison between theory and experiment,
  \emph{Phys. Rev. B}. {\bf 92}, \penalty0 075410 (Aug, 2015).
\newblock \doi{10.1103/PhysRevB.92.075410}.
\newblock URL \url{http://link.aps.org/doi/10.1103/PhysRevB.92.075410}.

\bibitem{Leadley97}
D.~R. Leadley, R.~J. Nicholas, D.~K. Maude, A.~N. Utjuzh, J.~C. Portal, J.~J.
  Harris, and C.~T. Foxon, Fractional quantum {Hall} effect measurements at
  zero $\sim g$ factor, \emph{Phys. Rev. Lett.} {\bf 79}, \penalty0 4246--4249
  (Nov, 1997).
\newblock \doi{10.1103/PhysRevLett.79.4246}.
\newblock URL \url{http://link.aps.org/doi/10.1103/PhysRevLett.79.4246}.

\bibitem{Xu09}
X.~Du, I.~Skachko, F.~Duerr, A.~Luican, and E.~Y. Andrei, Fractional quantum
  {Hall} effect and insulating phase of {Dirac} electrons in graphene,
  \emph{Nature}. {\bf 462}, \penalty0 192--195 (Nov, 2009).

\bibitem{Bolotin09}
K.~Bolotin, F.~Ghahari, M.~D. Shulman, H.~Stormer, and P.~Kim, Observation of
  the fractional quantum {Hall} effect in graphene, \emph{Nature}. {\bf 462},
  \penalty0 196--199,  (2009).
\newblock \doi{10.1038/nature08582}.

\bibitem{Dean11}
C.~R. Dean, A.~F. Young, P.~Cadden-Zimansky, L.~Wang, H.~Ren, K.~Watanabe,
  T.~Taniguchi, P.~Kim, J.~Hone, and K.~L. Shepard, Multicomponent fractional
  quantum {Hall} effect in graphene, \emph{Nature Physics}. {\bf 7}, \penalty0
  693--696,  (2011).

\bibitem{Feldman12}
B.~E. Feldman, B.~Krauss, J.~H. Smet, and A.~Yacoby, Unconventional sequence of
  fractional quantum {Hall} states in suspended graphene, \emph{Science}. {\bf
  337}\penalty0 (6099), \penalty0 1196--1199,  (2012).
\newblock \doi{10.1126/science.1224784}.
\newblock URL \url{http://www.sciencemag.org/content/337/6099/1196.abstract}.

\bibitem{Amet15}
F.~Amet, A.~J. Bestwick, J.~R. Williams, L.~Balicas, K.~Watanabe, T.~Taniguchi,
  and D.~Goldhaber-Gordon, Composite fermions and broken symmetries in
  graphene, \emph{Nat. Commun.} {\bf 6}, \penalty0 5838 (Jan, 2015).
\newblock \doi{10.1038/ncomms6838}.
\newblock URL \url{http://dx.doi.org/10.1038/ncomms6838}.

\bibitem{Kott14}
T.~M. Kott, B.~Hu, S.~H. Brown, and B.~E. Kane, Valley-degenerate
  two-dimensional electrons in the lowest {Landau} level, \emph{Phys. Rev. B}.
  {\bf 89}, \penalty0 041107 (Jan, 2014).
\newblock \doi{10.1103/PhysRevB.89.041107}.
\newblock URL \url{http://link.aps.org/doi/10.1103/PhysRevB.89.041107}.

\bibitem{Balram15}
A.~C. Balram, C.~T\"oke, A.~W\'ojs, and J.~K. Jain, Phase diagram of fractional
  quantum {Hall} effect of composite fermions in multicomponent systems,
  \emph{Phys. Rev. B}. {\bf 91}, \penalty0 045109 (Jan, 2015).
\newblock \doi{10.1103/PhysRevB.91.045109}.
\newblock URL \url{http://link.aps.org/doi/10.1103/PhysRevB.91.045109}.

\bibitem{Shayegan07}
M.~Shayegan.
\newblock Case for the magnetic-field-induced two-dimensional {Wigner} crystal.
\newblock In \emph{Perspectives in Quantum Hall Effects}, p. 343–384.
  Wiley-VCH Verlag GmbH,  (2007).
\newblock ISBN 9783527617258.
\newblock \doi{10.1002/9783527617258.ch10}.
\newblock URL \url{http://dx.doi.org/10.1002/9783527617258.ch10}.

\bibitem{Fertig07}
H.~A. Fertig.
\newblock Properties of the electron solid.
\newblock In \emph{Perspectives in Quantum {Hall} Effects}, p. 71–108.
  Wiley-VCH Verlag GmbH,  (2007).
\newblock ISBN 9783527617258.
\newblock \doi{10.1002/9783527617258.ch10}.
\newblock URL \url{http://dx.doi.org/10.1002/9783527617258.ch10}.

\bibitem{Jiang90}
H.~W. Jiang, R.~L. Willett, H.~L. Stormer, D.~C. Tsui, L.~N. Pfeiffer, and
  K.~W. West, Quantum liquid versus electron solid around $\nu=1/5$
  {Landau}-level filling, \emph{Phys. Rev. Lett.} {\bf 65}, \penalty0 633--636
  (Jul, 1990).
\newblock \doi{10.1103/PhysRevLett.65.633}.

\bibitem{Goldman90}
V.~J. Goldman, M.~Santos, M.~Shayegan, and J.~E. Cunningham, Evidence for
  two-dimensional quantum {Wigner} crystal, \emph{Phys. Rev. Lett.} {\bf 65},
  \penalty0 2189--2192 (Oct, 1990).
\newblock \doi{10.1103/PhysRevLett.65.2189}.

\bibitem{Williams91}
F.~I.~B. Williams, P.~A. Wright, R.~G. Clark, E.~Y. Andrei, G.~Deville, D.~C.
  Glattli, O.~Probst, B.~Etienne, C.~Dorin, C.~T. Foxon, and J.~J. Harris,
  Conduction threshold and pinning frequency of magnetically induced {Wigner}
  solid, \emph{Phys. Rev. Lett.} {\bf 66}, \penalty0 3285--3288 (Jun, 1991).
\newblock \doi{10.1103/PhysRevLett.66.3285}.
\newblock URL \url{https://link.aps.org/doi/10.1103/PhysRevLett.66.3285}.

\bibitem{Paalanen92}
M.~A. Paalanen, R.~L. Willett, R.~R. Ruel, P.~B. Littlewood, K.~W. West, and
  L.~N. Pfeiffer, Electrical conductivity and {Wigner} crystallization,
  \emph{Phys. Rev. B}. {\bf 45}, \penalty0 13784--13787 (Jun, 1992).
\newblock \doi{10.1103/PhysRevB.45.13784}.
\newblock URL \url{http://link.aps.org/doi/10.1103/PhysRevB.45.13784}.

\bibitem{Santos92}
M.~B. Santos, Y.~W. Suen, M.~Shayegan, Y.~P. Li, L.~W. Engel, and D.~C. Tsui,
  Observation of a reentrant insulating phase near the $1/3$ fractional quantum
  {Hall} liquid in a two-dimensional hole system, \emph{Phys. Rev. Lett.} {\bf
  68}, \penalty0 1188--1191 (Feb, 1992).
\newblock \doi{10.1103/PhysRevLett.68.1188}.

\bibitem{Santos92b}
M.~B. Santos, J.~Jo, Y.~W. Suen, L.~W. Engel, and M.~Shayegan, Effect of
  {Landau}-level mixing on quantum-liquid and solid states of two-dimensional
  hole systems, \emph{Phys. Rev. B}. {\bf 46}, \penalty0 13639--13642 (Nov,
  1992).
\newblock \doi{10.1103/PhysRevB.46.13639}.
\newblock URL \url{https://link.aps.org/doi/10.1103/PhysRevB.46.13639}.

\bibitem{Manoharan94a}
H.~C. Manoharan and M.~Shayegan, {Wigner} crystal versus {Hall} insulator,
  \emph{Phys. Rev. B}. {\bf 50}, \penalty0 17662--17665 (Dec, 1994).
\newblock \doi{10.1103/PhysRevB.50.17662}.

\bibitem{Engel97}
L.~Engel, C.-C. Li, D.~Shahar, D.~Tsui, and M.~Shayegan, Microwave resonances
  in low-filling insulating phases of two-dimensional electron and hole
  systems, \emph{Physica E}. {\bf 1}\penalty0 (14), \penalty0 111 -- 115,
  (1997).
\newblock \doi{http://dx.doi.org/10.1016/S1386-9477(97)00025-8}.

\bibitem{Li00}
C.-C. Li, J.~Yoon, L.~W. Engel, D.~Shahar, D.~C. Tsui, and M.~Shayegan,
  Microwave resonance and weak pinning in two-dimensional hole systems at high
  magnetic fields, \emph{Phys. Rev. B}. {\bf 61}, \penalty0 10905--10909 (Apr,
  2000).
\newblock \doi{10.1103/PhysRevB.61.10905}.

\bibitem{Ye02}
P.~D. Ye, L.~W. Engel, D.~C. Tsui, R.~M. Lewis, L.~N. Pfeiffer, and K.~West,
  Correlation lengths of the {Wigner}-crystal order in a two-dimensional
  electron system at high magnetic fields, \emph{Phys. Rev. Lett.} {\bf 89},
  \penalty0 176802 (Oct, 2002).
\newblock \doi{10.1103/PhysRevLett.89.176802}.

\bibitem{Chen04}
Y.~P. Chen, R.~M. Lewis, L.~W. Engel, D.~C. Tsui, P.~D. Ye, Z.~H. Wang, L.~N.
  Pfeiffer, and K.~W. West, Evidence for two different solid phases of
  two-dimensional electrons in high magnetic fields, \emph{Phys. Rev. Lett.}
  {\bf 93}, \penalty0 206805 (Nov, 2004).
\newblock \doi{10.1103/PhysRevLett.93.206805}.

\bibitem{Csathy05}
G.~A. Cs\'{a}thy, H.~Noh, D.~C. Tsui, L.~N. Pfeiffer, and K.~W. West,
  Magnetic-field-induced insulating phases at large $r_s$, \emph{Phys. Rev.
  Lett.} {\bf 94}, \penalty0 226802 (Jun, 2005).
\newblock \doi{10.1103/PhysRevLett.94.226802}.

\bibitem{Pan05}
W.~Pan, G.~A. Cs\'athy, D.~C. Tsui, L.~N. Pfeiffer, and K.~W. West, Transition
  from a fractional quantum {Hall} liquid to an electron solid at {Landau}
  level filling $\ensuremath{\nu}=\frac{1}{3}$ in tilted magnetic fields,
  \emph{Phys. Rev. B}. {\bf 71}, \penalty0 035302 (Jan, 2005).
\newblock \doi{10.1103/PhysRevB.71.035302}.
\newblock URL \url{https://link.aps.org/doi/10.1103/PhysRevB.71.035302}.

\bibitem{Sambandamurthy06}
G.~Sambandamurthy, Z.~Wang, R.~Lewis, Y.~P. Chen, L.~Engel, D.~Tsui,
  L.~Pfeiffer, and K.~West, Pinning mode resonances of new phases of 2d
  electron systems in high magnetic fields, \emph{Solid State Commun.} {\bf
  140}\penalty0 (2), \penalty0 100 -- 106,  (2006).

\bibitem{Chen06}
Y.~P. Chen, G.~Sambandamurthy, Z.~H. Wang, R.~M. Lewis, L.~W. Engel, D.~C.
  Tsui, P.~D. Ye, L.~N. Pfeiffer, and K.~W. West, Melting of a 2d quantum
  electron solid in a high magnetic field, \emph{Nature Phys.} {\bf 2},
  \penalty0 452--455 (July, 2006).
\newblock \doi{10.1038/nphys322}.

\bibitem{Liu14a}
Y.~Liu, D.~Kamburov, S.~Hasdemir, M.~Shayegan, L.~N. Pfeiffer, K.~W. West, and
  K.~W. Baldwin, Fractional quantum {Hall} effect and {Wigner} crystal of
  interacting composite fermions, \emph{Phys. Rev. Lett.} {\bf 113}, \penalty0
  246803 (Dec, 2014).
\newblock \doi{10.1103/PhysRevLett.113.246803}.
\newblock URL \url{http://link.aps.org/doi/10.1103/PhysRevLett.113.246803}.

\bibitem{Zhang15c}
C.~Zhang, R.-R. Du, M.~J. Manfra, L.~N. Pfeiffer, and K.~W. West, Transport of
  a sliding {Wigner} crystal in the four flux composite fermion regime,
  \emph{Phys. Rev. B}. {\bf 92}, \penalty0 075434 (Aug, 2015).
\newblock \doi{10.1103/PhysRevB.92.075434}.
\newblock URL \url{https://link.aps.org/doi/10.1103/PhysRevB.92.075434}.

\bibitem{Deng16}
H.~Deng, Y.~Liu, I.~Jo, L.~N. Pfeiffer, K.~W. West, K.~W. Baldwin, and
  M.~Shayegan, Commensurability oscillations of composite fermions induced by
  the periodic potential of a {Wigner} crystal, \emph{Phys. Rev. Lett.} {\bf
  117}, \penalty0 096601 (Aug, 2016).
\newblock \doi{10.1103/PhysRevLett.117.096601}.
\newblock URL \url{http://link.aps.org/doi/10.1103/PhysRevLett.117.096601}.

\bibitem{Jang17}
J.~Jang, B.~M. Hunt, L.~N. Pfeiffer, K.~W. West, and R.~C. Ashoori, Sharp
  tunnelling resonance from the vibrations of an electronic {Wigner} crystal,
  \emph{Nature Physics}. {\bf 13}\penalty0 (4), \penalty0 340--344 (APR, 2017).
\newblock ISSN 1745-2473.
\newblock \doi{10.1038/NPHYS3979}.

\bibitem{Chen19}
S.~Chen, R.~Ribeiro-Palau, K.~Yang, K.~Watanabe, T.~Taniguchi, J.~Hone, M.~O.
  Goerbig, and C.~R. Dean, Competing fractional quantum {Hall} and electron
  solid phases in graphene, \emph{Phys. Rev. Lett.} {\bf 122}, \penalty0 026802
  (Jan, 2019).
\newblock \doi{10.1103/PhysRevLett.122.026802}.
\newblock URL \url{https://link.aps.org/doi/10.1103/PhysRevLett.122.026802}.

\bibitem{Maryenko18}
D.~Maryenko, A.~McCollam, J.~Falson, Y.~Kozuka, J.~Bruin, U.~Zeitler, and
  M.~Kawasaki, Composite fermion liquid to a {Wigner} solid transition in the
  lowest {Landau} level of zinc oxide, \emph{Nature Commun.} {\bf 9}, \penalty0
  4356,  (2018).
\newblock \doi{10.1038/s41467-018-06834-6}.
\newblock URL \url{https://www.nature.com/articles/s41467-018-06834-6}.

\bibitem{Lam84}
P.~K. Lam and S.~M. Girvin, Liquid-solid transition and the fractional
  quantum-{Hall} effect, \emph{Phys. Rev. B}. {\bf 30}, \penalty0 473--475
  (Jul, 1984).
\newblock \doi{10.1103/PhysRevB.30.473}.

\bibitem{Levesque84}
D.~Levesque, J.~J. Weis, and A.~H. MacDonald, Crystallization of the
  incompressible quantum-fluid state of a two-dimensional electron gas in a
  strong magnetic field, \emph{Phys. Rev. B}. {\bf 30}, \penalty0 1056--1058
  (Jul, 1984).
\newblock \doi{10.1103/PhysRevB.30.1056}.

\bibitem{Zhu93}
X.~Zhu and S.~G. Louie, Wigner crystallization in the fractional quantum {Hall}
  regime: A variational quantum {Monte} {Carlo} study, \emph{Phys. Rev. Lett.}
  {\bf 70}, \penalty0 335--338 (Jan, 1993).
\newblock \doi{10.1103/PhysRevLett.70.335}.

\bibitem{Price93}
R.~Price, P.~M. Platzman, and S.~He, Fractional quantum {Hall} liquid, {Wigner}
  solid phase boundary at finite density and magnetic field, \emph{Phys. Rev.
  Lett.} {\bf 70}, \penalty0 339--342 (Jan, 1993).
\newblock \doi{10.1103/PhysRevLett.70.339}.

\bibitem{Platzman93}
P.~M. Platzman and R.~Price, Quantum freezing of the fractional quantum {Hall}
  liquid, \emph{Phys. Rev. Lett.} {\bf 70}, \penalty0 3487--3489 (May, 1993).
\newblock \doi{10.1103/PhysRevLett.70.3487}.

\bibitem{He05}
W.~J. He, T.~Cui, Y.~M. Ma, C.~B. Chen, Z.~M. Liu, and G.~T. Zou, Phase
  boundary between the fractional quantum {Hall} liquid and the {Wigner}
  crystal at low filling factors and low temperatures: A path integral {Monte}
  {Carlo} study, \emph{Phys. Rev. B}. {\bf 72}, \penalty0 195306 (Nov, 2005).
\newblock \doi{10.1103/PhysRevB.72.195306}.

\bibitem{Yi98}
H.~Yi and H.~A. Fertig, Laughlin-{{Jastrow}}-correlated {Wigner} crystal in a
  strong magnetic field, \emph{Phys. Rev. B}. {\bf 58}, \penalty0 4019--4027
  (Aug, 1998).
\newblock \doi{10.1103/PhysRevB.58.4019}.

\bibitem{Narevich01}
R.~Narevich, G.~Murthy, and H.~A. Fertig, Hamiltonian theory of the
  composite-fermion {Wigner} crystal, \emph{Phys. Rev. B}. {\bf 64}, \penalty0
  245326 (Dec, 2001).
\newblock \doi{10.1103/PhysRevB.64.245326}.

\bibitem{Chang05}
C.-C. Chang, G.~S. Jeon, and J.~K. Jain, Microscopic verification of
  topological electron-vortex binding in the lowest {Landau}-level crystal
  state, \emph{Phys. Rev. Lett.} {\bf 94}, \penalty0 016809 (Jan, 2005).
\newblock \doi{10.1103/PhysRevLett.94.016809}.

\bibitem{Archer13}
A.~C. Archer, K.~Park, and J.~K. Jain, Competing crystal phases in the lowest
  {Landau} level, \emph{Phys. Rev. Lett.} {\bf 111}, \penalty0 146804 (Oct,
  2013).
\newblock \doi{10.1103/PhysRevLett.111.146804}.

\bibitem{Yannouleas03}
C.~Yannouleas and U.~Landman, Two-dimensional quantum dots in high magnetic
  fields: Rotating-electron-molecule versus composite-fermion approach,
  \emph{Phys. Rev. B}. {\bf 68}, \penalty0 035326 (Jul, 2003).
\newblock \doi{10.1103/PhysRevB.68.035326}.
\newblock URL \url{http://link.aps.org/doi/10.1103/PhysRevB.68.035326}.

\bibitem{Lee01}
S.-Y. Lee, V.~W. Scarola, and J.~K. Jain, Stripe formation in the fractional
  quantum {Hall} regime, \emph{Phys. Rev. Lett.} {\bf 87}, \penalty0 256803
  (Nov, 2001).
\newblock \doi{10.1103/PhysRevLett.87.256803}.
\newblock URL \url{http://link.aps.org/doi/10.1103/PhysRevLett.87.256803}.

\bibitem{Archer11}
A.~C. Archer and J.~K. Jain, Static and dynamic properties of type-ii composite
  fermion {Wigner} crystals, \emph{Phys. Rev. B}. {\bf 84}, \penalty0 115139
  (Sep, 2011).
\newblock \doi{10.1103/PhysRevB.84.115139}.

\bibitem{Thomson04}
J.~J. Thomson, On the structure of the atom: an investigation of the stability
  and periods of oscillation of a number of corpuscles arranged at equal
  intervals around the circumference of a circle; with application of the
  results to the theory of atomic structure, \emph{Phil. Mag.} {\bf 7},
  \penalty0 237,  (1904).

\bibitem{Wales06}
D.~J. Wales and S.~Ulker, Structure and dynamics of spherical crystals
  characterized for the {Thomson} problem, \emph{Phys. Rev. B}. {\bf 74},
  \penalty0 212101 (Dec, 2006).
\newblock \doi{10.1103/PhysRevB.74.212101}.
\newblock URL \url{https://link.aps.org/doi/10.1103/PhysRevB.74.212101}.

\bibitem{Wales09}
D.~J. Wales, H.~McKay, and E.~L. Altschuler, Defect motifs for spherical
  topologies, \emph{Phys. Rev. B}. {\bf 79}, \penalty0 224115 (Jun, 2009).
\newblock \doi{10.1103/PhysRevB.79.224115}.
\newblock URL \url{https://link.aps.org/doi/10.1103/PhysRevB.79.224115}.

\bibitem{Thomson}
The minimum energy locations can be found at \url{http://thomson.phy.syr.edu/}.

\bibitem{Rhim15}
J.-W. Rhim, J.~K. Jain, and K.~Park, Analytical theory of strongly correlated
  {Wigner} crystals in the lowest {Landau} level, \emph{Phys. Rev. B}. {\bf
  92}, \penalty0 121103 (Sep, 2015).
\newblock \doi{10.1103/PhysRevB.92.121103}.
\newblock URL \url{https://link.aps.org/doi/10.1103/PhysRevB.92.121103}.

\bibitem{Zhu10}
H.~Zhu, Y.~P. Chen, P.~Jiang, L.~W. Engel, D.~C. Tsui, L.~N. Pfeiffer, and
  K.~W. West, Observation of a pinning mode in a {Wigner} solid with
  $\ensuremath{\nu}=1/3$ fractional quantum {Hall} excitations, \emph{Phys.
  Rev. Lett.} {\bf 105}, \penalty0 126803 (Sep, 2010).
\newblock \doi{10.1103/PhysRevLett.105.126803}.
\newblock URL \url{http://link.aps.org/doi/10.1103/PhysRevLett.105.126803}.

\bibitem{Shi18}
J.~Shi and W.~Ji, Dynamics of the {Wigner} crystal of composite particles,
  \emph{Phys. Rev. B}. {\bf 97}, \penalty0 125133 (Mar, 2018).
\newblock \doi{10.1103/PhysRevB.97.125133}.
\newblock URL \url{https://link.aps.org/doi/10.1103/PhysRevB.97.125133}.

\bibitem{Hu19}
Y.~Hu and J.~K. Jain, Kohn-{Sham} theory of the fractional quantum {Hall}
  effect, \emph{Phys. Rev. Lett.} {\bf 123}, \penalty0 176802 (Oct, 2019).
\newblock \doi{10.1103/PhysRevLett.123.176802}.
\newblock URL \url{https://link.aps.org/doi/10.1103/PhysRevLett.123.176802}.

\bibitem{Giuliani08}
G.~Giuliani and G.~Vignale, \emph{Quantum Theory of the Electron Liquid}.
  (Cambridge University Press, The Edinburgh Building, Cambridge CB2 2RU, UK,
  2008).

\bibitem{Ferconi95}
M.~Ferconi, M.~R. Geller, and G.~Vignale, Edge structure of fractional quantum
  {Hall} systems from density-functional theory, \emph{Phys. Rev. B}. {\bf 52},
  \penalty0 16357--16360 (Dec, 1995).
\newblock \doi{10.1103/PhysRevB.52.16357}.
\newblock URL \url{http://link.aps.org/doi/10.1103/PhysRevB.52.16357}.

\bibitem{Heinonen95}
O.~Heinonen, M.~I. Lubin, and M.~D. Johnson, Ensemble density functional theory
  of the fractional quantum {Hall} effect, \emph{Phys. Rev. Lett.} {\bf 75},
  \penalty0 4110--4113 (Nov, 1995).
\newblock \doi{10.1103/PhysRevLett.75.4110}.
\newblock URL \url{http://link.aps.org/doi/10.1103/PhysRevLett.75.4110}.

\bibitem{Zhao17}
J.~Zhao, M.~Thakurathi, M.~Jain, D.~Sen, and J.~K. Jain, Density-functional
  theory of the fractional quantum {Hall} effect, \emph{Phys. Rev. Lett.} {\bf
  118}, \penalty0 196802 (May, 2017).
\newblock \doi{10.1103/PhysRevLett.118.196802}.
\newblock URL \url{https://link.aps.org/doi/10.1103/PhysRevLett.118.196802}.

\bibitem{Zhang14}
Y.~Zhang, G.~J. Sreejith, N.~D. Gemelke, and J.~K. Jain, Fractional angular
  momentum in cold-atom systems, \emph{Phys. Rev. Lett.} {\bf 113}, \penalty0
  160404 (Oct, 2014).
\newblock \doi{10.1103/PhysRevLett.113.160404}.
\newblock URL \url{http://link.aps.org/doi/10.1103/PhysRevLett.113.160404}.

\bibitem{Grayce94}
C.~J. Grayce and R.~A. Harris, Magnetic-field density-functional theory,
  \emph{Physical Review A}. {\bf 50}\penalty0 (4), \penalty0 3089,  (1994).

\bibitem{kohn04}
W.~Kohn, A.~Savin, and C.~A. Ullrich, Hohenberg--kohn theory including spin
  magnetism and magnetic fields, \emph{International journal of quantum
  chemistry}. {\bf 100}\penalty0 (1), \penalty0 20--21,  (2004).

\bibitem{Tellgren12}
E.~I. Tellgren, S.~Kvaal, E.~Sagvolden, U.~Ekstr\"om, A.~M. Teale, and
  T.~Helgaker, Choice of basic variables in current-density-functional theory,
  \emph{Phys. Rev. A}. {\bf 86}, \penalty0 062506 (Dec, 2012).
\newblock \doi{10.1103/PhysRevA.86.062506}.
\newblock URL \url{https://link.aps.org/doi/10.1103/PhysRevA.86.062506}.

\bibitem{Tellgren18b}
E.~I. Tellgren, A.~Laestadius, T.~Helgaker, S.~Kvaal, and A.~M. Teale, Uniform
  magnetic fields in density-functional theory, \emph{The Journal of chemical
  physics}. {\bf 148}\penalty0 (2), \penalty0 024101,  (2018).

\bibitem{Levy79}
M.~Levy, Universal variational functionals of electron densities, first-order
  density matrices, and natural spin-orbitals and solution of the
  v-representability problem, \emph{Proceedings of the National Academy of
  Sciences}. {\bf 76}\penalty0 (12), \penalty0 6062--6065,  (1979).

\bibitem{Lieb83}
E.~H. Lieb, Density functionals for {Coulomb} systems, \emph{Int. J. Quantum
  Chem.} {\bf 24}, \penalty0 243,  (1983).

\bibitem{Seidl96}
A.~Seidl, A.~G\"orling, P.~Vogl, J.~A. Majewski, and M.~Levy, Generalized
  {Kohn}-{Sham} schemes and the band-gap problem, \emph{Phys. Rev. B}. {\bf
  53}, \penalty0 3764--3774 (Feb, 1996).
\newblock \doi{10.1103/PhysRevB.53.3764}.
\newblock URL \url{https://link.aps.org/doi/10.1103/PhysRevB.53.3764}.

\bibitem{Kummel08}
S.~K\"ummel and L.~Kronik, Orbital-dependent density functionals: Theory and
  applications, \emph{Rev. Mod. Phys.} {\bf 80}, \penalty0 3--60 (Jan, 2008).
\newblock \doi{10.1103/RevModPhys.80.3}.
\newblock URL \url{https://link.aps.org/doi/10.1103/RevModPhys.80.3}.

\bibitem{Bonsall77}
L.~Bonsall and A.~A. Maradudin, Some static and dynamical properties of a
  two-dimensional {Wigner} crystal, \emph{Phys. Rev. B}. {\bf 15}, \penalty0
  1959--1973 (Feb, 1977).
\newblock \doi{10.1103/PhysRevB.15.1959}.

\bibitem{Tsiper01}
E.~V. Tsiper and V.~J. Goldman, Formation of an edge striped phase in the
  $\ensuremath{\nu}=\frac{1}{3}$ fractional quantum {Hall} system, \emph{Phys.
  Rev. B}. {\bf 64}, \penalty0 165311 (Oct, 2001).
\newblock \doi{10.1103/PhysRevB.64.165311}.
\newblock URL \url{https://link.aps.org/doi/10.1103/PhysRevB.64.165311}.

\bibitem{Jain95}
J.~K. Jain and T.~Kawamura, Composite fermions in quantum dots, \emph{EPL
  (Europhysics Letters)}. {\bf 29}\penalty0 (4), \penalty0 321,  (1995).
\newblock URL \url{http://stacks.iop.org/0295-5075/29/i=4/a=009}.

\bibitem{Kaduk12}
B.~Kaduk, T.~Kowalczyk, and T.~Van~Voorhis, Constrained density functional
  theory, \emph{Chemical reviews}. {\bf 112}\penalty0 (1), \penalty0 321--370,
  (2011).

\bibitem{Jain89b}
J.~K. Jain, Incompressible quantum {Hall} states, \emph{Phys. Rev. B}. {\bf
  40}, \penalty0 8079--8082 (Oct, 1989).
\newblock \doi{10.1103/PhysRevB.40.8079}.
\newblock URL \url{http://link.aps.org/doi/10.1103/PhysRevB.40.8079}.

\bibitem{Jain90}
J.~K. Jain, Theory of the fractional quantum {Hall} effect, \emph{Phys. Rev.
  B}. {\bf 41}, \penalty0 7653--7665 (Apr, 1990).
\newblock \doi{10.1103/PhysRevB.41.7653}.

\bibitem{Wen91}
X.~G. Wen, Non-abelian statistics in the fractional quantum {Hall} states,
  \emph{Phys. Rev. Lett.} {\bf 66}, \penalty0 802--805 (Feb, 1991).
\newblock \doi{10.1103/PhysRevLett.66.802}.
\newblock URL \url{http://link.aps.org/doi/10.1103/PhysRevLett.66.802}.

\bibitem{Willett87}
R.~Willett, J.~P. Eisenstein, H.~L. St\"ormer, D.~C. Tsui, A.~C. Gossard, and
  J.~H. English, Observation of an even-denominator quantum number in the
  fractional quantum {Hall} effect, \emph{Phys. Rev. Lett.} {\bf 59}, \penalty0
  1776--1779 (Oct, 1987).
\newblock \doi{10.1103/PhysRevLett.59.1776}.
\newblock URL \url{http://link.aps.org/doi/10.1103/PhysRevLett.59.1776}.

\bibitem{Morf98}
R.~H. Morf, Transition from quantum {Hall} to compressible states in the second
  {Landau} level: New light on the $\nu=5/2$ enigma, \emph{Phys. Rev. Lett.}
  {\bf 80}, \penalty0 1505--1508 (Feb, 1998).
\newblock \doi{10.1103/PhysRevLett.80.1505}.
\newblock URL \url{http://link.aps.org/doi/10.1103/PhysRevLett.80.1505}.

\bibitem{Balram18}
A.~C. Balram, M.~Barkeshli, and M.~S. Rudner, Parton construction of a wave
  function in the anti-{Pfaffian} phase, \emph{Phys. Rev. B}. {\bf 98},
  \penalty0 035127 (Jul, 2018).
\newblock \doi{10.1103/PhysRevB.98.035127}.
\newblock URL \url{https://link.aps.org/doi/10.1103/PhysRevB.98.035127}.

\bibitem{Wu17b}
Y.~Wu, T.~Shi, and J.~K. Jain, Non-abelian parton fractional quantum {Hall}
  effect in multilayer graphene, \emph{Nano Letters}. {\bf 17}\penalty0 (8),
  \penalty0 4643--4647,  (2017).
\newblock \doi{10.1021/acs.nanolett.7b01080}.
\newblock URL \url{http://dx.doi.org/10.1021/acs.nanolett.7b01080}.
\newblock PMID: 28649831.

\bibitem{Blok90}
B.~Blok and X.~G. Wen, Effective theories of the fractional quantum {Hall}
  effect: Hierarchy construction, \emph{Phys. Rev. B}. {\bf 42}, \penalty0
  8145--8156 (Nov, 1990).
\newblock \doi{10.1103/PhysRevB.42.8145}.
\newblock URL \url{http://link.aps.org/doi/10.1103/PhysRevB.42.8145}.

\bibitem{Blok90b}
B.~Blok and X.~G. Wen, Effective theories of the fractional quantum {Hall}
  effect at generic filling fractions, \emph{Phys. Rev. B}. {\bf 42}, \penalty0
  8133--8144 (Nov, 1990).
\newblock \doi{10.1103/PhysRevB.42.8133}.
\newblock URL \url{http://link.aps.org/doi/10.1103/PhysRevB.42.8133}.

\bibitem{Wen92b}
X.-G. Wen, Theory of the edge states in fractional quantum {Hall} effects,
  \emph{International Journal of Modern Physics B}. {\bf 06}\penalty0 (10),
  \penalty0 1711--1762,  (1992).
\newblock \doi{10.1142/S0217979292000840}.
\newblock URL
  \url{http://www.worldscientific.com/doi/abs/10.1142/S0217979292000840}.

\bibitem{Balram18a}
A.~C. Balram, S.~Mukherjee, K.~Park, M.~Barkeshli, M.~S. Rudner, and J.~K.
  Jain, Fractional quantum {Hall} effect at $\ensuremath{\nu}=2+6/13$: The
  parton paradigm for the second {Landau} level, \emph{Phys. Rev. Lett.} {\bf
  121}, \penalty0 186601 (Nov, 2018).
\newblock \doi{10.1103/PhysRevLett.121.186601}.
\newblock URL \url{https://link.aps.org/doi/10.1103/PhysRevLett.121.186601}.

\bibitem{Wen91b}
X.~Wen, Edge excitations in the fractional quantum {Hall} states at general
  filling fractions, \emph{Modern Physics Letters B}. {\bf 05}\penalty0 (01),
  \penalty0 39--46,  (1991).
\newblock \doi{10.1142/S0217984991000058}.
\newblock URL
  \url{https://www.worldscientific.com/doi/abs/10.1142/S0217984991000058}.

\bibitem{Moore98}
J.~E. Moore and X.-G. Wen, Classification of disordered phases of quantum
  {Hall} edge states, \emph{Phys. Rev. B}. {\bf 57}, \penalty0 10138--10156
  (Apr, 1998).
\newblock \doi{10.1103/PhysRevB.57.10138}.
\newblock URL \url{http://link.aps.org/doi/10.1103/PhysRevB.57.10138}.

\bibitem{Kane97}
C.~L. Kane and M.~P.~A. Fisher, Quantized thermal transport in the fractional
  quantum {Hall} effect, \emph{Phys. Rev. B}. {\bf 55}, \penalty0 15832--15837
  (Jun, 1997).
\newblock \doi{10.1103/PhysRevB.55.15832}.
\newblock URL \url{http://link.aps.org/doi/10.1103/PhysRevB.55.15832}.

\bibitem{Wen99}
X.-G. Wen, Projective construction of non-abelian quantum {Hall} liquids,
  \emph{Phys. Rev. B}. {\bf 60}, \penalty0 8827--8838 (Sep, 1999).
\newblock \doi{10.1103/PhysRevB.60.8827}.
\newblock URL \url{https://link.aps.org/doi/10.1103/PhysRevB.60.8827}.

\bibitem{Barkeshli10}
M.~Barkeshli and X.-G. Wen, Effective field theory and projective construction
  for ${Z}_{k}$ parafermion fractional quantum {Hall} states, \emph{Phys. Rev.
  B}. {\bf 81}, \penalty0 155302 (Apr, 2010).
\newblock \doi{10.1103/PhysRevB.81.155302}.
\newblock URL \url{http://link.aps.org/doi/10.1103/PhysRevB.81.155302}.

\bibitem{Barkeshli14}
M.~Barkeshli and J.~McGreevy, Continuous transition between fractional quantum
  {Hall} and superfluid states, \emph{Phys. Rev. B}. {\bf 89}, \penalty0 235116
  (Jun, 2014).
\newblock \doi{10.1103/PhysRevB.89.235116}.
\newblock URL \url{http://link.aps.org/doi/10.1103/PhysRevB.89.235116}.

\bibitem{Repellin15}
C.~Repellin, T.~Neupert, B.~A. Bernevig, and N.~Regnault, Projective
  construction of the ${\mathbb{z}}_{k}$ {Read}-{Rezayi} fractional quantum
  {Hall} states and their excitations on the torus geometry, \emph{Phys. Rev.
  B}. {\bf 92}, \penalty0 115128 (Sep, 2015).
\newblock \doi{10.1103/PhysRevB.92.115128}.
\newblock URL \url{https://link.aps.org/doi/10.1103/PhysRevB.92.115128}.

\bibitem{Goldman19}
H.~Goldman, R.~Sohal, and E.~Fradkin, Landau-{Ginzburg} theories of non-abelian
  quantum {Hall} states from non-abelian bosonization, \emph{Phys. Rev. B}.
  {\bf 100}, \penalty0 115111 (Sep, 2019).
\newblock \doi{10.1103/PhysRevB.100.115111}.
\newblock URL \url{https://link.aps.org/doi/10.1103/PhysRevB.100.115111}.

\bibitem{Bandyopadhyay18}
S.~Bandyopadhyay, L.~Chen, M.~T. Ahari, G.~Ortiz, Z.~Nussinov, and A.~Seidel,
  Entangled {Pauli} principles: The {DNA} of quantum {Hall} fluids, \emph{Phys.
  Rev. B}. {\bf 98}, \penalty0 161118 (Oct, 2018).
\newblock \doi{10.1103/PhysRevB.98.161118}.
\newblock URL \url{https://link.aps.org/doi/10.1103/PhysRevB.98.161118}.

\bibitem{Wojs09}
A.~W\'ojs, Transition from abelian to non-abelian quantum liquids in the second
  {Landau} level, \emph{Phys. Rev. B}. {\bf 80}, \penalty0 041104 (Jul, 2009).
\newblock \doi{10.1103/PhysRevB.80.041104}.
\newblock URL \url{http://link.aps.org/doi/10.1103/PhysRevB.80.041104}.

\bibitem{Kumar10}
A.~Kumar, G.~A. Cs\'athy, M.~J. Manfra, L.~N. Pfeiffer, and K.~W. West,
  Nonconventional odd-denominator fractional quantum {Hall} states in the
  second {Landau} level, \emph{Phys. Rev. Lett.} {\bf 105}, \penalty0 246808
  (Dec, 2010).
\newblock \doi{10.1103/PhysRevLett.105.246808}.
\newblock URL \url{http://link.aps.org/doi/10.1103/PhysRevLett.105.246808}.

\bibitem{Shingla18}
V.~Shingla, E.~Kleinbaum, A.~Kumar, L.~N. Pfeiffer, K.~W. West, and G.~A.
  Cs\'athy, Finite-temperature behavior in the second {Landau} level of the
  two-dimensional electron gas, \emph{Phys. Rev. B}. {\bf 97}, \penalty0 241105
  (Jun, 2018).
\newblock \doi{10.1103/PhysRevB.97.241105}.
\newblock URL \url{https://link.aps.org/doi/10.1103/PhysRevB.97.241105}.

\bibitem{Read99}
N.~Read and E.~Rezayi, Beyond paired quantum {Hall} states: Parafermions and
  incompressible states in the first excited {Landau} level, \emph{Phys. Rev.
  B}. {\bf 59}, \penalty0 8084--8092 (Mar, 1999).
\newblock \doi{10.1103/PhysRevB.59.8084}.
\newblock URL \url{http://link.aps.org/doi/10.1103/PhysRevB.59.8084}.

\bibitem{Rezayi09}
E.~H. Rezayi and N.~Read, Non-abelian quantized {Hall} states of electrons at
  filling factors 12/5 and 13/5 in the first excited {Landau} level,
  \emph{Phys. Rev. B}. {\bf 79}, \penalty0 075306 (Feb, 2009).
\newblock \doi{10.1103/PhysRevB.79.075306}.
\newblock URL \url{http://link.aps.org/doi/10.1103/PhysRevB.79.075306}.

\bibitem{Sreejith13}
G.~J. Sreejith, Y.-H. Wu, A.~W\'ojs, and J.~K. Jain, Tripartite composite
  fermion states, \emph{Phys. Rev. B}. {\bf 87}, \penalty0 245125 (Jun, 2013).
\newblock \doi{10.1103/PhysRevB.87.245125}.
\newblock URL \url{http://link.aps.org/doi/10.1103/PhysRevB.87.245125}.

\bibitem{Zhu15}
W.~Zhu, S.~S. Gong, F.~D.~M. Haldane, and D.~N. Sheng, Fractional quantum
  {Hall} states at $\ensuremath{\nu}=13/5$ and $12/5$ and their non-abelian
  nature, \emph{Phys. Rev. Lett.} {\bf 115}, \penalty0 126805 (Sep, 2015).
\newblock \doi{10.1103/PhysRevLett.115.126805}.
\newblock URL \url{http://link.aps.org/doi/10.1103/PhysRevLett.115.126805}.

\bibitem{Mong15}
R.~S.~K. Mong, M.~P. Zaletel, F.~Pollmann, and Z.~Papi\ifmmode~\acute{c}\else
  \'{c}\fi{}, Fibonacci anyons and charge density order in the 12/5 and 13/5
  quantum {Hall} plateaus, \emph{Phys. Rev. B}. {\bf 95}, \penalty0 115136
  (Mar, 2017).
\newblock \doi{10.1103/PhysRevB.95.115136}.
\newblock URL \url{http://link.aps.org/doi/10.1103/PhysRevB.95.115136}.

\bibitem{Pakrouski16}
K.~Pakrouski, M.~Troyer, Y.-L. Wu, S.~Das~Sarma, and M.~R. Peterson, Enigmatic
  12/5 fractional quantum {Hall} effect, \emph{Phys. Rev. B}. {\bf 94},
  \penalty0 075108 (Aug, 2016).
\newblock \doi{10.1103/PhysRevB.94.075108}.
\newblock URL \url{http://link.aps.org/doi/10.1103/PhysRevB.94.075108}.

\bibitem{Zibrov16}
A.~A. {Zibrov}, C.~R. {Kometter}, H.~{Zhou}, E.~M. {Spanton}, T.~{Taniguchi},
  K.~{Watanabe}, M.~P. {Zaletel}, and A.~F. {Young}, Tunable interacting
  composite fermion phases in a half-filled bilayer-graphene {Landau} level,
  \emph{Nature}. {\bf 549}, \penalty0 360--364 (Sep, 2017).
\newblock \doi{10.1038/nature23893}.
\newblock URL
  \url{http://www.nature.com/nature/journal/v549/n7672/full/nature23893.html}.

\bibitem{Bid10}
A.~Bid, N.~Ofek, H.~Inoue, M.~Heiblum, C.~L. Kane, V.~Umansky, and D.~Mahalu,
  Observation of neutral modes in the fractional quantum {Hall} regime,
  \emph{Nature}. {\bf 466}\penalty0 (7306), \penalty0 585--590 (JUL 29, 2010).
\newblock ISSN 0028-0836.
\newblock \doi{10.1038/nature09277}.

\bibitem{Dolev11}
M.~Dolev, Y.~Gross, R.~Sabo, I.~Gurman, M.~Heiblum, V.~Umansky, and D.~Mahalu,
  Characterizing neutral modes of fractional states in the second {Landau}
  level, \emph{Phys. Rev. Lett.} {\bf 107}, \penalty0 036805 (Jul, 2011).
\newblock \doi{10.1103/PhysRevLett.107.036805}.
\newblock URL \url{http://link.aps.org/doi/10.1103/PhysRevLett.107.036805}.

\bibitem{Gross12}
Y.~Gross, M.~Dolev, M.~Heiblum, V.~Umansky, and D.~Mahalu, Upstream neutral
  modes in the fractional quantum {Hall} effect regime: Heat waves or coherent
  dipoles, \emph{Phys. Rev. Lett.} {\bf 108}, \penalty0 226801 (May, 2012).
\newblock \doi{10.1103/PhysRevLett.108.226801}.
\newblock URL \url{http://link.aps.org/doi/10.1103/PhysRevLett.108.226801}.

\bibitem{Inoue14}
H.~Inoue, A.~Grivnin, Y.~Ronen, M.~Heiblum, V.~Umansky, and D.~Mahalu,
  Proliferation of neutral modes in fractional quantum {Hall} states,
  \emph{Nature Communications}. {\bf 5}, \penalty0 4067 (JUN, 2014).
\newblock ISSN 2041-1723.
\newblock \doi{10.1038/ncomms5067}.

\bibitem{Banerjee17}
M.~Banerjee, M.~Heiblum, A.~Rosenblatt, Y.~Oreg, D.~E. Feldman, A.~Stern, and
  V.~Umansky, Observed quantization of anyonic heat flow, \emph{Nature}. {\bf
  545}\penalty0 (7652), \penalty0 75+ (MAY 4, 2017).
\newblock ISSN 0028-0836.
\newblock \doi{10.1038/nature22052}.

\bibitem{Banerjee17b}
M.~Banerjee, M.~Heiblum, V.~Umansky, D.~E. Feldman, Y.~Oreg, and A.~Stern,
  Observation of half-integer thermal {Hall} conductance, \emph{Nature}. {\bf
  559}, \penalty0 205--210,  (2018).
\newblock ISSN 1476-4687.
\newblock \doi{10.1038/s41586-018-0184-1}.
\newblock URL \url{https://doi.org/10.1038/s41586-018-0184-1}.

\bibitem{Levin09a}
M.~Levin and B.~I. Halperin, Collective states of non-abelian quasiparticles in
  a magnetic field, \emph{Phys. Rev. B}. {\bf 79}, \penalty0 205301 (May,
  2009).
\newblock \doi{10.1103/PhysRevB.79.205301}.
\newblock URL \url{http://link.aps.org/doi/10.1103/PhysRevB.79.205301}.

\bibitem{Kim18}
Y.~Kim, A.~C. Balram, T.~Taniguchi, K.~Watanabe, J.~K. Jain, and J.~H. Smet,
  Even denominator fractional quantum {Hall} states in higher {Landau} levels
  of graphene, \emph{Nature Physics}. {\bf 15}\penalty0 (2), \penalty0
  154--158,  (2019).
\newblock ISSN 1745-2481.
\newblock \doi{10.1038/s41567-018-0355-x}.
\newblock URL \url{https://doi.org/10.1038/s41567-018-0355-x}.

\bibitem{Trugman85}
S.~A. Trugman and S.~Kivelson, Exact results for the fractional quantum {Hall}
  effect with general interactions, \emph{Phys. Rev. B}. {\bf 31}, \penalty0
  5280--5284 (Apr, 1985).
\newblock \doi{10.1103/PhysRevB.31.5280}.
\newblock URL \url{http://link.aps.org/doi/10.1103/PhysRevB.31.5280}.

\bibitem{Rezayi91}
E.~H. Rezayi and A.~H. MacDonald, Origin of the $\nu${}=2/5 fractional quantum
  {Hall} effect, \emph{Phys. Rev. B}. {\bf 44}, \penalty0 8395--8398 (Oct,
  1991).
\newblock \doi{10.1103/PhysRevB.44.8395}.
\newblock URL \url{http://link.aps.org/doi/10.1103/PhysRevB.44.8395}.

\bibitem{McCann06}
E.~McCann and V.~I. Fal'ko, {Landau}-level degeneracy and quantum {Hall} effect
  in a graphite bilayer, \emph{Phys. Rev. Lett.} {\bf 96}, \penalty0 086805
  (Mar, 2006).
\newblock \doi{10.1103/PhysRevLett.96.086805}.
\newblock URL \url{http://link.aps.org/doi/10.1103/PhysRevLett.96.086805}.

\bibitem{Barlas12}
Y.~Barlas, K.~Yang, and A.~H. MacDonald, Quantum {Hall} effects in
  graphene-based two-dimensional electron systems, \emph{Nanotechnology}. {\bf
  23}\penalty0 (5), \penalty0 052001 (jan, 2012).
\newblock \doi{10.1088/0957-4484/23/5/052001}.
\newblock URL \url{https://doi.org/10.1088%2F0957-4484%2F23%2F5%2F052001}.

\bibitem{Kim15}
Y.~Kim, D.~S. Lee, S.~Jung, V.~Skákalová, T.~Taniguchi, K.~Watanabe, J.~S.
  Kim, and J.~H. Smet, Fractional quantum {Hall} states in bilayer graphene
  probed by transconductance fluctuations, \emph{Nano Letters}. {\bf
  15}\penalty0 (11), \penalty0 7445--7451,  (2015).
\newblock \doi{10.1021/acs.nanolett.5b02876}.
\newblock URL \url{http://dx.doi.org/10.1021/acs.nanolett.5b02876}.
\newblock PMID: 26479836.

\bibitem{Cote10}
R.~C\^ot\'e, W.~Luo, B.~Petrov, Y.~Barlas, and A.~H. MacDonald, Orbital and
  interlayer skyrmion crystals in bilayer graphene, \emph{Phys. Rev. B}. {\bf
  82}, \penalty0 245307 (Dec, 2010).
\newblock \doi{10.1103/PhysRevB.82.245307}.

\bibitem{Apalkov11}
V.~M. Apalkov and T.~Chakraborty, Stable {Pfaffian} state in bilayer graphene,
  \emph{Phys. Rev. Lett.} {\bf 107}, \penalty0 186803 (Oct, 2011).
\newblock \doi{10.1103/PhysRevLett.107.186803}.
\newblock URL \url{http://link.aps.org/doi/10.1103/PhysRevLett.107.186803}.

\bibitem{Snizhko12}
K.~Snizhko, V.~Cheianov, and S.~H. Simon, Importance of interband transitions
  for the fractional quantum {Hall} effect in bilayer graphene, \emph{Phys.
  Rev. B}. {\bf 85}, \penalty0 201415 (May, 2012).
\newblock \doi{10.1103/PhysRevB.85.201415}.
\newblock URL \url{http://link.aps.org/doi/10.1103/PhysRevB.85.201415}.

\bibitem{Bao10}
W.~Bao, Z.~Zhao, H.~Zhang, G.~Liu, P.~Kratz, L.~Jing, J.~Velasco, D.~Smirnov,
  and C.~N. Lau, Magnetoconductance oscillations and evidence for fractional
  quantum {Hall} states in suspended bilayer and trilayer graphene, \emph{Phys.
  Rev. Lett.} {\bf 105}, \penalty0 246601 (Dec, 2010).
\newblock \doi{10.1103/PhysRevLett.105.246601}.
\newblock URL \url{https://link.aps.org/doi/10.1103/PhysRevLett.105.246601}.

\bibitem{Luhman08}
D.~R. Luhman, W.~Pan, D.~C. Tsui, L.~N. Pfeiffer, K.~W. Baldwin, and K.~W.
  West, Observation of a fractional quantum {Hall} state at
  $\ensuremath{\nu}=1/4$ in a wide {Ga}{As} quantum well, \emph{Phys. Rev.
  Lett.} {\bf 101}, \penalty0 266804 (Dec, 2008).
\newblock \doi{10.1103/PhysRevLett.101.266804}.
\newblock URL \url{https://link.aps.org/doi/10.1103/PhysRevLett.101.266804}.

\bibitem{Shabani09a}
J.~Shabani, T.~Gokmen, and M.~Shayegan, Correlated states of electrons in wide
  quantum wells at low fillings: The role of charge distribution symmetry,
  \emph{Phys. Rev. Lett.} {\bf 103}, \penalty0 046805 (Jul, 2009).
\newblock \doi{10.1103/PhysRevLett.103.046805}.
\newblock URL \url{https://link.aps.org/doi/10.1103/PhysRevLett.103.046805}.

\bibitem{Faugno19}
W.~N. Faugno, A.~C. Balram, M.~Barkeshli, and J.~K. Jain, Prediction of a
  non-abelian fractional quantum {Hall} state with $f$-wave pairing of
  composite fermions in wide quantum wells, \emph{Phys. Rev. Lett.} {\bf 123},
  \penalty0 016802 (Jul, 2019).
\newblock \doi{10.1103/PhysRevLett.123.016802}.
\newblock URL \url{https://link.aps.org/doi/10.1103/PhysRevLett.123.016802}.

\bibitem{Shabani09b}
J.~Shabani, T.~Gokmen, Y.~T. Chiu, and M.~Shayegan, Evidence for developing
  fractional quantum {Hall} states at even denominator $1/2$ and $1/4$ fillings
  in asymmetric wide quantum wells, \emph{Phys. Rev. Lett.} {\bf 103},
  \penalty0 256802 (Dec, 2009).
\newblock \doi{10.1103/PhysRevLett.103.256802}.
\newblock URL \url{https://link.aps.org/doi/10.1103/PhysRevLett.103.256802}.

\bibitem{Shabani13}
J.~Shabani, Y.~Liu, M.~Shayegan, L.~N. Pfeiffer, K.~W. West, and K.~W. Baldwin,
  Phase diagrams for the stability of the $\ensuremath{\nu}=\frac{1}{2}$
  fractional quantum {Hall} effect in electron systems confined to symmetric,
  wide {Ga}{As} quantum wells, \emph{Phys. Rev. B}. {\bf 88}, \penalty0 245413
  (Dec, 2013).
\newblock \doi{10.1103/PhysRevB.88.245413}.
\newblock URL \url{https://link.aps.org/doi/10.1103/PhysRevB.88.245413}.

\bibitem{Balram19}
A.~C. Balram, M.~Barkeshli, and M.~S. Rudner, Parton construction of
  particle-hole-conjugate {Read}-{Rezayi} parafermion fractional quantum {Hall}
  states and beyond, \emph{Phys. Rev. B}. {\bf 99}, \penalty0 241108 (Jun,
  2019).
\newblock \doi{10.1103/PhysRevB.99.241108}.
\newblock URL \url{https://link.aps.org/doi/10.1103/PhysRevB.99.241108}.

\end{thebibliography}
\bibliographystyle{ws-rv-van}

\end{document}